\documentclass[reprint,amssymb,amsmath,aps,superscriptaddress,floatfix,pra,longbibliography,twocolumn]{revtex4-2}

\usepackage{bbold}
\usepackage{upgreek}
\usepackage{graphicx}
\usepackage{amssymb}
\usepackage{epstopdf}
\usepackage{upgreek}
\usepackage{dcolumn}
\usepackage{bm}
\usepackage{textcomp, gensymb}
\usepackage{braket}
\usepackage{amsmath}
\usepackage{verbatim}
\usepackage{mathdots}
\usepackage{float}
\usepackage{sidecap}
\usepackage{amsmath}
\usepackage[T1]{fontenc}
\usepackage{multirow}
\usepackage{array}
\usepackage{makecell}

\expandafter\ifx\csname package@font\endcsname\relax\else
 \expandafter\expandafter
 \expandafter\usepackage
 \expandafter\expandafter
 \expandafter{\csname package@font\endcsname}%
\fi

\newcommand{\be}{\begin{eqnarray}}
\newcommand{\ee}{\end{eqnarray}}
\newcommand{\bfig}{\begin{figure}}
\newcommand{\efig}{\end{figure}}

\makeatletter
\newcommand{\thickhline}{%
  \noalign {\ifnum 0=`}\fi \hrule height 2pt
  \futurelet \reserved@a \@xhline
}
\newcolumntype{"}{@{\hskip\tabcolsep\vrule width 2pt\hskip\tabcolsep}}
\makeatother

\begin{document}
\title{A high on-off ratio beamsplitter interaction for gates on bosonically encoded qubits}

\author{Benjamin J. Chapman}
\thanks{These authors contributed equally to this work.\\ benchapman@microsoft.com \\ stijn.degraaf@yale.edu \\ }
\affiliation{Departments of Physics and Applied Physics, Yale University, New Haven, Connecticut 06511, USA}
\affiliation{Yale Quantum Institute, Yale University, New Haven, Connecticut 06511, USA}

\author{Stijn J. de Graaf}
\thanks{These authors contributed equally to this work.\\ benchapman@microsoft.com \\ stijn.degraaf@yale.edu \\ }
\affiliation{Departments of Physics and Applied Physics, Yale University, New Haven, Connecticut 06511, USA}
\affiliation{Yale Quantum Institute, Yale University, New Haven, Connecticut 06511, USA}
\author{Sophia H. Xue}
\affiliation{Departments of Physics and Applied Physics, Yale University, New Haven, Connecticut 06511, USA}
\affiliation{Yale Quantum Institute, Yale University, New Haven, Connecticut 06511, USA}
\author{Yaxing Zhang}
\affiliation{Departments of Physics and Applied Physics, Yale University, New Haven, Connecticut 06511, USA}
\affiliation{Yale Quantum Institute, Yale University, New Haven, Connecticut 06511, USA}
\author{James Teoh}
\affiliation{Departments of Physics and Applied Physics, Yale University, New Haven, Connecticut 06511, USA}
\affiliation{Yale Quantum Institute, Yale University, New Haven, Connecticut 06511, USA}
\author{Jacob C. Curtis}
\affiliation{Departments of Physics and Applied Physics, Yale University, New Haven, Connecticut 06511, USA}
\affiliation{Yale Quantum Institute, Yale University, New Haven, Connecticut 06511, USA}
\author{Takahiro Tsunoda}
\affiliation{Departments of Physics and Applied Physics, Yale University, New Haven, Connecticut 06511, USA}
\affiliation{Yale Quantum Institute, Yale University, New Haven, Connecticut 06511, USA}
\author{Alec Eickbusch}
\affiliation{Departments of Physics and Applied Physics, Yale University, New Haven, Connecticut 06511, USA}
\affiliation{Yale Quantum Institute, Yale University, New Haven, Connecticut 06511, USA}
\author{Alexander P. Read}
\affiliation{Departments of Physics and Applied Physics, Yale University, New Haven, Connecticut 06511, USA}
\affiliation{Yale Quantum Institute, Yale University, New Haven, Connecticut 06511, USA}
\author{Akshay Koottandavida}
\affiliation{Departments of Physics and Applied Physics, Yale University, New Haven, Connecticut 06511, USA}
\affiliation{Yale Quantum Institute, Yale University, New Haven, Connecticut 06511, USA}
\author{Shantanu O. Mundhada}
\affiliation{Departments of Physics and Applied Physics, Yale University, New Haven, Connecticut 06511, USA}
\affiliation{Yale Quantum Institute, Yale University, New Haven, Connecticut 06511, USA}
\author{Luigi Frunzio}
\affiliation{Departments of Physics and Applied Physics, Yale University, New Haven, Connecticut 06511, USA}
\affiliation{Yale Quantum Institute, Yale University, New Haven, Connecticut 06511, USA}
\author{M. H. Devoret}
\affiliation{Departments of Physics and Applied Physics, Yale University, New Haven, Connecticut 06511, USA}
\affiliation{Yale Quantum Institute, Yale University, New Haven, Connecticut 06511, USA}
\author{S. M. Girvin}
\affiliation{Departments of Physics and Applied Physics, Yale University, New Haven, Connecticut 06511, USA}
\affiliation{Yale Quantum Institute, Yale University, New Haven, Connecticut 06511, USA}
\author{R. J. Schoelkopf}
\affiliation{Departments of Physics and Applied Physics, Yale University, New Haven, Connecticut 06511, USA}
\affiliation{Yale Quantum Institute, Yale University, New Haven, Connecticut 06511, USA}

\begin{abstract}
Encoding a qubit in a high-quality superconducting microwave cavity offers the opportunity to perform the first layer of error-correction in a single device, but presents a challenge: how can quantum oscillators be controlled while introducing a minimal number of additional error channels? We focus on the two-qubit portion of this control problem by using a three-wave mixing coupling element to engineer a programmable beamsplitter interaction between two bosonic modes separated by more than an octave in frequency, without introducing major additional sources of decoherence. Combining this with single-oscillator control provided by a dispersively coupled transmon provides a framework for quantum control of multiple encoded qubits. The beamsplitter interaction $g_\text{bs}$ is fast relative to the time scale of oscillator decoherence, enabling over $10^3$ beamsplitter operations per coherence time, and approaching the typical rate of the dispersive coupling $\chi$ used for individual oscillator control. Further, the programmable coupling is engineered without adding unwanted interactions between the oscillators, as evidenced by the high on-off ratio of the operations, which can exceed $10^5$. We then introduce a new protocol to realize a hybrid controlled-SWAP operation in the regime $g_{bs}\approx\chi$, in which a transmon provides the control bit for the SWAP of two bosonic modes. Finally, we use this gate in a SWAP test to project a pair of bosonic qubits into a Bell state with measurement-corrected fidelity of $95.5\% \pm 0.2\%$.
\end{abstract}
\maketitle

\section{Introduction}
\label{sec:Intro}

Bosonic codes~\cite{cochrane:1999,gottesman:2001,mirrahimi:2014,michael:2016,puri:2017,Chuang:1997}---quantum error-correction schemes in which a qubit is encoded in the multiple levels of an oscillator---offer the chance to perform error-correction with a single quantum mode. Along with the long lifetimes of superconducting microwave cavities~\cite{reagor:2013,romanenko:2020} and their simple error models, this hardware-efficiency suggests their use as a first layer of error-correction for future quantum processors~\cite{Cai:2021}.  

Storing quantum information in highly coherent oscillators presents a control challenge: the requisite nonlinearity must be introduced without adding significant error channels or unwanted Hamiltonian terms.
In circuit QED implementations~\cite{blais:2004}, universal control of a single bosonic qubit can be accomplished with resonant drives on an oscillator and a dispersively-coupled transmon~\cite{leghtas:2013,heeres:2017,fosel:2020,kudra:2021,eickbusch:2021}, 
but making this control first-order insensitive to errors on the ancilla~\cite{rosenblum:2018,reinhold:2020}, or developing alternative control methods based on noise-biased ancillas~\cite{puri:2019,grimm:2020}, is an area of active research. For example, when the ancilla is excited, ancilla decay can dephase the oscillator. In practice, this is frequently the dominant error in these systems~\cite{hu:2019,campagne:2020,sivak:2022,ni:2022}.

To perform two-qubit operations, another ancilla can be dispersively coupled to a pair of oscillators. Manipulating the state of the ancilla provides an immediate path toward operations between the bosonic qubits~\cite{wang:2016,rosenblum:2018}, but introduces additional error pathways when it decays or dephases, due to its entanglement with the oscillators.

An alternative approach for generating two-qubit operations is to engineer a purely bilinear oscillator-oscillator coupling with a circuit in which the quantum degrees of freedom need not be explicitly excited nor entangled with the oscillators~\footnote{A third approach is to engineer a noise-biased coupler~\cite{pietikainen:2021}.}. While such a scheme only allows Gaussian interactions between the oscillators, it is largely immune to decay or dephasing errors on the coupler, since ideally the coupler is only virtually excited. Heating of the coupler is then the only coupler error which causes a cavity error.
The non-Gaussian resource needed for this approach can be provided by the ancillas used for single-oscillator control, and therefore allows two-qubit gates to be performed without adding additional error-prone control hardware. This approach is analogous to well-known paradigms in linear optics in which universal control of continuous-variable systems is accomplished with Gaussian two-mode interactions and a non-Gaussian single-mode resource~\cite{lloyd:1999,knill:2001}, but here benefits from the comparatively strong nonlinearity of the dispersive shift.

Such a recipe for two-qubit operations on bosonically encoded qubits has been recently employed in the form of controlled SWAP (cSWAP) and exponential SWAP (eSWAP) gates~\cite{gao:2019}. The performance of these demonstrations, however, has been limited by the speed and fidelity of the engineered beamsplitter interaction~\cite{gao:2018}. Recent work suggests that spurious Hamiltonian interactions of the transmon coupling element have been responsible for these limitations~\cite{zhang:2019}. 

At the same time, efforts throughout the superconducting qubit community have shown the promise of designing coupling elements (sometimes called tunable couplers) to mediate inter-qubit interactions. These devices enable strong coupling while suppressing spurious $ZZ$ interactions between qubits, and can achieve on-off ratios exceeding $10^3$~\cite{chen:2014,yan:2018}. Suppressing such unwanted interactions is critical for quantum error-correction, which often relies on the assumption that errors are uncorrelated.

With this in mind, we improve on the efforts of Refs.~\cite{pfaff:2017,gao:2018,gao:2019} by designing a coupler based on a superconducting nonlinear asymmetric element (SNAIL) dipole element~\cite{frattini:2017}.  The SNAIL's potential energy can be made asymmetric with a static external flux to allow a pump-controlled beamsplitter interaction via 3-wave mixing~\cite{zhou:2021}, similar to that achieved with SQUIDs driven by a monochromatic flux-pump~\cite{zakka:2011,sirois:2015}.  Fine-tuning the the potential of the SNAIL coupler also allows in-situ suppression of higher-order terms in the coupler and oscillator Hamiltonians, such as cross-Kerr ($ZZ$-like) interactions between the oscillators. SNAIL-based couplers constructed in this way are therefore complementary to SQUID-based couplers driven by bichromatic flux-pumps, which offer less control of the static coupler Hamiltonian but suppress many unwanted pumped interactions with symmetry~\cite{lu:2023}.

Here, we use a SNAIL-based coupler to realize a programmable beamsplitter interaction between two bosonic modes with over an octave separation in frequency (Fig~\ref{fig:coupler}a). The interaction is fast relative to the time scale of oscillator decoherence, enabling over $10^3$ beamsplitter operations per coherence time. Further, the programmable coupling is engineered without adding unwanted interactions between the oscillators, as evidenced by the large ratio between the beamsplitter rate and the magnitude of the strongest spurious cavity-cavity interactions during idle periods, here called the on-off ratio. 
We find that the on-off ratio is always above $2\times10^3$ and can exceed $10^5$ at some operating points. To demonstrate the utility of the beamsplitter interaction as a primitive for constructing two-qubit gates, we combine it with a dispersively coupled transmon to realize a hybrid controlled-SWAP gate. We use this to create a Bell state with measurement-corrected fidelity of $95.5\%\,\pm\,0.2\%$.

\section{Coupler implementation and characterization}
\label{sec:coupler}
\begin{figure}[!thb]
\begin{center}
\includegraphics[width=0.95\linewidth]{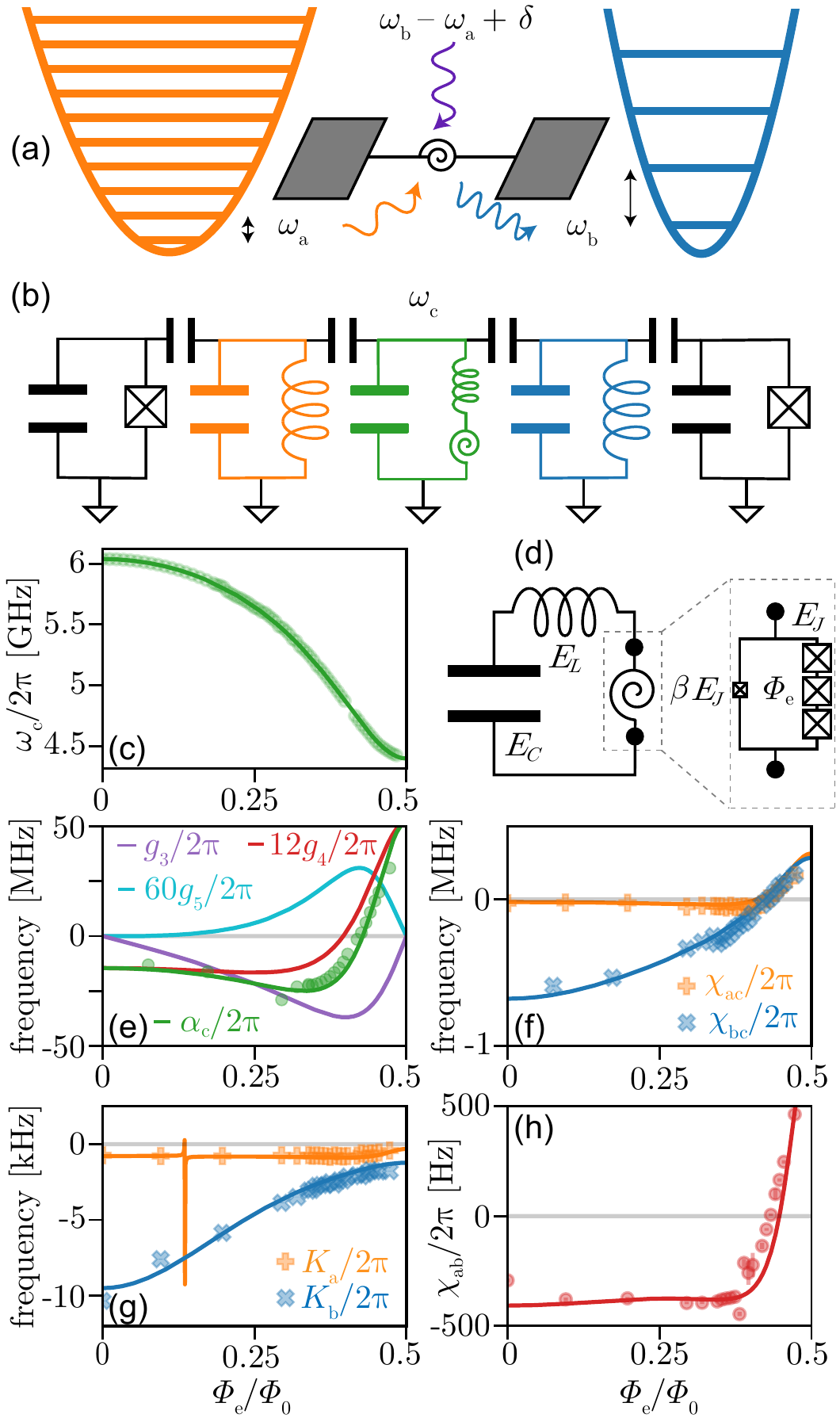}
\caption {\textbf{Concept and coupler characterization}. (a) A SNAIL-based~\cite{frattini:2017} three-wave mixing coupler exchanges photons between two high-quality superconducting microwave cavity modes with resonant frequencies of $3$ and $7$ GHz, at a rate proportional to the pump amplitude and with a phase set by the pump.  The process is enabled by a pump photon (purple) with frequency equal to the $4$ GHz detuning between the oscillators. (b) Effective circuit diagram. The cavity modes are each dispersively coupled to a transmon for readout and control.  Readout resonators and Purcell filters for the transmons are not pictured. 
(c) Measurements of the coupler frequency $\omega_c$ (circles) as a function of the external flux bias in units of $\Phi_0 = h/2e$. Solid line is a fit to the circuit model in (d)~\cite{frattini:2018}. The extracted parameters are: $E_J/h = 90.0\pm0.3$~GHz, $E_L/h = 64\pm2$~GHz, $E_C/h=177\pm2$~MHz, and SNAIL junction asymmetry, $\beta = 0.147\pm 0.001$. (e) Predicted Hamiltonian parameters, $g_3$, $g_4$, $g_5$ and anharmonicity of the coupler $\alpha_c$ (lines), and measured anharmonicity (circles). (f-h)
Measurements (markers) and predictions (lines, see App.~\ref{app:kerr_prediction}) of the cavity-coupler cross Kerrs (f), cavity self-Kerrs (g), and cavity-cavity cross-Kerr (h). The resonance in (g) occurs when $\omega_c = 2\omega_a$.
}
\label{fig:coupler}
\end{center}
\end{figure}

The beamsplitter Hamiltonian in the interaction picture
\begin{eqnarray}
\frac{\hat{\mathcal{H}}_\text{bs}}{\hbar} = g_\text{bs}(t)\left(e^{i\theta} \hat{a}^\dagger \hat{b} + e^{-i\theta} \hat{a} \hat{b}^\dagger\right),
\label{Hbs}
\end{eqnarray}
generates the coherent exchange of excitations between two bosonic modes with annihilation operators $\hat{a}$ and $\hat{b}$. We refer to these modes as Alice and Bob.
To realize this interaction between bosonically encoded qubits, we link a pair of high-quality microwave cavities with a three-wave mixing coupling element, and control them by coupling each cavity to its own transmon ancilla. Fig.~\ref{fig:coupler}b shows an effective circuit diagram of the system. 
For a simplified model of the superconducting package, see Fig.~\ref{fig:DeviceModel} and App.~\ref{app:exp_setup}.

Drawing inspiration from the quantum amplifier community~\cite{frattini:2018,liu:2017} and the insights of Refs.~\cite{gao:2018,zhang:2019}, we adopt a design philosophy that the coupler should act more like a switch than a qubit and strive to minimize all Hamiltonian interactions which are not needed to generate the desired beamsplitter interaction in Eq.~(\ref{Hbs}). Constructing the coupler from a SNAIL dipole element~\cite{frattini:2017} serves this goal, as the Hamiltonian of the coupler may be tuned in-situ to suppress parasitic fourth-order interactions~\cite{frattini:2018,sivak:2019} such as self- and cross-Kerr interactions that distort multiphoton cavity states, while preserving the third-order nonlinearity~$g_3$ that enables the pumped 3-wave mixing beamsplitter interaction $g_\text{bs}\propto g_3$ (see App.~\ref{app:H_bs}). 
We emphasize that the amplitude $g_\text{bs}$ and phase $\theta$ of the beamsplitter interaction are controlled by the amplitude and phase of the microwave pump, making the beamsplitter interaction in Eq.~(\ref{Hbs}) fully programmable.

To demonstrate that the SNAIL can be used to suppress parasitic quartic interactions while preserving the desired cubic interaction, we characterize the coupler Hamiltonian following Ref.~\cite{frattini:2018}. We first measure the resonant frequency of the coupler $\omega_c$ as a function of the external flux bias (Fig.~\ref{fig:coupler}c), which we deliver to the SNAIL via a superconducting flux transformer loop (see App.~\ref{app:flux}). We then fit a circuit model (Fig.~\ref{fig:coupler}d) to those measurements \footnote{We constrain that fit by probing the DC resistance of the coupler at room temperature, and using the Ambegaokar-Baratoff relation \cite{Ambegaokar:1963} to infer the kinetic inductance of the SNAIL from that measurement.}. 
Second quantization of the circuit Hamiltonian and an expansion to fourth order then yields
\begin{eqnarray}
\frac{\hat{\mathcal{H}_\text{c}}}{\hbar} \approx \omega_c \hat{c}^\dagger \hat{c} + g_3 \left(\hat{c} + \hat{c}^\dagger \right)^3 + g_4 \left(\hat{c} + \hat{c}^\dagger \right)^4,
\label{Hc}
\end{eqnarray}
where $\hat{c}$ is the annihilation operator associated with the coupler mode. 
Here $g_3$ and $g_4$ are the third and fourth order nonlinearities of the coupler, both of which are functions of external flux and the fitted circuit parameters. Fig.~\ref{fig:coupler}e shows the resulting prediction for their dependence on external flux. (See App.~\ref{app:design_considerations} for discussion on the design choices of SNAIL circuit parameters.) 

Crucially, it is possible to choose the external flux to 
suppress $g_4$ while $g_3$ remains large. 
Suppressing the fourth-order interactions provides several advantages: it reduces the self- and cross-Kerr interactions of the cavities, which distort multiphoton cavity states and can therefore be a source of coherent errors; it reduces the anharmonicity of the coupler, which mitigates pump-induced decoherence effects~\cite{zhang:2019}; and it reduces the cross-Kerr interaction between the cavities and the coupler itself, which mitigates cavity dephasing from thermal noise in the coupler.

To verify these predictions, we measure the anharmonicity $\alpha_\text{c}$ of the coupler, defined as the difference between its $\ket{g}\leftrightarrow\ket{e}$ and $\ket{e}\leftrightarrow\ket{f}$ transition frequencies. (Here $\ket{g}, \ket{e},$ and $\ket{f}$ denote the first three energy eigenstates of the coupler mode.) Fig.~\ref{fig:coupler}e shows these measurements agree well with predictions made with perturbation theory in the small parameter $(p\phi_c)$~\cite{frattini:2021}:
\begin{eqnarray}
\alpha_\text{c} = 12\left(g_4 - 5\frac{g_3^2}{\omega_c}\right) + \mathcal{O}\left(p \phi_c\right)^4,
\label{Kerr}
\end{eqnarray}
where $\phi_c$ is the zero-point phase fluctuation of the coupler mode and $p$ is the inductive participation of the SNAIL junctions in this mode  (i.e. the fraction of the inductance arising from the junctions rather than the geometric inductance of the leads, see App.~\ref{app:design_considerations}).

Conveniently, the nonlinear parameters in the coupler Hamiltonian $\hat{\mathcal{H}}_c$ can also be used to predict the dependence of the joint cavity Hamiltonian $\hat{\mathcal{H}}$ on the external flux applied to the SNAIL, provided the linear coupling rates of each oscillator to the coupler $g_a$ and $g_b$ are known. We infer these by measuring the dependence of the cavity frequencies on the external flux (see App.~\ref{app:lincoupling}). Expanded to fourth order, the joint cavity Hamiltonian takes the form
\begin{eqnarray}
\label{eqn:H}
\hat{\mathcal{H}}/\hbar &\approx&~\omega_a \hat{a}^\dagger\hat{a} + \chi_{\text{a}} \hat{a}^\dagger\hat{a} \left(\ket{e}\bra{e}\right)_\text{a} \\ \nonumber
&& + \omega_b \hat{b}^\dagger\hat{b} + \chi_{\text{b}} \hat{b}^\dagger\hat{b} \left(\ket{e}\bra{e}\right)_\text{b} + \hat{\mathcal{H}}_\text{bs}/\hbar\\ \nonumber
&& + \frac{K_\text{a}}{2} \hat{a}^{\dagger^2}\hat{a}^2 + \frac{K_\text{b}}{2} \hat{b}^{\dagger^2}\hat{b}^2 + \hat{\mathcal{H}}_\text{Ss} \\ \nonumber
&& + \chi_\text{ab} \hat{a}^\dagger\hat{a}\hat{b}^\dagger\hat{b} + \left(\chi_{\text{ac}} \hat{a}^\dagger\hat{a} + \chi_{\text{bc}} \hat{b}^\dagger\hat{b} \right) \left(\ket{e}\bra{e}\right)_\text{c}.
\end{eqnarray}
Here the first two lines contain the desired terms: a pair of oscillators dispersively coupled to transmon ancillas for individual cavity control, and a beamsplitter interaction to couple the oscillators together. The cavity self-Kerr and Stark-shift interactions $\hat{\mathcal{H}}_{Ss}$ on the third line are parasitic, as are the cross-Kerr interactions on the fourth line. 

To visualize the flux dependence of $\hat{\mathcal{H}}$, the predicted and measured self- and cross-Kerrs are shown in Fig.~\ref{fig:coupler}f-h. The cavity self-Kerr interactions (Fig.~\ref{fig:coupler}g) are determined by the nonlinearity they inherit from both the coupler and from their transmon ancillas (see App.~\ref{app:kerr_prediction}). In our system there is no value of external flux that nulls them, but future designs could accomplish this by neutralizing the negative anharmonicity from the transmons with a positive anharmonicity from the coupler. To first order, the cavity Stark shifts are proportional to the self-Kerrs, and could be nulled in the same way.

In contrast, the cavity cross-Kerr interactions (Fig.~\ref{fig:coupler}f,h) are dictated solely by the coupler Hamiltonian. To first order, they are proportional to $g_4$, as evident in the change of their sign when the external flux approaches $\approx0.4 \Phi_0$, a feature that allows any of these interactions to be nulled via fine-tuning of the external flux. Higher-order corrections, however, make the precise value of external flux which nulls these interactions slightly different from the external flux at which $g_4$ vanishes~\cite{frattini:2018}, as well as the external flux at which the other interactions vanish. These effects mean that many of the quartic interactions can be suppressed simultaneously, but only one of the interactions can be finely nulled with a single value of the external flux.

In summary, the external flux bias of the SNAIL coupler serves a dual purpose. First, it breaks the symmetry of the confining potential of the SNAIL, enabling three-wave mixing and thus a switchable coupling between the cavities. At the same time, it provides a means of suppressing the undesired quartic interactions induced by the coupler, which give rise to the always-on cross-Kerr interaction $\chi_\text{ab}$ between the cavities, as well as the anharmonicity of the coupler.  

\section{The beamsplitter interaction}
\label{sec:uncond}
Ideally, a tunable coupler creates a hierarchy of rates. When the coupler is `on' the coupling should be strong, both in absolute terms (to reduce algorithm run-times) and relative to the time scale of decoherence. Conversely, when the coupler is `off', it should provide ample isolation; the interaction rates between the coupled modes should be small relative to the time scale of decoherence. This last condition ensures that errors due to residual coupling are subdominant. 

In our system, the cavities are strongly detuned, so the largest cavity-cavity Hamiltonian interaction when the coupler is `off' is the cross-Kerr interaction $\chi_\text{ab}$, which describes a $ZZ$-like interaction between the cavities. The desired hierarchy of rates is then
\begin{eqnarray}
\chi_\text{ab} \ll \tau_\text{bs}^{-1} \ll g_\text{bs},
\label{hierarchy}
\end{eqnarray}
where $\tau_\text{bs}$ is a time scale for cavity decoherence in the presence of the beamsplitter interaction. We can quantify the degree to which this hierarchy is enforced with two dimensionless figures of merit. Anticipating its use in pulsed operations, we note that the 50:50 beamsplitter time (half the time required to swap a photon between the cavities) sets a representative time-scale for the interaction: $t_\text{bs} \equiv \frac{\pi}{4 g_\text{bs}}$. The first figure of merit is therefore the number of beamsplitter operations that can be performed in one coherence time, $\tau_\text{bs}/t_\text{bs}$. The second is the on-off ratio $g_\text{bs}/\chi_\text{ab}$.

To measure these performance metrics, we turn on the beamsplitter coupling by applying a microwave pump tone with a frequency near to the detuning between the Stark-shifted resonance frequencies of the oscillators. As illustrated in Fig.~\ref{fig:coupler}a, the pump supplies (or removes) the energy required to convert a photon from one oscillator to the other. When the pump frequency is precisely equal to the difference of the two Stark-shifted oscillator frequencies, we call it resonant.

\begin{figure}[!tb]
\begin{center}
\includegraphics[width=1\linewidth]{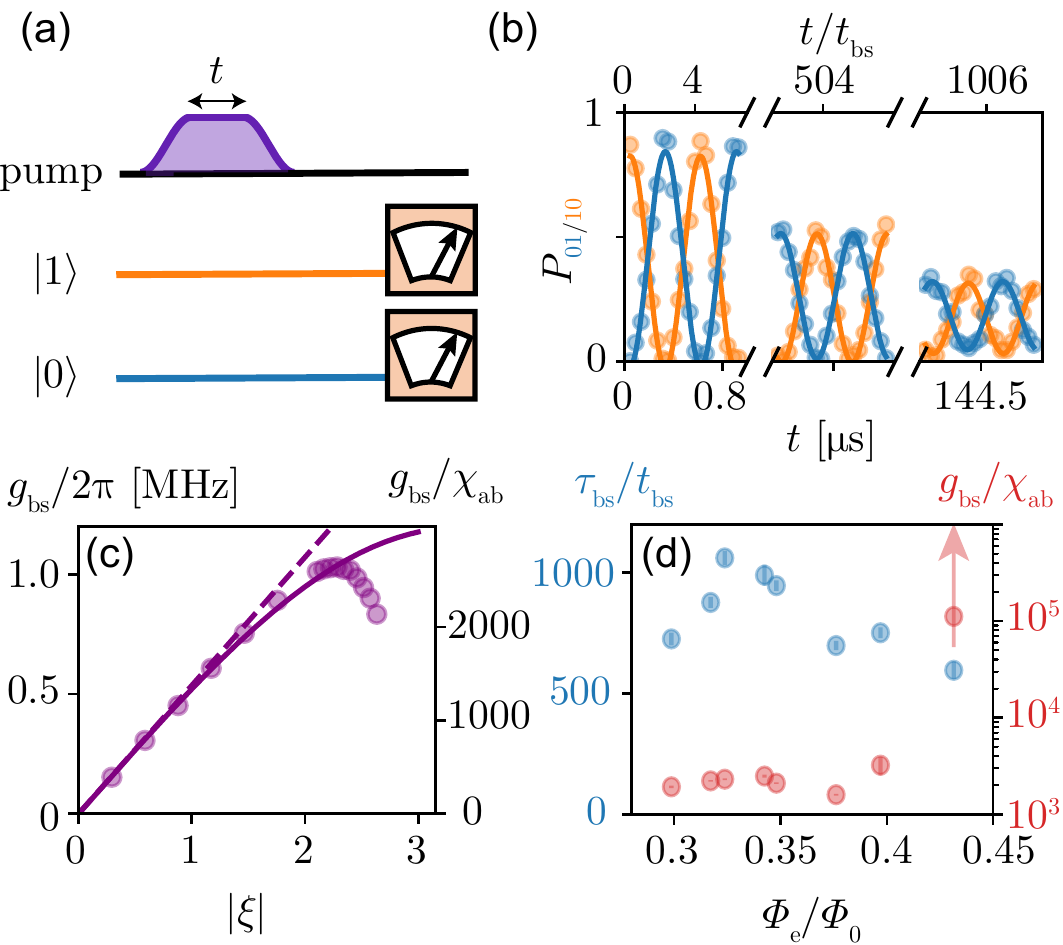}
\caption {\textbf{3-wave mixing beamsplitter}.
(a) Pulse sequence for measuring population exchange between the cavities. With an external flux applied to the coupler, the cavities are prepared in $\ket{0,1}$ and the pump is applied resonantly for a variable time $t$. The cavities are then measured to determine if they are in vacuum and the outcomes are correlated shot-by-shot to infer the joint probability distribution of the photon population in the oscillators (see App.~\ref{app:joint_prob_dist}). (b) The probability of finding one photon in Alice and none in Bob $P_{10}$ (orange), and vice-versa $P_{01}$ (blue), as a function of time. Solid lines are fits to Eq.~(\ref{P10}). See Fig.~\ref{fig:taus}c for full dataset without axis breaks. (c) The fitted beamsplitter rate $g_\text{bs}$ (left axis) and the on-off ratio (right axis), as a function of dimensionless pump amplitude $\xi$, with external flux $\Phi_\text{e}/\Phi_0 = 0.35$. Circles show experimental data, the dashed line shows a prediction which is perturbative in the drive strength and treats the coupler Hamiltonian to third order, and the solid line shows a perturbation theory prediction that accounts for resonant odd-order nonlinear interactions in the coupler Hamiltonian up to $g_{17}$~\cite{miano:2023} (see App.~\ref{app:H_bs} for details). (d) The maximum number of beamsplitters that can be achieved per coherence time (left axis) and the corresponding on-off ratio (right axis) at each value of the external flux $\Phi_\text{e}$.
}
\label{fig:concept}
\end{center}
\end{figure}

We measure the rate of the beamsplitter interaction and the decoherence of the cavities while the beamsplitter is `on' with the sequence in Fig.~\ref{fig:concept}a. To avoid coherent distortions of the cavity states caused by self- and cross-Kerr interactions, we characterize it in the single-photon manifold of the cavities. The cavities are first prepared in Fock state $\ket{0,1}$ and a resonant pump is applied for a variable time $t$. The overlap of each cavity state with vacuum is then measured and the outcomes are correlated shot-by-shot to infer the photon number probability distribution of the joint cavity state (see App.~\ref{app:joint_prob_dist} for details).  Fig.~\ref{fig:concept}b shows the probability $P_{10}$ of finding a photon in Alice and none in Bob (orange), or vice-versa ($P_{01}$, blue). The two traces have equal amplitude and opposite phase, indicating that the prepared photon is exchanged between the two bosonic modes. The expected dynamics are~\cite{gao:2018}
\begin{eqnarray}
P_{01/10} &=& \frac{1}{2}e^{-t/\tau}\left( 1 \pm e^{-t/\tau_\phi} \cos{\left(2 g_{bs} t + \theta\right)}\right) \nonumber \\ && + \mathcal{O}\left(\frac{\kappa_\text{a} - \kappa_\text{b}}{2g_\text{bs}}\right),
\label{P10}
\end{eqnarray}
where $\kappa_{\text{a}}$ and $\kappa_{\text{b}}$ are the decay rates of Alice and Bob, respectively.
Fitting to Eq.~(\ref{P10}) (see App.~\ref{app:pumped_decoherence} for details), we extract the beamsplitter rate $g_\text{bs}$, the time-scale for decay from the single-photon manifold $\tau$, and the time-scale for dephasing $\tau_{\phi}$ in the presence of the beamsplitter drive. 

Fig.~\ref{fig:concept}c shows the beamsplitter rates extracted from a series of these measurements with different dimensionless pump amplitudes $\xi$, which can be interpreted as the classical displacement of the coupler by the pump. As expected for a three-wave mixing process, initially the beamsplitter rate scales linearly with the pump amplitude, $g_\text{bs}\propto g_3 \xi$, in agreement with the perturbation theory estimate (dashed line). As the amplitude increases, however, the expansion of the coupler Hamiltonian becomes less accurate, and $g_\text{bs}$ deviates from this linear scaling. 

One possible explanation for the observed turn-around in $g_\text{bs}$ is the presence of odd higher-order nonlinearities in the Hamiltonian of the coupler that can give rise to resonant beamsplitter interactions that compete with $g_3$. For example, the first correction of this kind adjusts the beamsplitter interaction by an amount proportional to $g_5 |\xi|^2 \xi$. This correction reduces the magnitude of the beamsplitter rate because $g_3$ and $g_5$ have opposite sign. The solid line in Fig.~\ref{fig:concept}c shows a prediction of $g_\text{bs}$ accounting for odd-order rotating wave approximation (RWA) processes up to $g_{17}$ (see Eq.~(\ref{full_gbs})), and indicates that while this explanation is consistent with a slow-down, it fails to predict the sharp turn-around, and suggests the importance of capturing non-RWA terms and treating the coupler Hamiltonian non-perturbatively~\cite{Venkatraman:2022}, for example in the charge basis.   

Nevertheless, at its maximum, $g_\text{bs}/2\pi$ is greater than 1 MHz. This is approximately a factor of 50 faster than the beamsplitter rate used in Ref.~\cite{gao:2019}, and allows a 50:50 beamsplitter (or half of a SWAP) to occur in $t_\text{bs}\approx 125$ ns. 

To determine the on-off ratio, we compare this beamsplitter rate to the cross-Kerr interaction between the cavities, $\chi_\text{ab}$. When the external flux is biased to $\Phi_\text{e}/\Phi_0 = 0.32$, the measured cavity-cavity cross-Kerr is $\chi_\text{ab}/2\pi=-390\pm10$ Hz (see App.~\ref{app:crosskerr} for measurement details), allowing the on-off ratio to exceed $2\times10^3$. 

Fits to Eq.~(\ref{P10}) also allow extraction of the decay times $\tau$ and dephasing times $\tau_\phi$ (see Fig.~\ref{fig:taus} in App.~\ref{app:pumped_decoherence}). 
The fitted values of $\tau$ are relatively independent of the pump amplitude and near the inverse of the average cavity decay rate. The dephasing times $\tau_{\phi}$ are also relatively independent of pump amplitude, and are much longer than the decay times $\tau$. The decoherence in the presence of the beamsplitter is therefore cavity-decay limited, which is significant because photon loss is typically the error that bosonic codes are designed to correct~\cite{cochrane:1999,mirrahimi:2014,michael:2016,puri:2017}. The dominant error during the beamsplitter gate is therefore correctable.

For times $t\ll \text{min}\left(\tau,\tau_\phi\right)$, the fidelity of the time-evolved state diminishes at the composite time-scale $\tau_{bs}^{-1} \equiv \frac{1}{\tau} + \frac{1}{2\tau_{\phi}}$~\cite{lu:2023}.
The unconventional weighting given here to decay and dephasing can be understood from the perspective that unlike a qubit which cannot decay from its ground state, the single-photon manifold of the cavities is always subject to decay. We take $\tau_\text{bs}$ as a representative time scale for cavity decoherence, and compare it to the beamsplitter time $t_\text{bs}$ to quantify the number of beamsplitters that can be performed in a single coherence time. With $\Phi_\text{e}/\Phi_0 = 0.32$ and a strong drive, $\tau_\text{bs}/t_\text{bs}>10^3$, indicating that high fidelity gates could be achieved with this beamsplitter interaction.

To determine the flux bias which best enforces the hierarchy in Eq.~(\ref{hierarchy}), we repeat this procedure for many values of the external flux. Fig.~\ref{fig:concept}d shows the number of beamsplitters per coherence time (left axis), and the on-off ratio (right axis) as a function of external flux. For a given value of the external flux, we choose the pump amplitude that maximizes the beamsplitters per coherence time and plot the on-off ratio achieved at that pump amplitude. The quoted figures of merit are therefore achieved simultaneously.

Two clear operation points are visible. The first, near $\Phi_\text{e}/\Phi_0 = 0.32$, yields the performance presented in Fig.~\ref{fig:concept}b and optimizes $\tau_{\text{bs}}/ t_{\text{bs}}$. The second operation point occurs at the external flux $\Phi_\text{e}/\Phi_0 = 0.43$ which nulls $\chi_\text{ab}$, and optimizes the on-off ratio $g_\text{bs} / \chi_\text{ab}$. 
To locate it, we resolve the cross-Kerr far below either of the individual linewidths of the cavities by making a large displacement in the other cavity (see App.~\ref{app:crosskerr} for details). With this technique, we null $|\chi_\text{ab}|/2\pi<10$~Hz. Choosing $\Phi_\text{e}/\Phi_0 = 0.43$ still allows over $700$ beamsplitters per coherence time, and due to the suppression of $\chi_\text{ab}$, elevates the on-off ratio to over $10^5$. In particular, we expect such an operating point to find use in applications with special sensitivity to self and cross-Kerrs, for example CNOT gates between qubits encoded in grid states~\cite{gottesman:2001}. More broadly, this exceptional on-off ratio is highly-attractive from a quantum error-correction standpoint, as uncorrelated errors are an assumption for many forms of error-correction. 


 
\section{Hybrid controlled-SWAP of bosonically encoded qubits}
\label{sec:cond}

The fast beamsplitter rate engineered with this 3-wave mixing coupler brings the system into a regime where $t_{bs}$ is short not only compared to the cavity coherence times but also to those of the transmon ancillas (see App.~\ref{app:params}). This regime, in which $T_2/t_{bs}\approx 400\gg 1$, enables high fidelity two-qubit gates constructed from the combination of the beamsplitter interaction and single-cavity dispersive control. 

To demonstrate this we perform a hybrid controlled-SWAP (cSWAP) gate on two bosonic qubits, in which the transmon coupled to Bob encodes the control bit:
\begin{eqnarray}
\text{cSWAP} \equiv \ket{g}\bra{g}\otimes \mathbb{1} + \ket{e}\bra{e}\otimes \text{SWAP}.
\label{cSWAP}
\end{eqnarray}
(The beamsplitter interaction can also be used to realize a SWAP controlled on the state of a bosonically encoded qubit---see App.~\ref{app:3cav_cSWAP} for a proposed sequence for this gate). cSWAP, also known as a quantum Fredkin gate~\cite{milburn:1989}, is a key operation for quantum implementations of random access memory (QRAM)~\cite{giovannetti:2008a,giovannetti:2008b} and fundamental state-comparison and purification algorithms such as the SWAP test~\cite{ekert:2002,abanin:2012,nguyen:2021,carrasco:2021}. Previously, it has been demonstrated probabilistically in optical platforms~\cite{patel:2016} and deterministically in circuit QED~\cite{gao:2019}, but performance there was limited by a less favorable ratio between transmon coherence and the beamsplitter time $T_2/t_{bs}\approx7$. 

\begin{figure*}[!thb]
\begin{center}
\includegraphics[width=1\linewidth]{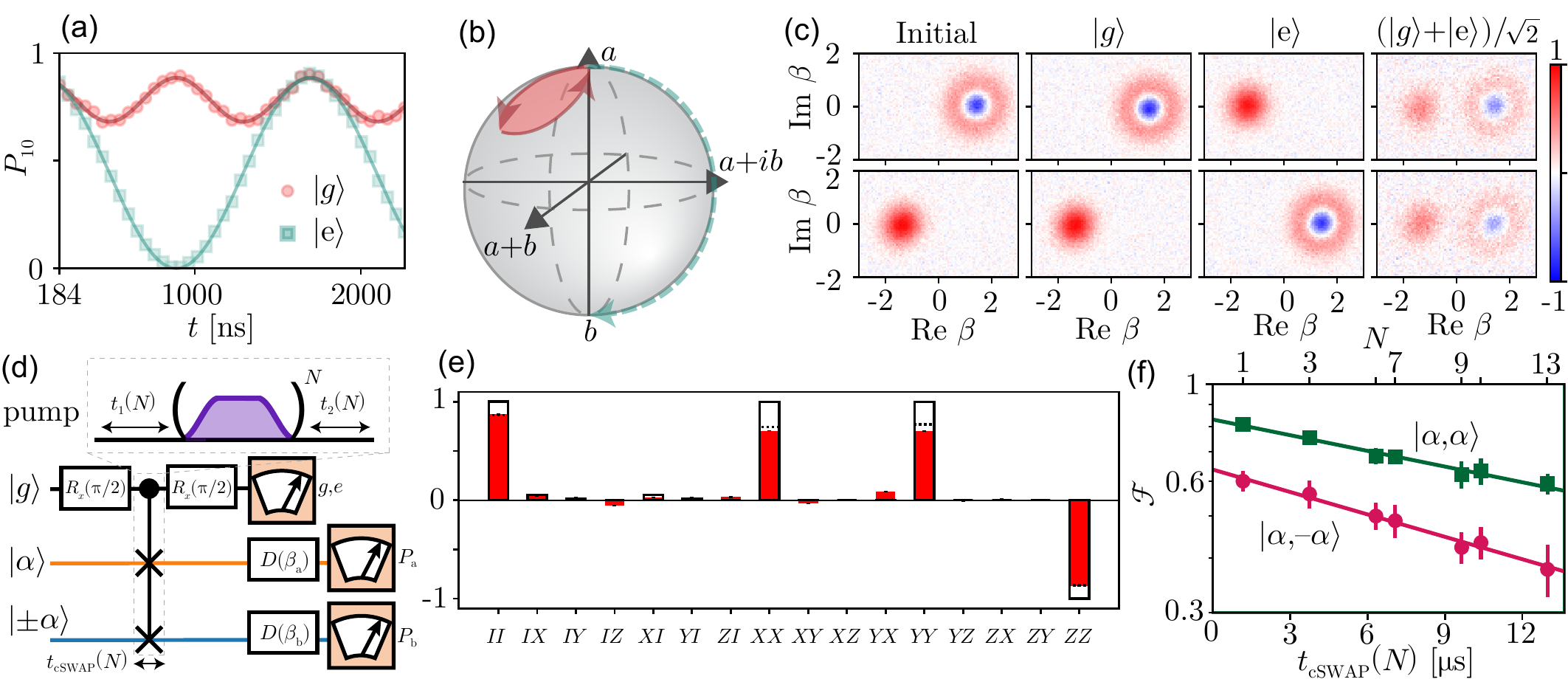}
\caption {\textbf{Concept and performance of the conditional SWAP}. (a) Measured probability $P_{10}$ after applying the beamsplitter pump with $\delta = -|\chi|$ and $g_\text{bs} = \frac{\chi}{2\sqrt{3}}$ for time $t$, with the control transmon prepared in $\ket{g}$ (red circles) or $\ket{e}$ (turquoise squares). Lines are fits to Eq.~(\ref{P10}). (b) Illustration of the evolution of the oscillator operator $\hat{a}$ in the Heisenberg picture during the demonstrated cSWAP protocol. The SU(2) algebra generated by the detuned beamsplitter Hamiltonian $\mathcal{H}_{\Delta bs}$ allows these dynamics to be illustrated on a sphere.  When the control is in $\ket{g}$, $\hat{a}$ traces a complete loop and $\hat{a}\to\hat{a}$ (solid red). When the control is in $\ket{e}$, the cavity operators exchange and $\hat{a}\to\hat{b}$ (dashed turquoise). Evolution of the $\hat{b}$ operator follows the same trajectories inverted through the origin. (c) Wigner tomography of Alice (top row) and Bob (bottom row) initialized in $D(\alpha) \ket{1}$ and $\ket{-\alpha}$, and after performing a cSWAP with the control in $\ket{g}$, $\ket{e}$, and $\left(\ket{g} + \ket{e}\right)/\sqrt{2}$. (d) Gate sequence for a SWAP test followed by Wigner tomography, with a variable number $N$ of cSWAPs. App.~\ref{app:delay_calibration} details how to calibrate the delays $t_1(N)$ and $t_2(N)$ required to account for the control-state-dependent phase. (e) Joint Pauli measurements on $\ket{\Psi^+}$ (red) with $N=1$, compared to their ideal (solid) and expected (dashed) values. See App.~\ref{app:Paulibars} for details on the tomography and rescaling of $YY$ bar, and App.~\ref{app:error_budget} for calculation of the expected values in the presence of decoherence. (f) Bell fidelity $\mathcal{F}$ (log scale) as a function of the odd number $N$ of cSWAPs used in the SWAP test (upper $x$-axis) and the duration of the $N$ cSWAPs including pre- and post-delays (lower $x$-axis), with the oscillators prepared in $\ket{\alpha,-\alpha}$ (circles). The upper $x$-axis is unevenly spaced because the delays $t_1(N)$ and $t_2(N)$ depend on $N$. Squares show fidelity to $\ket{\alpha, \alpha}$ when oscillators are prepared in $\ket{\alpha, \alpha}$, as a control. Solid lines are linear fits; their slope is used to infer the SPAM-corrected infidelity of the cSWAP, in analogy with randomized benchmarking.
}
\label{fig:cSWAP}
\end{center}
\end{figure*}

In our system, the control for the cSWAP arises from the dependence of the beamsplitter resonance condition on the frequency of each oscillator (see Fig.~\ref{fig:coupler}a), which in turn depends on the state of its transmon ancilla through dispersive coupling. For example, when Bob's ancilla is excited, the frequency at which the beamsplitter interaction is resonant shifts by $\chi_{b}$ (hereafter simply $\chi$). Pumping at $\omega_b-\omega_a+\chi$ for $t_{\text{gate}} = 2t_{\text{bs}}$ enacts a SWAP when the ancilla is in $\ket{e}$. The oscillator dynamics when the ancilla is in $\ket{g}$, however, depend on the relative strengths of $g_{\text{bs}}$ and $\chi$. A photon evolving under the detuned beamsplitter Hamiltonian $\hat{\mathcal{H}}_{\Delta\text{bs}} \equiv \hat{\mathcal{H}}_\text{bs} +\Delta \hat{a}^\dagger \hat{a}$ will swap between the oscillators at a rate $\Omega$ and with contrast $C\equiv\text{max}(P_{10}) - \text{min}(P_{10})$ given by
\begin{eqnarray}
  \Omega &=& \sqrt{g_{bs}^2 + \frac{\Delta^2}{4}}, \label{eqn:omega} \\ \nonumber 
  C &=& \frac{g_{bs}^2}{\Omega^2},
\end{eqnarray}
in analogy to a detuned Rabi drive on a qubit. $\Delta$ is the detuning of the beamsplitter drive from the ancilla-state-dependent resonance condition (for cSWAP, $\Delta = \chi$ when the ancilla is in $\ket{g}$). If $g_{bs}\ll|\chi|$, the contrast is approximately zero, and the desired identity operation is achieved, albeit slowly~\cite{gao:2019}. A new approach is required, however, when $g_\text{bs}$ approaches $|\chi|$, as now the cavity populations will exchange significantly when the control is in $\ket{g}$ (see Fig.~\ref{fig:cSWAP}a). 

This problem can be partially resolved by fine-tuning the beamsplitter rate to ensure that when the ancilla is in $\ket{g}$, the cavities undergo exactly two detuned SWAPs during $t_{\text{gate}}$. To give a geometric interpretation of this process, we formalize the analogy to Rabi oscillations by moving to the Heisenberg picture. In that representation, evolution under the detuned beamsplitter Hamiltonian
can be expressed as a unitary matrix acting on the vector of operators $\left(\begin{smallmatrix}\hat{a} \\ \hat{b}\end{smallmatrix}\right)$. As the operator evolution obeys an SU(2) algebra~\cite{schwinger:1965}, it can be visualized with trajectories on a Bloch sphere, like those shown in Fig.~\ref{fig:cSWAP}b~\cite{tsunoda:2022}. Under $\hat{\mathcal{H}}_{\Delta\text{bs}}$, the operators rotate at a rate $\Omega$ about an axis
\begin{eqnarray}
\vec{n} = \frac{1}{\Omega}\begin{pmatrix}
 g_{bs} \cos \theta \\
 g_{bs} \sin \theta \\
 \Delta/2
\end{pmatrix}.
\label{eqn:n_vector}
\end{eqnarray}
To ensure that the operator execute precisely two detuned SWAPs during $t_{\text{gate}}$ -- equivalent to the identity up to a phase space rotation on both cavities -- when the ancilla is in $\ket{g}$, we need to set the rate at which photon evolving under the detuned beamsplitter Hamiltonian $\hat{\mathcal{H}}_{\Delta\text{bs}}$ to be twice the rate it under the resonant beamslitter Hamiltonian $\hat{\mathcal{H}}_\text{bs}$, i.e. $\Omega = 2g_{bs}$. This condition is satisfied by setting $g_{bs} = |\chi|/\left(2\sqrt{3}\right)$ \cite{tsunoda:2022}. This is illustrated in Fig.~\ref{fig:cSWAP}a when $t_\text{gate} = 750$ ns. The phase space rotation, whose angle corresponds to half the solid angle enclosed by the red trajectory in Fig.~\ref{fig:cSWAP}b, is only imparted to the oscillators when the control is in $\ket{g}$. To account for this, we add a delay before and after the beamsplitter drive during which the dispersive shift acts, imparting an equal phase space rotation to both oscillators when the control is in $\ket{e}$. This results in an overall phase space rotation on the oscillators that is independent of the control state and can thus be removed with local operations (see App.~\ref{app:delay_calibration} for details).

In this experiment, the dispersive coupling $\chi/2\pi = -1.104$ MHz and so the required beamsplitter rate $g_{bs} = |\chi|/(2\sqrt{3}) = 2\pi\times319$ kHz. Accounting for the pre- and post-delays, as well as pulse ramp times to ensure finite bandwidth, the total gate time is $1.3~\upmu$s and is limited by the dispersive shift $\chi$.

To ensure that the phase correction is properly calibrated, we perform this fast cSWAP protocol on the displaced Fock states $D(\alpha)\ket{1}\otimes D(-\alpha)\ket{0}$ with $\alpha = \sqrt{2}$ (see left column of Fig.~\ref{fig:cSWAP}c). These multiphoton initial states are both sensitive to phase space rotations and distinguishable under arbitrary rotation, such that a phase space rotation on each oscillator cannot be mistaken for a SWAP (as would be the case for $\ket{\alpha, -\alpha}$). 

The central columns of Fig.~\ref{fig:cSWAP}c show local Wigner tomography on each oscillator after enacting a cSWAP with the control initialized in $\ket{g}$ and $\ket{e}$, and serves as a truth table for the gate. The oscillator states emerge unchanged when the control is in $\ket{g}$, but are exchanged without distortion when the control is in $\ket{e}$. 

Importantly, the quantum cSWAP gate is not merely a controlled SWAP on two quantum bits with a classical control; both targets and the control are quantum bits. To demonstrate this, we perform a SWAP test by preparing the control in a superposition state $(\ket{g} + \ket{e})/\sqrt{2}$, and applying cSWAP followed by a second $\pi/2$ pulse and a measurement of the control (see Fig.~\ref{fig:jw}a in App.~\ref{app:jw}). This protocol projects the oscillators into an eigenstate of SWAP, which is entangled with the state of the control. Measuring the control in $\ket{g}$ means that the oscillators are in the $+1$ eigenstate of SWAP; the unitary $(\mathbb{1} + \text{SWAP})/\sqrt{2}$ has acted on the joint oscillator state. Local Wigner tomography, unable to reveal entanglement between oscillators, shows mixed states (right column of Fig.~\ref{fig:cSWAP}c) composed equally of the initial states of Alice and Bob, with half the contrast of the prior columns. In both oscillators, the two displaced Fock states appear on opposite sides of the tomogram, confirming that there is no residual control-state-dependent phase on the oscillator states.

To observe the inter-oscillator entanglement, we replace local Wigner tomography with a measurement of the joint Wigner function of the oscillators, in which the joint parity is obtained by correlating local parity measurements shot-by-shot:
\begin{align}
  W(\beta_A, \beta_B) = &\braket{D(\beta_A, \beta_B) P_A P_B D^{\dag}(\beta_A, \beta_B)}.
\end{align}
Here $P_{A,B}\equiv e^{i\pi\hat{n}_\text{A,B}}$ is the photon number parity operator in each oscillator. Having validated the phase calibration, we encode a qubit in the oscillator using the mapping $\{\ket{\alpha},\ket{-\alpha}\}$ to represent logical states $\{\ket{0_L},\ket{1_L}\}$. In this basis it is possible to measure two-qubit Pauli operators without resorting to density matrix reconstruction methods~\cite{vlastakis:2015,wang:2016} (see App.~\ref{app:Paulibars} for details). We prepare the oscillators in $\ket{\alpha,-\alpha}$, such that a SWAP test produces a Bell state in this basis $\ket{\Psi^{\pm}} \equiv \mathcal{N}\left(\ket{\alpha,-\alpha} \pm \ket{-\alpha,\alpha}\right)$ whose parity is conditioned on the measurement of the control. (Here $\mathcal{N}\approx 1/\sqrt{2}$ is a normalization factor). Post-selecting on measuring $\ket{g}$ yields $\ket{\Psi^+}$.

Visualizing the full joint Wigner function of $\ket{\Psi^+}$ is difficult because it depends on two complex variables and is thus 4-dimensional, but the signatures of entanglement can be observed in slices in the real-real or imaginary-imaginary planes (see Fig.~\ref{fig:jw}b-c in App.~\ref{app:jw}). These match qualitatively with the ideal simulated case (Fig.~\ref{fig:jw}d-e in App.~\ref{app:jw}). The interference fringes in the imaginary-imaginary slice reveal the inter-cavity entanglement generated by the SWAP test. 

To quantify this entanglement, we directly measure the expectation values of all 16 two-qubit Pauli operators $\langle \Psi^+|\sigma_{i,\text{A}} \sigma_{j,\text{B}}|\Psi^+\rangle$ ($i,j=1,2,3$ or $4$) in the coherent-state basis by sampling the joint Wigner function at 16 pairs of displacements $(\beta_\text{A},\beta_\text{B})$~\cite{vlastakis:2015,wang:2016}. For an ideal Bell state, simultaneous measurements of the same single-qubit Pauli operator on each qubit are perfectly positively or negatively correlated, while all other pairs of Pauli measurements are uncorrelated: $|\langle \Psi^+|\sigma_{i,\text{A}} \sigma_{j,\text{B}}|\Psi^+\rangle|~=~\delta_{i,j}$ (the Kronecker-delta). In this experiment, where the basis states for the Bell pair are quasi-orthogonal coherent states with $\alpha = \sqrt{2}$, one caveat is that the measured value of $\braket{YY}$ is ideally only $|\braket{\frac{i\pi}{8\alpha}|-\frac{i\pi}{8\alpha}}|^2\approx 0.73$ (see App.~\ref{app:Paulibars}). This non-ideality arises because not all Pauli operators can be perfectly expressed as simple combinations of points in the measured Wigner function. It is the price paid for efficiently extracting the expectation values of the Pauli operators.

In Fig.~\ref{fig:cSWAP}e we compare these best-case outcomes for $\ket{\Psi^+}$ to our measurements. The fidelity to the ideal Bell state $\mathcal{F} \equiv \frac{1}{4}\left(\braket{II} + \braket{XX} + \braket{YY} - \braket{ZZ}\right)$ is 74.1\% without correcting for state-preparation and measurement (SPAM) errors, which exceeds the classical limit of $50\%$~\cite{wang:2016}.

To identify the origin of the infidelity, we estimate and simulate the effects of transmon and cavity decoherence using independently measured system parameters (see App.~\ref{app:error_budget} for details of the error budget). The expected values of joint Pauli measurements, indicated by the dashed lines in Fig.~\ref{fig:cSWAP}e, yield an overall estimated fidelity within 2\% of the measured value, and suggest that the cSWAP gate is decoherence-limited. Furthermore, the error budget suggests that the bulk of the errors arise from measurement of the joint oscillator state, and not from the cSWAP gate itself.

To confirm this, we perform the Bell state preparation sequence while replacing the cSWAP with an odd number $N$ of cSWAPs, as shown in Fig.~\ref{fig:cSWAP}d. Since cSWAP squares to the identity, this allows us to enhance the effect of gate errors while leaving the measurement unchanged. By fitting the exponential decay of the Bell fidelity as we increase the duration of the $N$ cSWAP sequence, we extract a measurement-corrected infidelity of $4.5\%\,\pm\, 0.2\%$. 
 This value, obtained by multiplying the fractional decrease in fidelity per unit time from the fitted slope by the total duration of 1 cSWAP (including delays), agrees with the predicted value of $4.2\%\,\pm\,0.1\%$ from our error budget, which suggests cavity and ancilla errors are approximately equally responsible for the infidelity.

As a control experiment, we run the same protocol with $\ket{\alpha, \alpha}$ as our initial cavity state. Since this state is already an eigenstate of SWAP and therefore ideally unchanged by a SWAP test, it is less prone to many of the error mechanisms that limited the fidelity of $\ket{\Psi^+}$, and results in a lower measurement-corrected infidelity to the target state of $3.2\%\,\pm\,0.2\%$, in agreement with our error budget. 

\section{Conclusion}
SNAIL-based couplers can expand oscillator-control capabilities without adding significant sources of error because they need only be virtually excited to facilitate the coupling. Here we show how such a coupler enables a fast and programmable three-wave mixing beamsplitter interaction which does not degrade oscillator coherence and allows over $10^3$ consecutive coherent beamsplitter operations. 

Importantly, the strong oscillator-oscillator coupling $g_\text{bs}$ is achieved without introducing unwanted interactions, as evidenced by its high on-off ratio, which can exceed $10^5$. The coupler therefore satisfies the desired hierarchy of rates $\chi_\text{ab}\ll \tau_{\text{bs}}^{-1} \ll g_\text{bs}$.

To demonstrate its utility as a primitive for two-qubit bosonic gates, we describe a new protocol that leverages the beamsplitter in a deterministic hybrid controlled-SWAP gate in which a transmon serves as the control qubit. We use this to perform a SWAP test on coherent states with opposite phase, projecting the oscillators into a Bell state with a measurement-corrected infidelity of $4.5\%\,\pm\,0.2\%$.

Looking forward, the hybrid controlled SWAP operation can be further optimized using established control techniques such as GRAPE~\cite{khaneja2005} or REBURP~\cite{geen1991}, or implemented with the canonical construction of beamsplitter, controlled parity, and inverse beamsplitter (see App.~\ref{app:cSWAP_comparison}). Extending it to allow the control qubit to be provided by a bosonically encoded qubit (for example with the proposed sequence in App.~\ref{app:3cav_cSWAP}) will benefit QRAM implementations with bosonic codes. The beamsplitter interaction engineered in this work is also the first step in performing a universal gate set in circuit-QED dual rail qubits, where dominant errors can be detected, leaving background Pauli errors that are orders of magnitude smaller \cite{teoh:2022}. Finally, the principles used to construct the hybrid cSWAP presented here can be generalized to design families of error-detectable two-qubit gates on bosonically encoded qubits~\cite{tsunoda:2022}.
As SNAIL-based couplers need not themselves add significant errors, such gates enable the long coherence times in superconducting cavities to translate to higher two-qubit gate fidelities with bosonic qubits, paving a path to hardware-efficient quantum error-correction.

\vspace{0.1in}
We acknowledge N. Ofek, P. Reinhold, and Y. Liu for their work building the FPGA firmware and software, V. Sivak for providing the Josephson Array Mode Parametric Amplifier (JAMPA), and V. Joshi, G. Liu, and M. Malnou for their work on the design and assembly of the lumped element SNAIL parametric amplifiers (LSPAs) used in the experiments. The LSPAs were fabricated by the NIST Quantum Sensors Group with support from the NIST program on scalable superconducting computing. We thank A. Miano for help computing nonlinear SNAIL parameters at higher orders. We also thank N. Frattini, C. Wang, C. Hann, R. Andrews, L. Burkhart, J. Venkatraman, X. Xiao, Y. Lu, A. Maiti, K. Chou, W. Kalfus, P. Lu, and C. Zhou for useful discussions. Use of facilities was supported by the Yale Institute for Nanoscience and Quantum Engineering (YINQE), Y. Sun, M. Rooks, and the Yale SEAS cleanroom. Research was sponsored by the Army Research Office (ARO), under Grant Number W911NF-18-1-0212. The views and conclusions contained in this document are those of the authors and should not be interpreted as representing the official parties, either expressed or implied, of the Army Research Office (ARO) or the U.S. Government. The U.S. Government is authorized to reproduce and distribute reprints for government purposes notwithstanding any copyright notation herein. Disclosure: L.F., M.H.D, and R.J.S. are founders of Quantum Circuits Inc. (QCI) and L.F. and R.J.S are shareholders of QCI.

%

\appendix

\section{Flux delivery in an all-superconducting package}
\label{app:flux}
\begin{figure}[!tb]
\begin{center}
\includegraphics[width=1\linewidth]{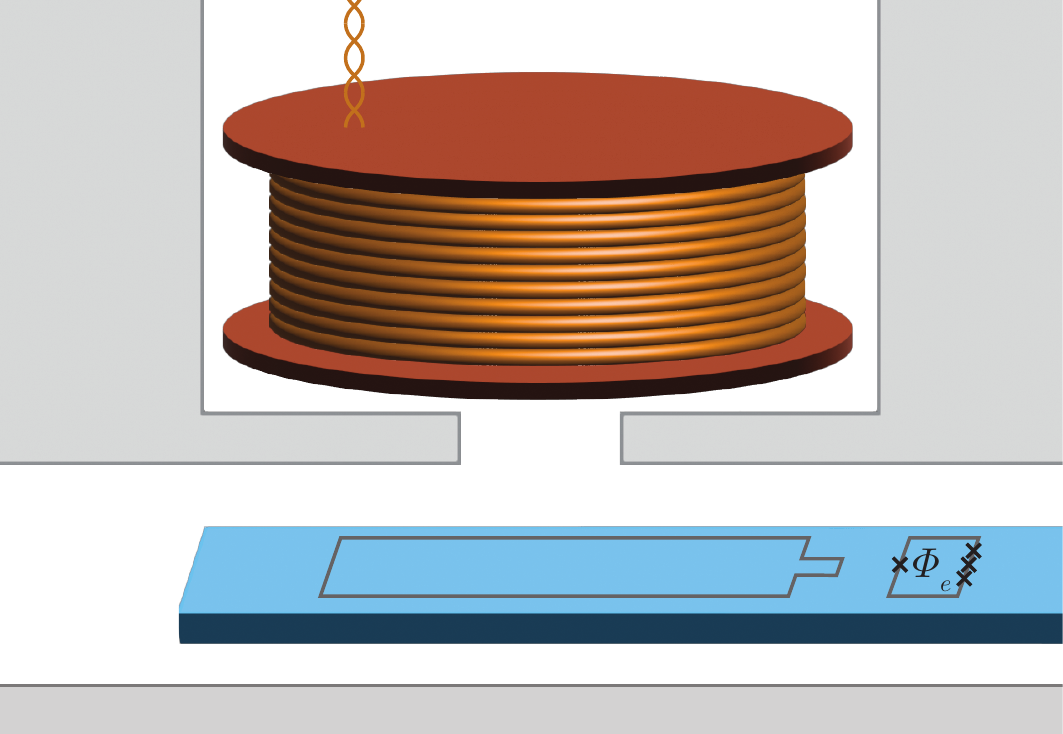}
\caption {\textbf{Flux delivery with a superconducting pick-up loop}. Cartoon schematic of the superconducting pick-up loop used to transfer magnetic flux from the superconducting coil to the SNAIL loop. Diagram is not to scale. For simplicity, the pads and leads of the SNAIL coupler circuit are omitted, and only the SNAIL loop is shown.
}
\label{fig:flux_transformer}
\end{center}
\end{figure}

\begin{figure}[!thb]
\begin{center}
\includegraphics[width=1\linewidth]{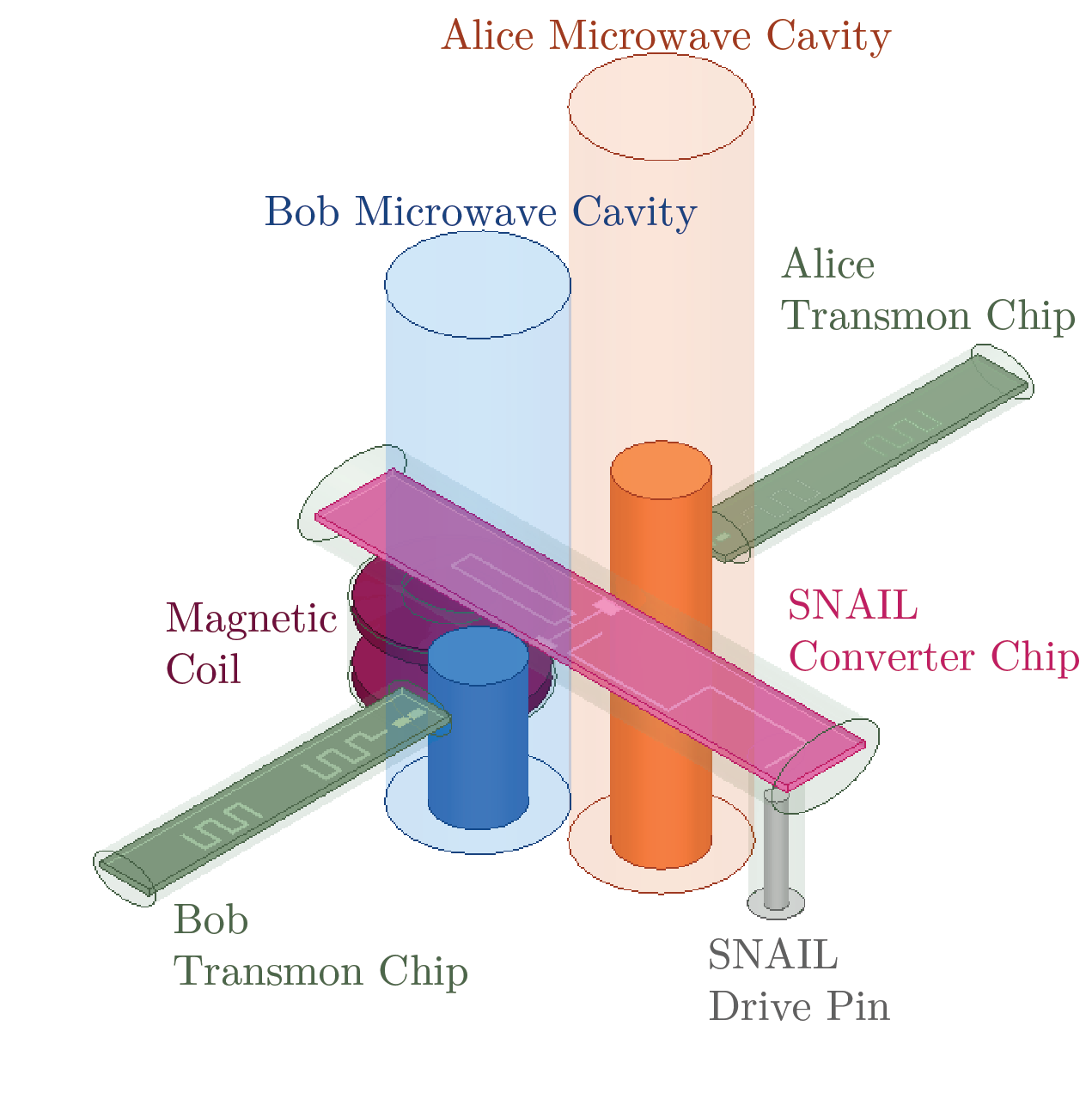}
\caption{\textbf{Experimental device model.} Schematic model of the superconducting device. The structure is machined out of high-purity aluminum. Each transmon chip houses a readout resonator and a Purcell filter, in addition to the transmon. Cavity drive pins, transmon drive pins and transmon readout pins are omitted in this model for simplicity.}
\label{fig:DeviceModel}
\end{center}
\end{figure}

Flux-biasing a SNAIL in an all-superconducting package is challenging because the Meissner effect~\cite{tinkham:2004} screens external magnetic fields and because oscillator coherence can be degraded if a magnetic coil wound on a spool of normal metal is located too near to the cavity. One solution to this problem is to use an all-superconducting pick-up loop~\cite{zimmerman:1971} as an intermediary between the magnetic coil and the SNAIL loop, as illustrated in Fig.~\ref{fig:flux_transformer}. 
To see how the magnetic coil integrates into the superconducting package, please refer to Fig. ~\ref{fig:DeviceModel}. When a DC magnetic field from the coil threads through the pick-up loop, a screening current flows around the loop to counter this flux. The magnetic field from this screening current then threads through the loop of the SNAIL, delivering the flux bias. This process is analogous to the operation of a transformer, except that the properties of superconductors allow it to occur at DC. 

Experimentally, we measure that between 20 and 40 mA are required to thread one flux quantum through the SNAIL using this method. To apply this large current without heating the mixing-chamber plate, we solder the superconducting magnet wire (SC-SW-M-0.127mm CuNi clad NbTi from Supercon) to a segment of copper cable and heatsink their joint (and about 10 cm of wire on either side) to an OFHC copper bobbin secured to the 4K plate. The other end of the copper wire is then routed out of the fridge and to a current source at room temperature. The absence of joints in the NbTi wire below 4K avoids placing a heat load on the dilution unit and allows us to send 200 mA through this line without any change in mixing chamber temperature. 

\section{System Design considerations}
\label{app:design_considerations}
Design of the experimental system is guided by enforcing the hierarchy of rates in Eq.~(\ref{hierarchy}). The parameter space of the design includes the bare oscillator frequencies $\omega_\text{A,B}$, the linear coupling rates between the coupler and the cavities $g_\text{a,b}$, and the circuit parameters of the coupler: $\beta, E_L, E_J, E_C$, the number of large junctions in the shunting array $M$, and the number of SNAILs $N$ arrayed in series in the coupler. For brevity we omit discussion here on choices regarding the transmon and readout parameters.

\subsection{Choice of coupler and bare oscillator frequencies}
When choosing the bare oscillator frequencies, the primary consideration is that the difference of the oscillator frequencies sets the frequency of the beamsplitter pump. It is therefore convenient to parameterize the problem in terms of $\omega_{\pm} \equiv \omega_\text{B} \pm \omega_\text{A}$. (Without loss of generality we take $\omega_\text{B}>\omega_\text{A}$.) To avoid low-order spurious interactions with waveguide modes in the three-dimensional package, $\omega_- + \omega_\text{B}$ should be much less than the waveguide cutoff $\omega_\text{wg}$ of the coupler or transmon tunnels. This consideration constrains $\omega_-$ from above. In our package the waveguide cutoff frequencies are at $\omega_\text{wg}/2\pi \approx 20$ GHz, meaning at least 5 pump photons are required to convert a photon in Bob into one of the waveguide modes. Ensuring that spurious interactions of this kind are high-order mitigates pump-induced decoherence on the cavities. As a general principle, reducing $\omega_-$ as much as possible helps to suppress unwanted pumped interactions.


The sum frequency $\omega_+$ should be chosen to allow for acceptable thermal populations in both cavities---the Bose-Einstein distribution is exponential in frequency, so modest increases in oscillator frequency rapidly suppress residual thermal populations. Additionally, unwanted hybridization with auxiliary modes can be avoided by ensuring they are not resonant with intermodulation products of the oscillator frequencies. 

\subsection{Choice of the linear coupling between the cavities and the coupler}
The beamsplitter rate is proportional to the hybridization between the cavities and the coupler, as are the decay and dephasing rates which the cavities inherit from the coupler. The linear coupling should therefore be chosen to maximize this coupling without meaningfully degrading the coherence of the cavities. 

We parameterize the hybridization between each oscillator and the coupler in terms of their detuning $\Delta_\text{a,b} \equiv \omega_\text{a,b}-\omega_\text{c}$ and their linear coupling rate $g_\text{a,b}$. The decay rate the oscillator inherits as a result of the hybridization is 
\begin{eqnarray}
\kappa_\text{a,b} \approx \left(\frac{g_\text{a,b}}{\Delta_\text{a,b}}\right)^2 \gamma_\text{c},
\label{kappa_a}
\end{eqnarray}
where $\gamma_c \equiv 1/T_{1,c}$ is the decay rate of the coupler. (This can be seen by considering a complex bare coupler frequency $\omega_\text{C}\to\omega_\text{C} + i\gamma_\text{c}$ in Eq.~(\ref{dressedmodes}).)

By the same mechanism, the oscillator inherits a dephasing rate from the coupler if the coupler frequency fluctuates stochastically, for example because of flux noise in the loop of the SNAIL. How this inherited dephasing rate relates to the fluctuations of the coupler frequency will depend on its noise spectrum.

In the case of white noise with a constant power spectral density $S_0$, we can identify the dephasing rate on the coupler as $\gamma_{\text{C}\phi} = \omega_\text{C}^2 S_0/4$, and the dephasing rate inherited by the oscillator 
\begin{eqnarray}
\kappa_\text{a,$\phi$} \approx \frac{g_\text{a}^4}{\Delta_\text{a}^4} \gamma_{\text{C}\phi},
\end{eqnarray}
where we have used the fact that $\omega_\text{C}-\omega_\text{A} \approx \Delta_\text{a}$.

It is more likely, however, that the noise spectrum is pink and of the form $S(f)=S_P/f$, in which case the dephasing rate of the coupler is given by $\gamma_{\text{C}\phi} = \omega_\text{C} \sqrt{S_\text{p} \ln{\frac{0.401}{f_\text{m}}}}$. Here, $f_\text{m}$ is a low-frequency cutoff which should be taken as the inverse of the longest time-scale in the system, and the value 0.401 arises from the Euler-Mascheroni constant~\cite{martinis:2003}. The dephasing rate inherited by the oscillator is then
\begin{eqnarray}
\kappa_\text{a,$\phi$} \approx \frac{g_\text{a}^2}{\Delta_\text{a}^2} \gamma_{\text{C}\phi}.
\label{kappa_aphi}
\end{eqnarray}
To be conservative, we use Eq.~(\ref{kappa_aphi}) to model the oscillator dephasing inherited from the coupler.

With the inherited decay and dephasing rates calculated, the optimal degree of hybridization can be calculated with knowledge of the bare oscillator and coupler decoherence rates (and the assumption that all coupler dephasing arises from flux noise). Assuming bare oscillator coherence at the level of 1 ms, we aim to make the inherited decay and dephasing rates much less than (1 ms)$^{-1}$. In our design, we assumed coupler coherence times of $\gamma_\text{C}^{-1} = 10~\upmu$s and $\gamma_{\text{C}\phi}^{-1} = 20~\upmu$s based on previous measurements of similar devices~\cite{grimm:2020}. With these assumptions and the detunings in our system, choosing $g_\text{a,b}/2\pi \sim 100$ MHz results in inherited decay and dephasing rates around (4 ms)$^{-1}$. 

As shown in Fig.~\ref{fig:coherences}a, the measured $T_1$ time of the coupler exceeds this estimate. As a consequence, we see no flux dependence in the oscillator decay times. The dephasing time of the coupler is comparable to our estimate,  resulting in a measurable variation in oscillator dephasing time with changing flux bias.  Nevertheless, decay remains the dominant cavity error. In future designs, enhancing the coupling rates $g_\text{a,b}$ will allow a factor of several improvement in the beamsplitter rate without degrading oscillator coherence. The cost for this change will be a commensurate increase in $\chi_\text{ab}$, but as this rate can be heavily suppressed by tuning the external flux this tradeoff seems favorable.

\subsection{Choice of the SNAIL coupler parameters}
As discussed in the main text, the coupler parameters are chosen to suppress quartic interactions in its Hamiltonian and enhance cubic interactions. Realizing this goal requires relating those interactions to the circuit parameters of the coupler.

Following Refs.~\cite{frattini:2017,frattini:2018}, we begin with general comments on the potential $U_\text{s}$ of the SNAIL dipole, parametrized as pictured in the inset of Fig.~\ref{fig:coupler}d.  When the number of large junctions in the array $M>1$ and the external flux is a DC bias~\cite{you:2019}, the potential may be written as
\begin{eqnarray}
\frac{U_\text{s}(\hat{\varphi_\text{s}})}{E_J} = -\beta \cos (\hat{\varphi}_\text{s} - \phi_\text{e}) - M \cos \frac{\hat{\varphi_\text{s}}}{M}.
\end{eqnarray}
Here $\phi_\text{e} \equiv \Phi_\text{e}/\Phi_0$ is the phase difference imparted by the external flux and $\hat{\varphi_\text{s}}$ is the phase drop across the SNAIL; $\beta<1$ is the ratio of Josephson energies for the primary and shunting Josephson junctions.

The position of the potential minimum $\varphi_\text{m}$ at a given external flux is found by solving 
\begin{eqnarray}
\frac{dU_\text{s}/d\hat{\varphi}_s}{E_J} = \beta \sin\left(\hat{\varphi_\text{s}} - \phi_\text{e}\right) + \sin\left(\frac{\hat{\varphi_\text{s}}}{M}\right)=0. 
\label{Umin}
\end{eqnarray}
When $\beta<1/M$, a single minimum exists, and so to avoid multi-stability we work in this regime. The position of the potential minimum reaches its largest value $\varphi_\text{m}^\text{max} = M \arcsin{(\beta^{-1})}$ when $\phi_\text{e} = \varphi_\text{m}^\text{max} + \pi/2$.

Near the minimum $\varphi_\text{m}$ the potential $U_\text{s}$ is well approximated by fifth-order expansion in $\hat{\varphi}_\text{s}-\varphi_\text{m}$:
\begin{eqnarray}
U_\text{s}\left(\hat{\varphi}_{\text{s}}\right) &\approx& \frac{c_2}{2!} (\hat{\varphi}_\text{s}-\varphi_\text{m})^2 \nonumber + \frac{c_3}{3!} (\hat{\varphi}_\text{s}-\varphi_\text{m})^3 \\
&& + \frac{c_4}{4!} (\hat{\varphi}_\text{s}-\varphi_\text{m})^4 + \frac{c_5}{5!} (\varphi_\text{s}-\varphi_\text{m})^5,
\end{eqnarray}
with 
\begin{eqnarray}
c_2 &=& \beta \cos(\varphi_\text{m} - \phi_\text{e}) + M^{-1} \cos{\frac{\varphi_\text{m}}{M}}, \nonumber \\
c_3 &=& \frac{M^2-1}{M^2} \sin{\frac{\varphi_\text{m}}{M}}, \nonumber \\
c_4 &=& -\beta \cos(\varphi_\text{m} - \phi_\text{e}) - M^{-3} \cos{\frac{\varphi_\text{m}}{M}}, \nonumber \\
c_5 &=& \frac{1-M^4}{M^4} \sin{\frac{\varphi_\text{m}}{M}},
\label{c}
\end{eqnarray}
where Eq.~(\ref{Umin}) has been used to simplify the expression for $c_3$ and $c_5$. The maximum magnitude of $c_3$ and $c_5$ are obtained at $\varphi_\text{m}^\text{max}$, at which point the ratio of their magnitudes is $|\frac{c_3}{c_5}| = M^2/(M^2+1)$. This suggests choosing $M>2$ to attain a more favorable ratio of the desired $c_3$ to the unwanted $c_5$, although the choice of $M$ will also be guided by fabrication considerations. For our coupler, we use $M=3$.

As discussed in Ref.~\cite{frattini:2018}, calculating the Hamiltonian of the full circuit in Fig.~\ref{fig:coupler}d, including a linear inductance, rescales the potential coefficients $c_j\to \tilde{c}_j$, but does not change the location of the potential minima. The result is of the form given in Eq.~(\ref{H_snail}). Second quantization with $\hat{\varphi} - \varphi_\text{m} = \phi_\text{c} (\hat{c} + \hat{c}^\dagger), \hat{N} = -i\phi_\text{c}^{-1} (\hat{c}-\hat{c}^\dagger)/2$ yields Eq.~(\ref{Hc}), with $\hat{\phi}$ replaced by $\hat{\varphi}$. The coefficients in $\mathcal{H}_\text{c}$ are
\begin{eqnarray}
\omega_\text{c} &=& \sqrt{8 \tilde{c}_2 E_C E_J}, \nonumber \\ 
\phi_\text{c} &=& \left(\frac{2 E_C}{\tilde{c}_2 E_J}\right)^{\frac{1}{4}}, \nonumber \\
\hbar g_3 &=& E_\text{J} \phi_\text{c}^3 \frac{\tilde{c}_3}{3!}, \nonumber \\
\hbar g_4 &=& E_\text{J} \phi_\text{c}^4
\frac{\tilde{c}_4}{4!}, \nonumber \\
\hbar g_5 &=& E_\text{J} \phi_\text{c}^5
\frac{\tilde{c}_5}{5!}, 
\label{g}
\end{eqnarray}
with
\begin{eqnarray}
\tilde{c}_2 &=& p c_2, \\
\tilde{c}_3 &=& p^3 c_3, \nonumber \\
\tilde{c}_4 &=& p^4 \left(c_4-3\frac{{c_3}^2}{c_2}(1-p)\right), \nonumber \\
\tilde{c}_5 &=& p^5 \left( c_5 -10\frac{c_4c_3}{c_2}(1-p) + 15 \frac{c_3^2}{c_2^2}(1-p)^2\right), \nonumber
\end{eqnarray}
and $p\equiv \frac{c_2 E_J}{E_L + c_2 E_J}$.
A rotating wave approximation on $\mathcal{H}_\text{c}$ yields a Kerr nonlinearity of the form $-\frac{\alpha_\text{c}}{2} \hat{c}^{\dagger^2} \hat{c}^2$ (see Eq.~(\ref{Kerr})). We are now in position to optimize the circuit parameters of the coupler for maximal $g_3/g_5$ and minimal quartic interaction such as the anharmonicity and cross Kerr $\chi_\text{ab}$. 
As the frequency $\omega_\text{c}$ is constrained by other considerations, choosing the $\phi_\text{c}$ (equivalent to choosing the impedance of the coupler) fixes the linear circuit parameters $E_L$ and $E_C$. Reducing $\phi_\text{c}$ is a straight-forward way to increase the ratio $g_3$ relative to higher order terms. Care should be taken, though, that the pump amplitude can be commensurately increased, to avoid reduction of the beamsplitter rate. In practice this means ensuring that the drive can be delivered to the SNAIL coupler while sufficiently isolating other nonlinear components (such as the transmons). See App.~\ref{app:pumps} for more details.

Replacing a single SNAIL with an array of $N$ SNAILs accomplishes much the same goal (and at the same cost), provided that $E_J$ is adjusted such that the total inductance of the array is equal to the inductance of the original SNAIL. We note that this is only possible when $E_J\ll E_L$~\cite{sivak:2020}, but this requirement is well satisfied for our device. 

We tested multiple devices with $N=3$ SNAILs in an array and found them to have excellent suppression of quartic interactions. Their performance, however, was limited by an undesired coupling of the strong beamsplitter pump to the transmons used for oscillator control. Reverting to a single SNAIL remedied this issue, as the same beamsplitter rate could be attained with a 9-fold reduction in pump power. Future designs, though, could recover the enhanced suppression of higher-order terms by engineering better isolation between the coupler and the transmons, as in Ref.~\cite{zhou:2021}.

The only remaining parameter in the coupler circuit is the ratio of the Josephson energies $\beta$. Broadly speaking, larger values of $\beta$ in the range $\beta<1/M$ increase the nonlinearities $g_3$ and $g_4$ and causes the potential coefficients $\tilde{c}_2$ to be more sensitive to external flux.  Assuming undesirable quartic interactions have already been suppressed through other means, the trade-off in choosing $\beta$ amounts to balancing the desire to increase $g_3$ against greater susceptibility to flux noise (via the increase of $d\omega_\text{C}/d\Phi_\text{e}$). 

To quantify this, we perform numerical simulations sweeping $\beta$ with $E_J$ adjusted to hold $\omega_\text{C}$ fixed. They show that $g_3$ and $d\omega_\text{C}/d\Phi_\text{e}$ increase fractionally at the same rate when $\beta$ is less than a critical value $\beta_\text{c}\approx 0.15$ (for our choices of the other parameters). Beyond this, i.e. when $\beta_\text{c} < \beta < 1/M$, the susceptibility to external flux increases more rapidly with $\beta$ than $g_3$. 

Assuming that the flux noise on the coupler is pink, the decay and dephasing rates inherited by the cavities both scale as $(g_\text{a,b}/\Delta_\text{a,b})^2$. The optimal choice of $\beta$ then depends on the expected coupler coherence, and can be found with the following procedure. Beginning with $\beta=\beta_\text{c}$, calculate the dressed coherence of the cavities. If the cavities are dephasing limited, increase $(g_\text{a,b}/\Delta_\text{a,b})$ as much as possible until the inherited decay rate becomes non-negligible. Alternatively, if the cavities are decay limited, increase $\beta$ until the decay and dephasing times are balanced. Then adjust $(g_\text{a,b}/\Delta_\text{a,b})$ as needed.

Note that the above procedure requires knowledge of the bare oscillator and coupler coherences, which are not always available in design. Nonetheless it illustrates the general consideration in the choice. 

\section{Determining the linear oscillator-SNAIL coupling}
\label{app:lincoupling}
An important parameter in predicting the nonlinear properties of the oscillators is the linear coupling strength $g$ between each oscillator and the SNAIL coupler. This appears when considering their bare modes,
\begin{eqnarray}
\hat{\mathcal{H}}_{AC}/\hbar &=& \omega_A \hat{A}^{\dag}\hat{A} + \omega_C \hat{C}^{\dag}\hat{C} \nonumber \\
 &&+ g_a \left(\hat{A}^{\dag}\hat{C} + \hat{C}^{\dag}\hat{A}\right),
\end{eqnarray}
where $\omega_A$ and $\omega_C$ are the bare frequencies of Alice and the coupler, and $g$ is taken to be real and independent of the external flux. The same expression holds for Bob.
The Hamiltonian can be rewritten in matrix form,
\begin{align}
\hat{\mathcal{H}}_{AC}/\hbar &=
\begin{pmatrix}
 \hat{A}^{\dag} & \hat{C}^{\dag}
\end{pmatrix}
\begin{pmatrix}
 \omega_A & g_a \\
 g_a & \omega_C
\end{pmatrix}
\begin{pmatrix}
 \hat{A} \\
 \hat{C}
\end{pmatrix}\nonumber, \\ 
 &= \begin{pmatrix}
 \hat{A}^{\dag} & \hat{C}^{\dag}
\end{pmatrix}
\mathbf{M}
\begin{pmatrix}
 \hat{A} \\
 \hat{C}
\end{pmatrix},
\end{align}
and $\mathbf{M}$ diagonalized to find the dressed frequencies
\begin{eqnarray}
\omega_a &\approx \omega_A - \frac{g_a^2}{\omega_C - \omega_A} \nonumber, \\ 
\omega_c &\approx \omega_C + \frac{g_a^2}{\omega_C - \omega_A},
\label{dressedmodes}
\end{eqnarray}
where we use the fact that $g_a \ll \omega_C - \omega_A$. The same approximation relates the dressed frequencies to one another:
\begin{align}
  \omega_a \approx \omega_A - \frac{g_a^2}{\omega_c - \omega_A}.
  \label{eqn:cavity_freqs}
\end{align}

\begin{figure}
\begin{center}
\includegraphics[width=1\linewidth]{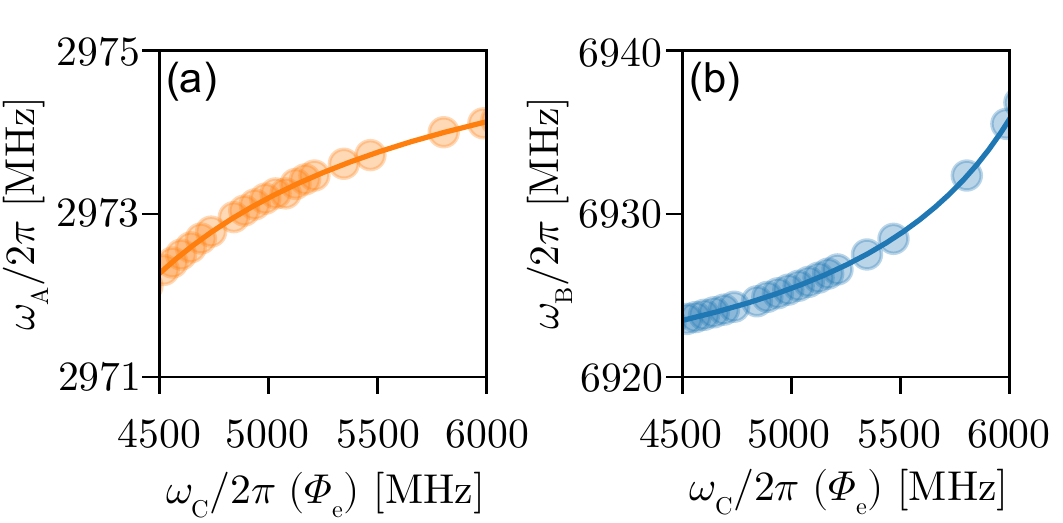}
\caption {\textbf{Cavity Frequencies}. Circles show measurements of the dressed oscillator frequencies for Alice (a) and Bob (b) at a range of external flux biases, plotted as a function of measured SNAIL frequency. Lines show fits to Eq.~(\ref{eqn:cavity_freqs}), from which we extract linear oscillator-SNAIL couplings of $g_a/2\pi = 75.6\pm0.2$ MHz and $g_b/2\pi = 134.9\pm0.1$ MHz, and bare mode frequencies of $\omega_A/2\pi = 2976.018(16)$ MHz and $\omega_B/2\pi = 6915.945(17)$ MHz.
}
\label{fig:cav_freqs}
\end{center}
\end{figure}

By observing how the oscillator frequencies depend on the SNAIL frequency as we tune external flux, we can fit to obtain the linear coupling $g$ and the bare oscillator frequency, as shown in Fig.~\ref{fig:cav_freqs}.

\section{Calculation of oscillator self and cross-Kerrs}
\label{app:kerr_prediction}

The analysis in App.~\ref{app:lincoupling} can be extended to account for two cavities (with dressed annihilation operators $\hat{a}$ and $\hat{b}$) coupled to the same SNAIL mode. In the dispersive limit, we approximate the bare SNAIL mode as
\begin{equation}
  \hat{C} \approx \hat{c} + \frac{g_a}{\Delta_a}\hat{a} + \frac{g_b}{\Delta_b}\hat{b},
  \label{eqn:dressed}
\end{equation}
where $\Delta_{a(b)} \equiv \omega_{a(b)} - \omega_c$. Inserting Eq.~(\ref{eqn:dressed}) into the Hamiltonian for the uncoupled SNAIL mode, Eq.~(\ref{Hc}), allows us to calculate the energy of the combined system perturbatively. 

The cross-Kerr between the cavities $\chi_{ab}$ can be defined as the frequency shift $\Delta E / \hbar$ proportional to $n_a n_b$. The $g_4$ nonlinearity yields a shift at 1st order in perturbation theory while $g_3$ yields a shift at 2nd order, and so the cross-Kerr can be approximated as
\begin{equation}
  \chi_{ab} \approx \left(24 g_4 + 36\frac{g_3^2}{\tilde{\omega}}\right) \left(\frac{g_a}{\Delta_a}\right)^2\left(\frac{g_b}{\Delta_b}\right)^2,
\end{equation}
with
\begin{align*}
  \tilde{\omega} \equiv (&\frac{1}{\omega_a-\omega_b-\omega_c} + \frac{1}{-\omega_a+\omega_b-\omega_c} + \\ &\frac{1}{\omega_a+\omega_b-\omega_c} + \frac{1}{-\omega_a-\omega_b-\omega_c})^{-1}.
\end{align*}
Here we assume $g_a \ll \Delta_a$ and $g_b \ll \Delta_b$, but do not make any assumptions about the relative sizes of the mode frequencies.

The same procedure can be used to estimate the cavity self-Kerr, which is proportional to $n^2$. To first order in $g_4$ and second order in $g_3$,
\begin{align}
  K_a \approx &\left(12g_4 - 18g_3^2\left(\frac{2\omega_c}{4\omega_a^2-\omega_c^2}+ \frac{4}{\omega_c}\right)\right)\left(\frac{g_a}{\Delta_a}\right)^4 \nonumber \\
  &+ \frac{\chi_{at}^2}{4\alpha_{at}},
\end{align}
where $\alpha_{at}$ is the anharmonicity of the transmon ancilla. The first term accounts for the coupler contribution to the cavity self-Kerr, while the second accounts for the ancilla contribution, which gives a fixed offset that is independent of external flux bias.


\section{Measurement of self- and cross-Kerr between modes}
\label{app:crosskerr}

To sensitively measure the oscillator-oscillator cross-Kerr interaction $\chi_\text{ab}\hat{a}^\dagger\hat{a}\hat{b}^\dagger\hat{b}$, we find the linear frequency shift of one oscillator while populating the other with a large and variable number of photons. The cross-Kerr interaction is symmetric under exchange of $\hat{a}$ and $\hat{b}$, so one can either displace Alice and detect a change in Bob's resonant frequency, or vice versa. Moreover, the frequency shift can be detected either interferometrically or spectroscopically, giving four possible measurement configurations for $\chi_\text{ab}$. 

\begin{figure}
\begin{center}
\includegraphics[width=1\linewidth]{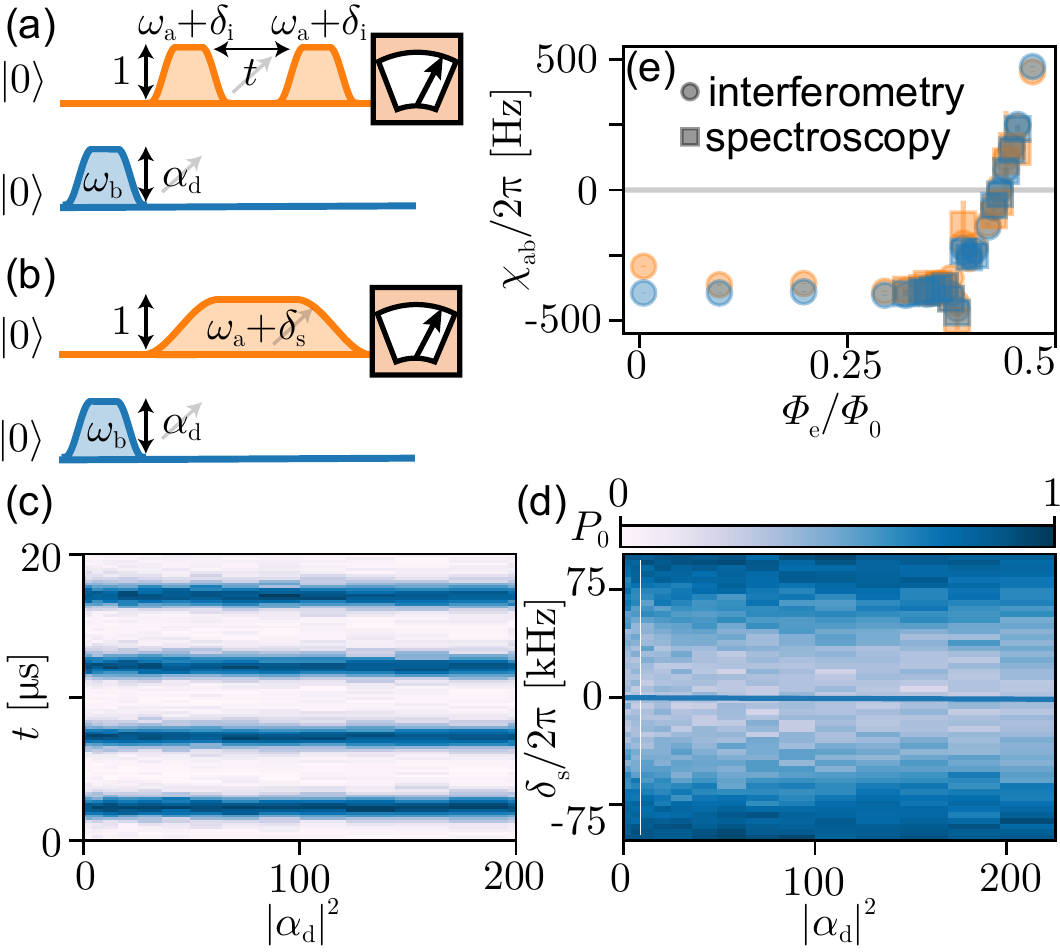}
\caption {\textbf{Measurement of the oscillator-oscillator cross Kerr} Pulse sequence for measuring the oscillator-oscillator cross-Kerr interferometrically (a) and spectroscopically (b) with Alice. Exchanging the cavity rails yields the complementary sequences used to measure with Bob. Both the interferometric and spectroscopic sequences start with a resonant and variable displacement of Bob which populates that cavity with an average of $n = |\alpha_d|^{2}$ photons. For the interferometric measurement in (a), the resulting frequency shift is detected by off-resonantly displacing Alice by $\alpha = 1$, waiting a variable time $t$, and then making a second displacement of the same size before measuring if Alice is in vacuum. For the spectroscopic measurement in (b), the frequency shift is detected by sweeping the frequency of a long displacement pulse on Alice and measuring if it's in vacuum. (c-d) The result of interferometric (c) and spectroscopic (d) measurements on Bob near the cross-Kerr free point $\Phi_\text{e}/\Phi_0 = 0.43$. Solid line in (d) is a linear fit to the centers of Gaussian fits on linecuts at each value of $\alpha_\text{d}$. (e) The inferred cross-Kerr $\chi_\text{ab}$ as a function of external flux, measured interferometrically (circles) and spectroscopically (squares), from Alice (orange) and Bob (blue). The measurements shown in Fig.~\ref{fig:coupler}h are the average of the two interferometric datasets. 
}
\label{fig:cross_kerrs}
\end{center}
\end{figure}

Fig.~\ref{fig:cross_kerrs}a illustrates the pulse sequence used to interferometrically measure Alice's frequency. The protocol starts by displacing Bob by a variable amount $\alpha_{\text{d}}$. We then apply a unit displacement to Alice with a detuned drive at frequency $\omega_\text{a} + \delta_\text{i}$ such that the resulting coherent state acquires phase at a rate $\delta_\text{i}/2\pi = 200$ kHz. In addition, the coherent state will pick up an extra phase due to the cross-Kerr at an average rate of $|\alpha_{\text{d}}|^2\chi_{ab}$. After a variable delay $t$, a second detuned unit displacement returns the coherent state to the origin if and only if it has acquired $\pi$ radians of phase during the delay. Fig.~\ref{fig:cross_kerrs}c shows data measured in this way near the cross-Kerr free point at $\Phi_\text{e}/\Phi_0 = 0.43$. By detecting the probability $P_0$ of finding Alice in vacuum, we can fit the measured revivals with the model
\begin{eqnarray}
P_0 = A_0 e^{1 + \cos\left( \left( \delta_\text{i} + |\alpha_\text{d}|^2 \chi_\text{ab} \right)t\right)} + C,
\end{eqnarray}
to infer $\chi_\text{ab}$. Here $A_0$ and $C$ capture measurement infidelity, and the detuning $\delta_\text{i}$ and displacement $\alpha_\text{d}$ are fixed by the protocol. The results of these fits are shown as circles in Fig.~\ref{fig:cross_kerrs}e. The model neglects the effects of cavity decay, but since measurements of Alice and Bob yield similar values for $\chi_\text{ab}$ despite their different decay rates, we deduce that corrections due to cavity decay are small.

Fig.~\ref{fig:cross_kerrs}b illustrates the pulse sequences used to spectroscopically measure Alice's frequency. After displacing Bob by $\alpha_\text{d}$, a long pulse with variable detuning $\delta_\text{s}$ is used to displace Alice, followed by a measurement to determine if Alice is in vacuum. Fig.~\ref{fig:cross_kerrs}d shows data measured in this way. Fitting a linear relation to the spectroscopy dip at each value of $n$, the slope gives a measurement of $\chi_\text{ab}$.

Cavity self-Kerr ($\frac{1}{2}K_a \hat{a}^{\dagger}\hat{a}^{\dagger}\hat{a}\hat{a}$) is measured very similarly to the interferometric measurement of cross-Kerr outlined above. Instead of displacing the other cavity, we vary the amplitude of the detuned displacement $\alpha$, such that the cavity acquires extra phase at a rate $\frac{1}{2}K_a |\alpha|^2$. We can then extract the self-Kerr by measuring whether the cavity is in vacuum and fitting it to
\begin{eqnarray}
P_0 = A_0 e^{1 + \cos\left( \left( \delta + \frac{1}{2} K_{a}|\alpha|^2 \right)t\right)} + C.
\end{eqnarray}

The cross-Kerr between the cavities and the SNAIL coupler ($\chi_{ac} \hat{a}^{\dagger}\hat{a} \left(\ket{e}\bra{e}\right)_c$) is measured similarly to the spectroscopic measurement of cavity-cavity cross-Kerr. The cavity is displaced by a variable displacement $\alpha_{\text{d}}$ before a long spectroscopy tone is applied to the coupler. We can then measure whether the coupler is in its ground state by performing a coupler-state-selective $\pi$ pulse on the transmon not coupled to the cavity and reading out the transmon state. The spectroscopy data shows a linear frequency shift proportional to $|\alpha_{\text{d}}|^2$, with the slope given by $\chi_{ac}$.

\section{Derivation of the beamsplitter Hamiltonian from a driven SNAIL}

\label{app:H_bs}
We consider a system of two oscillators with associated annihilation operators $\hat{a}$ and $\hat{b}$, which are jointly coupled to a conversion element with annihilation operator $\hat{c}$.

 The system Hamiltonian is 
\begin{eqnarray}
\frac{\hat{\mathcal{H_\text{s}}}}{\hbar} = \omega_a \hat{a}^{\dag}\hat{a} + \omega_b \hat{b}^{\dag}\hat{b} + \omega_c \hat{c}^{\dag}\hat{c} + \frac{\hat{\mathcal{H_\text{d}}}}{\hbar} + \frac{\hat{\mathcal{H_\text{cnl}}}}{\hbar}.
\end{eqnarray}
Here $\omega_c$ is the $g$-$e$ transition of the coupler, $\hat{\mathcal{H_\text{d}}}$ is the Hamiltonian of the drive, and $\hat{\mathcal{H}}_\text{cnl}$ is the non-quadratic part of the coupler Hamiltonian $\hat{\mathcal{H}}_\text{c}$. 

The Hamiltonian of the coupler circuit is~\cite{frattini:2018} (see also App.~\ref{app:design_considerations}):

\begin{eqnarray}
\hat{\mathcal{H_\text{c}}} &=& 4 E_C \hat{N}^2 + E_J\left(\frac{{\tilde{c}_2}}{2!} \hat{\phi}^2 + \frac{{\tilde{c}_3}}{3!} \hat{\phi}^3 + \frac{{\tilde{c}_4}}{4!} \hat{\phi}^4+\frac{{\tilde{c}_5}}{5!} \hat{\phi}^5\right) \nonumber \\ 
&&+ \mathcal{O}(\hat{\phi})^6,
\label{H_snail}
\end{eqnarray}
or equivalently
\begin{eqnarray}
\omega_c \hat{c}^{\dag}\hat{c} + \hat{\mathcal{H}}_\text{cnl} &=& E_J\left(\frac{{\tilde{c}_3}}{3!} \hat{\phi}^3 + \frac{{\tilde{c}_4}}{4!} \hat{\phi}^4 + \frac{{\tilde{c}_5}}{5!} \hat{\phi}^5\right) + \mathcal{O}(\hat{\phi})^6,\nonumber
\label{H_snailnl}
\end{eqnarray}
with $\hat{\mathcal{H}}_\text{cnl}$ the terms in the Hamiltonian with order higher than two. Second quantization on Eq.~(\ref{H_snail}) yields Eq.~(\ref{Hc}). As the coupler is capacitively connected to the two oscillators (see Fig.~\ref{fig:coupler}b), the phase operator $\hat{\phi}$ depends on the displacement of all three hybridized modes $\hat{\phi} \equiv \phi_a(\hat{a} + \hat{a}^\dagger) + \phi_b (\hat{b} + \hat{b}^{\dagger}) + \phi_c\left(\hat{c} + \hat{c}^\dagger\right)$. Here $\phi_i$ are the zero-point phases of the $i^\text{th}$ mode, with $i=a,b,c$. 

To actuate the mixing process, we apply a pump with frequency $\omega_{p}$ and strengths $\epsilon$. The Hamiltonian of the drive is 
\begin{eqnarray}
\hat{\mathcal{H_\text{d}}}/\hbar = -(\epsilon e^{-i\omega_p t}-\epsilon^*e^{i\omega_p t})(\hat{c} - \hat{c}^\dagger)
\label{eqn:drive}
\end{eqnarray}
We now move to the frame displaced by the drive~\cite{leghtas:2015}. The unitary for the displacement transformation is
\begin{eqnarray}
\hat{U}_\text{d} = e^{\tilde{\xi}(t) \hat{c}^\dagger - \tilde{\xi}^*(t) \hat{c}} 
\end{eqnarray}
with 
\begin{eqnarray}
\tilde{\xi} \equiv \xi e^{-i\omega_p t} \equiv \frac{\epsilon}{\omega_p - \omega_c} e^{-i\omega_p t}.
\label{eqn:xi}
\end{eqnarray}
In the above, we have neglected dissipation on the coupler mode and counter-rotating terms.

We then move into a frame co-rotating with the three modes via the unitary transformation 
\begin{eqnarray}
\hat{U}_\text{r} = e^{-i\left(\omega_a \hat{a}^{\dag}\hat{a} + \omega_b \hat{b}^{\dag}\hat{b} + \omega_c \hat{c}^{\dag}\hat{c}\right)t}.
\end{eqnarray}
The transformed phase operator is now
\begin{eqnarray}
\hat {\tilde \phi} &=& \hat U_r^\dagger \hat U_d^\dagger \hat \phi \hat U_d \hat U_r \\
&=& \phi_a(\hat{a} e^{-i\omega_at} + \hat{a}^\dagger e^{i\omega_at}) + \phi_b (\hat{b} e^{-i\omega_bt}+\hat{b}^\dagger e^{i\omega_b t}) \nonumber \\
&& + \phi_c\left(\hat{c} e^{-i\omega_ct} + \hat{c}^\dagger e^{i\omega_ct} + \xi e^{-i\omega_p t} + \xi^*e^{i\omega_pt}\right). \nonumber
\end{eqnarray}
When the frequency of the pump is $\Delta$ detuned from the difference $\omega_b-\omega_a$, a beamsplitter interaction of the form
\begin{eqnarray}
g_{bs} e^{i\theta} \hat{a}^\dagger \hat{b} e^{i\Delta t} + \textrm{h.c.}
\end{eqnarray}
emerges, with (to lowest-order)
\begin{eqnarray}
\label{gbs_pert}
g_{bs} &\approx& E_J \phi_a \phi_b \phi_c \xi \tilde{c}_3,  \\ \nonumber
&\approx& 6\frac{g_\text{a}}{\Delta_\text{a}} \frac{g_\text{b}}{\Delta_\text{b}}\xi g_3.
\end{eqnarray}

Greater accuracy can be obtained by including the effect of higher-order nonlinear terms in the Hamiltonian (beyond $g_3$). To leading order within the rotating wave approximation (but still treating the drive perturbatively) these modify the beamsplitter rate in Eq.~(\ref{gbs_pert}) to
\begin{equation}
  g_{bs} \approx \frac{g_\text{a}}{\Delta_\text{a}} \frac{g_\text{b}}{\Delta_\text{b}}\xi^*\left(\sum_{m=1} \frac{\left(2m+1\right)!}{m!\left(m-1\right)!} g_{2m+1} |\xi|^{2m-2}\right).
  \label{full_gbs}
\end{equation}
Terms that are higher order in $\frac{g_i}{\Delta_i}$ and those proportional to $g_{2m+1}\xi^*|\xi|^j$ with $j<2m-2$ are suppressed, since $(\frac{g_i}{\Delta_i})^2 \ll 1$ and $(p\phi_{c})^2 \ll 1$, respectively. 

This expression predicts that at sufficient pump amplitude the beamsplitter rate will deviate from a purely linear increase with amplitude. What physical considerations limit the pump amplitude itself?

The renormalized SNAIL potential (described in App.~\ref{app:design_considerations}) is $6\pi$-periodic in $\hat{\phi}$ with a single minimum within each period, provided that $\beta < 1/M$. Furthermore, adjacent maxima and minima are exactly $3\pi$ apart. If we approximate the maximum deviation from the potential minima due to the drive as $|\phi_{\text{max}}| = 2\phi_{c}|\xi|$, then the drive amplitude required to enter the adjacent well is approximately
\begin{equation}
  |\xi_{\text{crit}}| \approx \frac{3\pi}{2\phi_{c}}.
\end{equation}
For our system, this evaluates to $|\xi_{\text{crit}}| =12.9$, a value that agrees closely with other estimates for the critical photon number given in Ref. \cite{frattini:2021}. This calculation suggests that further optimization of the coupler may allow even faster beamsplitter rates than those demonstrated in the present work for which the $\xi<3$.  Such work will likely require more exact treatments of the coupler Hamiltonian.

\section{Measurement of the joint probability distribution of the photon population}
\label{app:joint_prob_dist}
The joint probability distribution of the photon population in the two cavities (shown in Fig.~\ref{fig:concept}b) is obtained by measuring the state of both cavities simultaneously and correlating the outcomes shot-by-shot. 

These measurements involve performing a 0-selective $\pi$-pulse on the ancilla for each oscillator (which succeeds only if the oscillator is in vacuum) and reading out the ancilla state. Ideally, if an oscillator is in $\ket{0}$, we will measure its ancilla in $\ket{e}$ with unit probability. Likewise, if the oscillator is in a state $\ket{\psi_{\Bar{0}}}$ where $\braket{0|\psi_{\Bar{0}}} = 0$, we will measure the ancilla in $\ket{g}$. We multiply the results of the simultaneous measurements shot-by-shot to infer the joint oscillator state (eg. ancillas measured in $\ket{e}_A\ket{g}_B$ registers as joint oscillator state $\ket{0}_A\ket{\psi_{\Bar{0}}}_B$). If we neglect oscillator heating, our cavities are confined to the $\{0,1\}$ subspace and we can replace $\ket{\psi_{\Bar{0}}}$ with $\ket{1}$. We will do so for ease of reading, but the analysis works equally well with $\ket{\psi_{\Bar{0}}}$ undetermined.

Averaging the results yields the uncorrected probabilities $Q(i,j)\ \text{where}\ i,j=\{e,g\}$. The uncorrected probabilities are the result of imperfect measurement arising from errors during our selective $\pi$-pulses and readout. These imperfections lead to false positives, where the ancilla is measured in $\ket{e}$ despite the oscillator being in $\ket{1}$, and false negatives, where the ancilla is measured in $\ket{g}$ despite the oscillator being in $\ket{0}$.

For a single oscillator measurement we construct a matrix $\mathbf{E}$ relating the probability of measuring the ancilla state to the true oscillator state:
\begin{equation*}
  \begin{pmatrix}
   Q(e) \\
   Q(g)
  \end{pmatrix}
  =
  \begin{pmatrix}
   P(e|0) & P(e|1) \\
   P(g|0) & P(g|1)
  \end{pmatrix}
  \begin{pmatrix}
   P(0) \\
   P(1)
  \end{pmatrix}
  =
  \mathbf{E}
  \begin{pmatrix}
   P(0) \\
   P(1)
  \end{pmatrix},
\end{equation*}
Here the matrix $\mathbf{E}$ can be written in terms of the false positive and true positive rates, $f$ and $t$. We can estimate these by taking the minimum and maximum points of our 0-selective Rabi oscillations, respectively.
\begin{equation*}
  \mathbf{E}
  =
  \begin{pmatrix}
   t & f \\
   1-t & 1-f
  \end{pmatrix}.
\end{equation*}

We can invert this matrix to get the oscillator state probabilities from our measurements:
\begin{equation*}
  \begin{pmatrix}
   P(0) \\
   P(1)
  \end{pmatrix}
  =
  \mathbf{E}^{-1}
  \begin{pmatrix}
   Q(e) \\
   Q(g)
  \end{pmatrix}.
\end{equation*}

The same procedure may be generalized for joint measurements, where we assume that measurement errors are uncorrelated between ancillae:
\begin{align*}
  Q(i,j) &= E(i,j|n,m)P(n,m), \\
  &= E_{(A)}(i|n)E_{(B)}(j|m)P(n,m), \\
  \mathbf{Q} &= \mathbf{E}_{(A)}\mathbf{P}\mathbf{E}_{(B)}^{T}.
\end{align*}
Inverting this we have
\begin{equation}
  \mathbf{P} = \mathbf{E}_{(A)}^{-1}\mathbf{Q}\left(\mathbf{E}_{(B)}^{T}\right)^{-1}.
  \label{eqn:Rabi_correction}
\end{equation}

\begin{figure}
\begin{center}
\includegraphics[width=1\linewidth]{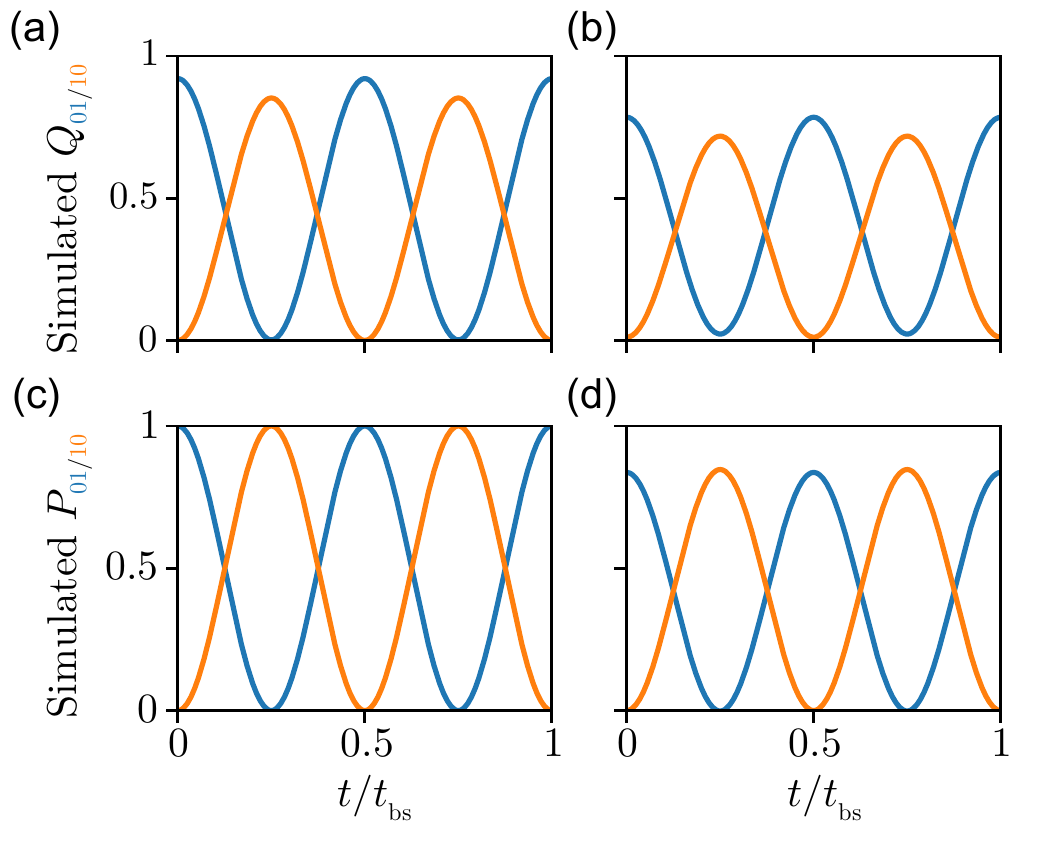}
\caption {\textbf{Correcting for measurement infidelity.} Top row shows mock data of the uncorrected probability $Q_{01/10}$ of measuring the oscillators in $\ket{0,1}$ or $\ket{1,0}$ during beamsplitter oscillations assuming a fixed measurement infidelity, with either (a) perfect or (b) imperfect preparation of initial state $\ket{0,1}$. Bottom row shows the inferred probability $P_{01/10}$ that the oscillators are in $\ket{0,1}$ or $\ket{1,0}$ given the uncorrected data with (c) perfect and (d) imperfect state preparation, after applying the correction in Eq.~\ref{eqn:Rabi_correction}. In practice, applying that correction requires populating the matrices $E_{(A)}$ and $E_{(B)}$, which can be done with measurements of Rabi oscillations. To generate the mock data, we take the conditional measurement probabilities to be $P_{(A)}(e|0) = 0.93$, $P_{(B)}(e|0) = 0.87$, $P_{(A)}(g|1) = 0.98$ and $P_B(g|1) = 0.99$. In the right column, we take the state preparation fidelities to be $P_{(A)}(0|0) = 0.99$, $P_{(B)}(0|0) = 0.98$, $P_{(A)}(1|1) = 0.87$ and $P_{(B)}(1|1) = 0.84$, where $P_{(A/B)}(i|j)$ is the probability of actually preparing Fock state $\ket{i}$ when we try to prepare Fock state $\ket{j}$. We emphasize that these values are not taken from the experiment, but are chosen to highlight the effect of the Rabi correction.
}
\label{fig:Rabi_correction}
\end{center}
\end{figure}

This method corrects the distortion in the measurement of the joint state of two oscillators when the measurement fidelity on each oscillator is different. We note that it does not normalize for imperfections in state preparation. Fig.~\ref{fig:Rabi_correction} shows how the correction protocol transforms mock uncorrected data, with and without perfect state preparation. In the case of imperfect state preparation, the oscillations of $P_{01}$ and $P_{10}$ should have the same amplitude, but not extend from 0 to 1.

To initialize the cavities in $\ket{1}_A\ket{0}_B$, we use a feedback routine whereby we displace Alice by $\alpha = 1$, and then simultaneously performing a 1-selective $\pi$-pulse on Alice's ancilla and a 0-selective $\pi$-pulse on Bob's ancilla. We then perform unselective $\pi$-pulses on both ancillas and read out their state. If the cavities are in $\ket{1}_A\ket{0}_B$, we should find the ancillas in $\ket{g}_A\ket{g}_B$. If successful, we repeat the $\pi$-pulses and readout once more, and continue if the result is once again $\ket{g}_A\ket{g}_B$. If either of these measurements fails, we wait long enough to let Alice relax to vacuum and restart from the beginning.

\section{Fitting procedure for coherence times}
\label{app:pumped_decoherence}
We characterize the beamsplitter interaction by fitting the model in Eq.~(\ref{P10}) to measurements like those shown in Fig.~\ref{fig:concept}a-b.  The procedure has several steps.  To begin, the duration and frequency of the pump is swept to produce a chevron pattern (as can be seen from Eq.~\ref{eqn:omega}) from which we determine the precise pump frequency for a resonant beamsplitter interaction.  Then, the pump frequency is fixed at resonance and measurements of $P_{01/10}$ are collected for a variety of unevenly spaced pump duration times $t$. Readout imperfections in the data are corrected as described in App.~\ref{app:joint_prob_dist}.  

As illustrated in Fig.~\ref{fig:concept}b, we collect an oscillation of both curves by finely sampling 21 time points in a time window of duration approximately $6 t_\text{bs}$.  This process is repeated with the fine sampling centered around 41 coarsely spaced times between 0 and about $3\tau_\text{bs}$ (estimated from the unpumped cavity lifetimes).  To extract $g_\text{bs}$ and the upper and lower envelopes of the probability curves, we fit each of the finely-sampled oscillations to a sinusoid, $A\sin(2g_{\text{bs}} t +\phi)+B$.  Together, the upper and lower envelopes contain complete information about the decay and dephasing: their mean, $B$, decays with the time scale $\tau$ and their amplitude, $A$, decays at the time scale $(1/\tau+1/\tau_\phi)^{-1}$.  We fit the mean and amplitude of both $P_{01}$ and $P_{10}$ to extract $\tau$ and $\tau_\phi$. 

The above procedure has several benefits relative to fitting Eq.~(\ref{P10}) directly or fitting $P_\Sigma \equiv P_{01} + P_{10}$ and $P_{01/10}/P_\Sigma$~\cite{gao:2018}.  First, unlike fitting Eq.~(\ref{P10}) directly, the decay and dephasing time can be extracted from separate traces, thereby reducing the number of decoherence time constants extracted from each fit from two to one.  Second, unlike fitting to $P_\Sigma$ and $P_{01/10}/P_\Sigma$, the two measured traces need not be combined.  This allows decay and dephasing times to be inferred separately from $P_{01}$ and $P_{10}$ and compared as a consistency check.  

\begin{figure}
\begin{center}
\includegraphics[width=1\linewidth]{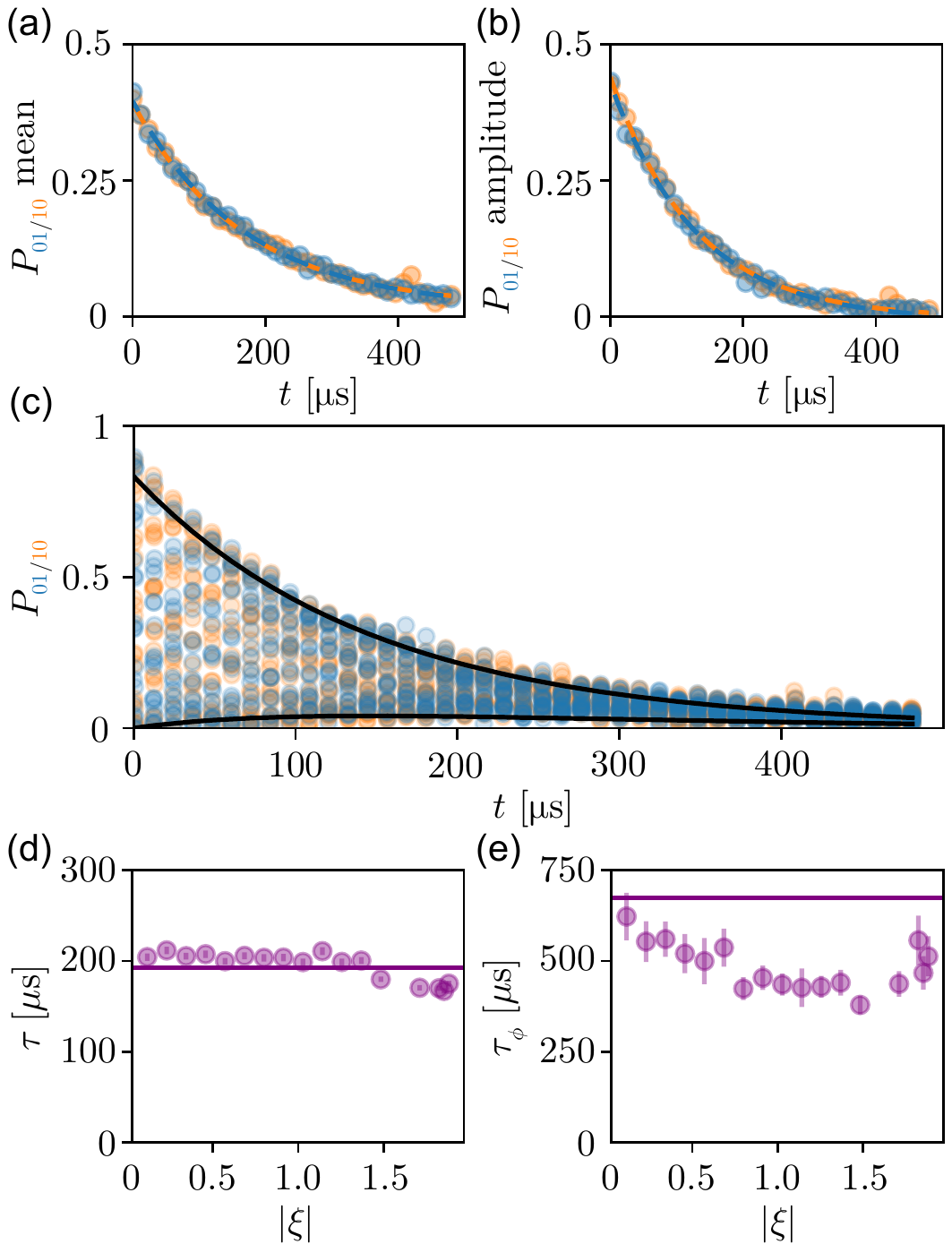}
\caption {\textbf{Fitting pumped decay and dephasing times}. (a) The decay of the mean of $P_{01}$ (blue circles) and $P_{10}$ (orange circles) is used to extract the decay time $\tau$. (b) The decay of the amplitude of $P_{01}$ (blue circles) and $P_{10}$ (orange circles) is used to extract $(1/\tau+1/\tau_\phi)^{-1}$, from which we obtain the dephasing time $\tau_\phi$. Dashed lines show the exponential decay fits. The decay constants for the mean of $P_{01/10}$ are $170\pm5~\mu$s and $167\pm7~\mu$s, respectively, while the decay constants for the amplitude are $132\pm4~\mu$s and $127\pm4~\mu$s. This yields an average fitted $\tau = 168\pm4~\mu$s and $\tau_\phi = 560\pm70~\mu$s. (c) The complete trace of the measured $P_{01/10}$ curves as a function of the variable delay $t$, with envelopes (black lines) obtained from the fitted decay and dephasing times. (d-e) Fitted decay times $\tau$ (d) and dephasing times $\tau_\phi$ (e) as a function of pump amplitude $\xi$. The solid lines shows the average unpumped decay and dephasing time of the cavities. Data are measured at the same flux point and pump amplitude as in Fig.~\ref{fig:concept}b.
}
\label{fig:taus}
\end{center}
\end{figure}

To illustrate the intermediate steps in the fitting process, Fig.~\ref{fig:taus}a shows the mean of $P_{01}$ and $P_{10}$, and Fig.~\ref{fig:taus}b shows their amplitude, measured at $\Phi_\text{e}/\Phi_0 = 0.32$ and with $|\xi| = 1.83$.  time scales inferred from separate fits agree within their standard error; quoted values for $\tau$ and $\tau_\phi$ are the average of the two.  To validate the results, we plot the readout-corrected traces for $P_{01/10}$ in Fig.~\ref{fig:taus}c, along with the envelopes inferred from the fitted time constants in Fig.~\ref{fig:taus}a-b.  

Fig.~\ref{fig:taus}d-e shows the time constants for decay from the single-photon manifold (d) and dephasing (e) extracted in this way, as a function of the pump amplitude.  Decay times are relatively independent of pump amplitude.  Dephasing times are slightly suppressed as pump amplitude increases, but remain much longer than the decay time $\tau$.

\section{System parameters}
\begin{figure}
\begin{center}
\includegraphics[width=1\linewidth]{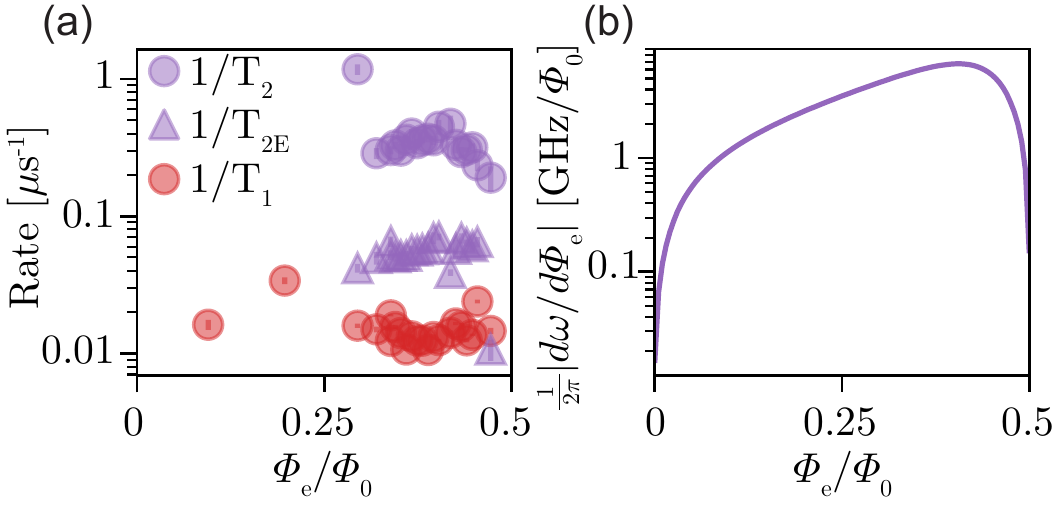}
\caption {\textbf{Dependence of SNAIL coupler coherence on external flux}. (a) SNAIL $1/T_1$ (red circles), $1/T_2$ (purple circles) and $1/T_{2E}$ (purple triangles) as a function of external flux through the SNAIL loop. (b) Simulated gradient of SNAIL frequency with respect to flux, which determines its sensitivity to dephasing due to flux noise through the loop. $T_{2E}$ is greatly increased near the flux sweet spot.
}
\label{fig:coherences}
\end{center}
\end{figure}

Tables~\ref{tab:time scale} and~\ref{tab:thermal_pops} list the coherence times and thermal populations for the experimental system. The coherence rates of the coupler, which depend on the external flux bias, are plotted in Fig.~\ref{fig:coherences}a. The decay times $T_1$ (red circles) are relatively independent of flux, but generally vary between 50 and 100 microseconds. The dephasing rates, however, exhibit a clear suppression near the flux sweet spot at $\Phi_0/2$, as expected from the flux-dependence of the coupler frequency $d\omega_c/d\Phi_\text{e}$ shown in Fig.~\ref{fig:coherences}b. Table~\ref{tab:couplings} lists the self and cross-Kerr interactions of the transmons which are not plotted in the main text.

In Table~\ref{tab:time scale}, we see that the relaxation times $T_{1c}$ of the two oscillators differ by a factor of 5. We first note the difference in oscillator resonant frequencies (2.976~GHz and 6.915~GHz), which would lead to a factor of 2.32 difference in oscillator lifetime given the same quality factor. Further, we measured the quality factor of the bare cavities (before any chips were inserted) to be 43 million and 20 million respectively, which contributes another factor of 
2.15 difference in oscillator relaxation time. Together, the differences in the oscillator resonant frequencies and bare quality factors are consistent with the factor of 5 difference in oscillator lifetime. However, these differences do not fully explain why the lifetime of either oscillator is lower than its bare lifetime.

This is not yet fully understood but there are a number of possible mechanisms that could degrade the quality factors. We estimate the contributions of Purcell decay due to the transmons and SNAIL from experimentally measured values of these modes' $T_1$ and their coupling to each oscillator, and find them to be negligible. Further, since we see no change in cavity $T_1$ times as we tune the external flux bias and thus the detuning between the SNAIL and the cavity, it is unlikely that the cavity $T_1$s is Purcell-limited by the SNAIL.

We also perform finite element simulations of the Purcell decay due to the Purcell filters, readout resonators and buffer mode, as well as direct radiative loss out of the qubit drive pins. These too only limit the cavity lifetimes to time scales that are an order of magnitude larger than the observed values.

The main contributions we find, besides the still-dominant bare loss rate of the cavities, are the radiative loss out of the cavity drive pin, whose coupling rate was designed to be slightly undercoupled, and the conductive losses in the magnet coil. Nevertheless, finite element simulation suggests these combined effects should not limit the oscillator lifetimes to less than 2.66~ms (Alice) and 0.81~ms (Bob).

Comparing $T_{1c}$ across cooldowns in which the experimental setup was changed can provide a way to study the sources of loss in this complex system. One notable data point is a cooldown in which both transmon chips were inserted but the SNAIL chip and the magnet coil were removed. Here we measured Bob’s $T_{1c}$ to be $333\pm18$~$\upmu$s, lower than the bare cavity $T_1$ of $457\pm7$~$\upmu$s, but higher than when the SNAIL and magnet are included ($91\pm4$~$\upmu$s). However, large run-to-run variations in cavity $T_1$ on cooldowns with the same package and chips makes it hard to precisely identify which changes are responsible. For example, on another cooldown with the SNAIL and magnet included, we measured Bob’s $T_{1c}$ to be $265\pm27$~$\upmu$s.

\label{app:params}
\begin{center}
\begin{table*}[ht]
\begin{tabular}{|c|c|c|c|}
\hline
 & & \makecell{Alice} & \makecell{Bob} \\
\hline
\makecell{Transmon $e \xrightarrow{} g$ relaxation time} & \makecell{$T_{1q}$} & \makecell{$ 127.2 \pm 1.9 ~\upmu s$} & \makecell{$57.1 \pm 0.6 ~\upmu s$}\\
\hline
\makecell{Transmon $T_2$} & \makecell{$T_{2q}$} & \makecell{$114.4 \pm 2.9 ~\upmu s$} & \makecell{$56.8 \pm 1.5 ~\upmu s$} \\
\hline
\makecell{Transmon dephasing time} & \makecell{$T_{\phi q}$} & \makecell{$208 \pm 10 ~\upmu s$} & \makecell{$113 \pm 6 ~\upmu s$} \\
\hline
\makecell{Oscillator relaxation time} & \makecell{$T_{1c}$} & \makecell{$482 \pm 16 ~\upmu s$} & \makecell{$91 \pm 4 ~\upmu s$} \\
\hline
\makecell{Oscillator dephasing time} & \makecell{$T_{\phi c}$} & \makecell{$2010 \pm 220 ~\upmu s$} & \makecell{$840 \pm 200 ~\upmu s$} \\
\hline
\makecell{Parity mapping time} & \makecell{$\tau_{p}$} & \makecell{$616$~ns} & \makecell{$432$~ns} \\
\hline
\makecell{Transmon readout time} & \makecell{$\tau_{RO}$} & \multicolumn{2}{c|}{$2.1 ~\upmu s$} \\
\hline
\makecell{cSWAP sequence length} & \makecell{$\tau_{cSWAP}$} & \multicolumn{2}{c|}{$1.3 ~\upmu s$} \\
\hline
\end{tabular}
\caption{\textbf{Experimental time scales.} All quoted coherence times were measured when the cSWAP experiment in Fig.~\ref{fig:cSWAP}e was performed. Coherence times drifted between then and the repeated cSWAP experiment in Fig.~\ref{fig:cSWAP}f. To compare with theoretical estimates for the repeated cSWAP experiment, we measured all coherence times again. The following list details those coherence times which changed by more than one standard deviation from the values in the table: $T_{1qA} = 112 \pm 8 \mu s$, $T_{1qB} = 45.6 \pm 9 \mu s$, $T_{2qA} = 97 \pm 10 \mu s$, $T_{\phi cA} = 1510 \pm 90 \mu s$. }
\label{tab:time scale}
\end{table*}
\end{center}

\begin{center}
\begin{table}[ht]
\begin{tabular}{|c|c|c|}
\hline
 & \makecell{Alice} & \makecell{Bob} \\
\hline
\makecell{Transmon $P_e$} & \makecell{$0.70 \pm 0.14~\%$} & \makecell{$1.02 \pm 0.20~\%$}\\
\hline
\makecell{Cavity $n_\text{thermal}$} & \makecell{$0.96 \pm 0.19~\%$} & \makecell{$0.11 \pm 0.02~\%$} \\
\hline
\makecell{SNAIL $P_e$} & \multicolumn{2}{c|}{$2.42 \pm 0.47~\%$} \\
\hline
\end{tabular}
\caption{\textbf{Mode thermal populations.} All values measured at $\Phi_\text{e} = 0.35~\Phi_0$, the same external flux as for the cSWAP data in Fig.~\ref{fig:cSWAP}.}
\label{tab:thermal_pops}
\end{table}
\end{center}

\begin{center}
\begin{table*}[ht]
\begin{tabular}{|c|c|c|c|}
\hline
 & & \makecell{Alice} & \makecell{Bob} \\
\hline
\makecell{Transmon-oscillator dispersive shift} & \makecell{$\chi_{qc}/2\pi$} & \makecell{$-766.4(5)$ kHz}& \makecell{$-1104.0(5)$ kHz}\\
\hline
\makecell{Transmon anharmonicity} & \makecell{$K_{q}/4\pi$} & \makecell{$-181.2537(4)$ MHz} & \makecell{$-184.2860(3)$ MHz} \\
\hline
\end{tabular}
\caption{\textbf{Mode Couplings}}
\label{tab:couplings}
\end{table*}
\end{center}

\section{Delivery of strong microwave drives}
\label{app:pumps}
Delivering a strong microwave drive to the SNAIL is necessary to achieve a fast beamsplitter rate. However, it is important that the drive port does not Purcell limit the adjacent high-Q cavity modes. Furthermore, the drive power we can deliver to the package is limited by the amount of power we can dissipate in the microwave attenuators before they start to heat the base temperature of the dilution refrigerator.

To satisfy these constraints, we filter the environment, as seen by the SNAIL and the adjacent cavity modes, by introducing a 3D superconducting stripline resonator between the drive port and the SNAIL, with a resonance frequency $\omega_f$ close to $\omega_b - \omega_a$. This allows drive photons near 4~GHz to reach the SNAIL while shielding oscillator photons at 3~GHz and 7~GHz from the lossy drive line. (Care is taken that the stripline's resonance frequency is sufficiently detuned from $\omega_\text{b} - \omega_\text{a}$ relative to its linewidth, to avoid unpumped conversion of photons in Bob).

To confirm that the drive port does not Purcell limit the oscillators, we perform a finite-element eigenmode simulation of the package with the 3D stripline resonator. From this we extract the coupling $Q$ of the oscillators and verify that they exceed $10^8$ (limiting their $T_1$ times to no less than 5.3~ms and 2.3~ms for Alice and Bob respectively).

\begin{figure}[!thb]
\begin{center}
\includegraphics[width=1\linewidth]{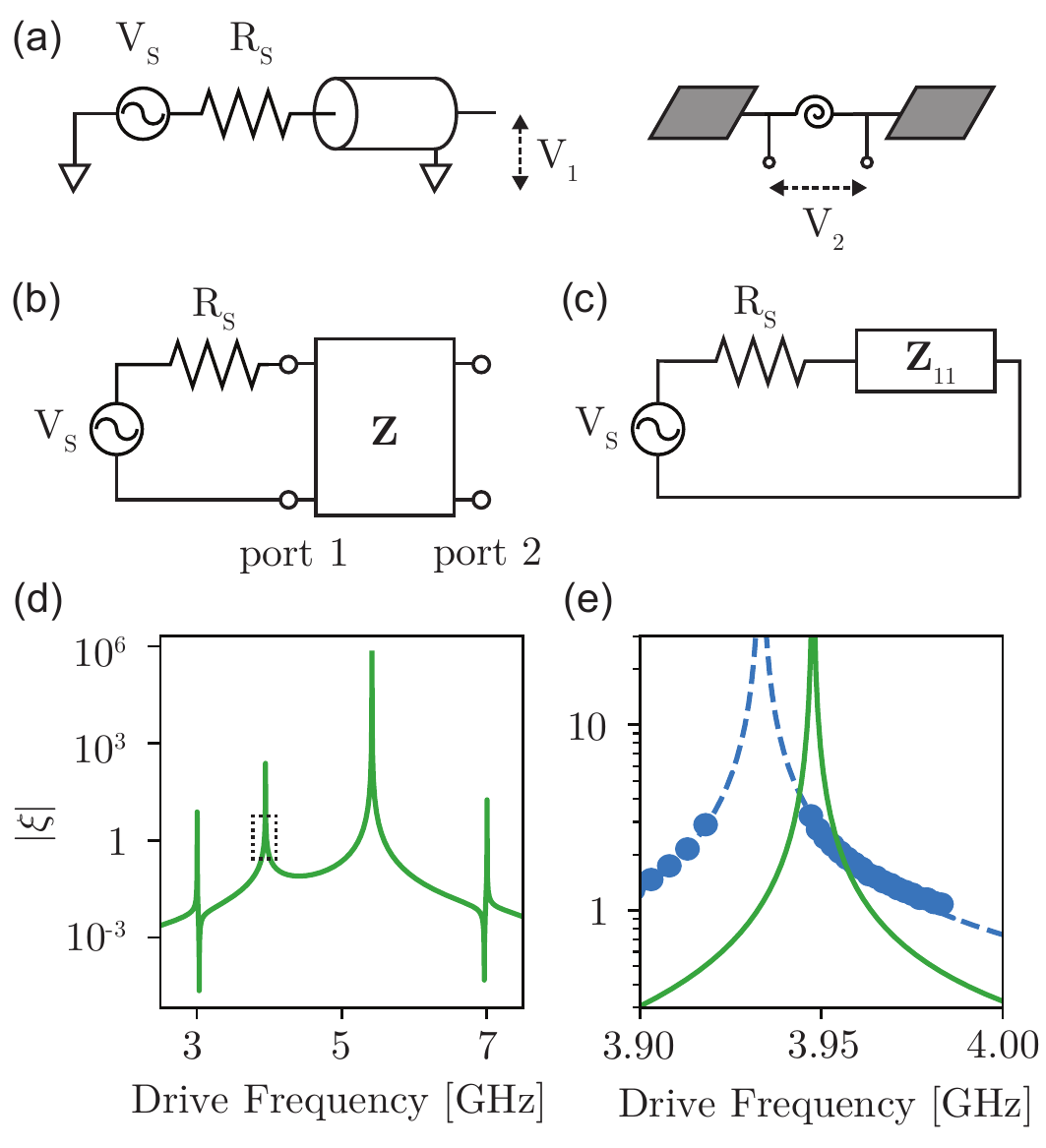}
\caption {\textbf{Network picture of driven system}. (a) A microwave source with a voltage $V_S$ and a resistance $R_S$ connected via a transmission line to the drive pin of the package. It generates a voltage $V_1$ at the drive pin (port 1) and a voltage $V_2$ across the SNAIL (port 2). (b) The same circuit with the package, including the SNAIL itself, expressed as a two-port network with an impedance matrix $\mathbf{Z}$. (c) When port 2 is open-circuited, we can treat the package as an impedance $Z_{11}$ in series with the microwave source. (d) The simulated normalized pump amplitude $|\xi|$ with 100~pW delivered to the drive port, as a function of drive frequency, when the external applied flux is 0.35$\Phi_0$. The resonances at 3 and 7~GHz are due to Alice and Bob's cavity modes, while the resonance at 5.2~GHz is due to the SNAIL mode. The peak near 4~GHz is due to the stripline resonator we introduce as a single-mode bandpass filter between the drive port and the SNAIL. (e) A zoom-in of the dotted region in (d), showing simulated prediction (green line), measured values extracted from measurements of Stark shift (blue dots), and simulated prediction shifted down in frequency by 14.3~MHz and up in amplitude by 2.84, to highlight the similar lineshape to the data (blue dashed line).}
\label{fig:Z11}
\end{center}
\end{figure}

The normalized pump amplitude $\xi$ that we can deliver for a given input power can be calculated from the impedance matrix of the embedding network of the SNAIL. 
Fig.~\ref{fig:Z11}a shows a simplified schematic of the microwave drive on the SNAIL coupler. A microwave generator sourcing a voltage $V_S$ generates a voltage $V_1$ at the coupling pin inside the experimental package, which in turn generates a voltage $V_2$ across the SNAIL. Their ratio is determined by the properties of the embedding network of the SNAIL.

If we consider just the linear part of the circuit, we can treat the entire package including the SNAIL itself as an impedance matrix $\mathbf{Z}$, as shown in Fig.~\ref{fig:Z11}b. This can be used to relate the voltage across the SNAIL to the current delivered from the source,
\begin{equation}
  Z_{21} = \frac{V_2}{I_1}.
\end{equation}

Meanwhile, the impedance presented to the source by the package is $Z_{11}$ (Fig.~\ref{fig:Z11}c). Current conservation indicates that the source voltage $V_S$ is related to the current $I_1$ by
\begin{equation}
  V_S = \left(R_S + Z_{11}\right)I_1.
\end{equation}
We can eliminate $I_1$ from the equations to obtain
\begin{equation}
  \frac{V_2}{V_S} = \frac{Z_{21}}{R_S + Z_{11}}.
\end{equation}
Microwave generators typically display the RMS power they would deliver to a matched load,
\begin{equation}
  P = \frac{|V_{\text{RMS}}|^2}{2R_S} = \frac{|V_S|^2}{4R_S},
\end{equation}
allowing us to reach an expression for $V_2$ in terms of source power
\begin{equation}
  |V_2| = \left|\frac{Z_{21}}{Z_{11}+R_S}\right|\sqrt{4PR_S}.
\end{equation}

Assuming a monochromatic source at $\omega_p$, the current passing through the SNAIL can be found by dividing $V_2$ by the impedance of the SNAIL at the pump frequency, $i\omega_pL_s$,
\begin{equation}
  I(t) = \text{Re}\left(\frac{V_2}{i\omega_pL_s}e^{i\omega_p t}\right).
\end{equation}

This couples to the flux of the coupler mode in the Hamiltonian
\begin{equation}
  \hat{H}_d = I(t)\hat{\Phi} = \text{Re}\left(\frac{ V_2}{i\omega_pL_s}e^{i\omega_p t}\right)\Phi^{\text{ZPF}}_c\left(\hat{c}+\hat{c}^{\dagger}\right).
\end{equation}
By comparing this to Eq.~(\ref{eqn:drive}), we can write
\begin{equation}
  |\epsilon| = \frac{\Phi^{\text{ZPF}}_c |V_2|}{2 \hbar \omega_p L_s}.
\end{equation}
From here, we can use the expression for $\xi$ from Eq.~(\ref{eqn:xi}) to finally get
\begin{equation}
  |\xi| = \left|\frac{Z_{21}}{Z_{11}+R_S}\right|\frac{\sqrt{PR_S}\Phi^{\text{ZPF}}_c}{\hbar\omega_pL_s\left(\omega_p-\omega_c\right)}.
\end{equation}

The impedance matrix $\mathbf{Z}$ can be obtained from a simulation of the package in an electromagnetic solver such as HFSS, while the zero point flux of the SNAIL mode can be found by combining an electromagnetic simulation with further processing, as performed for example in the pyEPR package~\cite{minev:2021}. Fig.~\ref{fig:Z11}d shows $|\xi|$ as a function of drive frequency for an input power of 100~pW at the drive port of the package. The inset shows how the bandpass filter near 4~GHz allows us to obtain $|\xi|>1$ over a 40MHz bandwidth near the designed pump frequency. The simulation matches measurements of $\xi$ (found by measuring coupler Stark shift and comparing to its anharmonicity) up to a shift of 14.3~MHz in frequency and a factor of 2.84 in power. We attribute the discrepancies to our uncertainty in the cold attenuation of the microwave lines and the precise location of the coupler chip after mounting.

\section{Calibrating delays during cSWAP}

\label{app:delay_calibration}
As described in Sec.~\ref{sec:cond}, during the cSWAP pulse the mode operators $\hat{a}$ and $\hat{b}$ will exchange if the ancilla is in $\ket{e}$ and stay the same if the ancilla is in $\ket{g}$, but they will also acquire a dynamical phase that differs depending on the ancilla state. This phase can be obtained by calculating half the solid angle enclosed by the trajectory of the mode operators on the Bloch sphere. A phase $\phi$ applied to a mode operator ($\hat{a}\rightarrow e^{i\phi}\hat{a}$) corresponds to a rotation of the phase space of that mode by an angle $\phi$. In order to perform just a cSWAP, we need to correct for this ancilla-state-dependent phase.

This can be done by adding delays to the sequence. Due to the dispersive interaction, Bob's state will rotate by an angle $\phi = \chi t$ during a delay of length $t$ only when Bob's transmon is excited. Since the cSWAP pulse exchanges the oscillator states when the transmon is in $\ket{e}$, a delay after the pulse rotates Bob's final state whereas a delay before the pulse rotates Alice's final state. Adding two delays, $t_1$ before the pulse and $t_2$ after the pulse, is sufficient to deterministically erase the ancilla-state-dependent phase. There is a remaining phase on the mode operators that is independent of the ancilla state, but this can be removed by applying a frame update to each cavity in software, by a pre-determined phase.

The delays and the frame updates are calibrated by initializing the cavities in $D(\alpha)\ket{1}\otimes D(-\alpha)\ket{0}$, performing the cSWAP pulse with the ancilla in either $\ket{g}$ or $\ket{e}$ and measuring Wigner functions of each mode, as shown in the center columns of Fig.~\ref{fig:cSWAP}c. The angle in phase space between the fitted center of the displaced Fock state and its expected position on the $x$-axis, for both initial ancilla states, allows us to determine the delays and frame updates required. We repeat this process with these delays and frame updates applied to confirm that we produce the expected Wigner function.

Empirically, the dynamical phase described above is not the only phase that needs to be accounted for. For example, the phase offset of the beamsplitter drive relative to the difference between the cavity reference signals (either due to different electrical lengths in the cables, or due to a phase offset we control in the DAC) can advance or retard the phase of the mode operators when a SWAP is performed. Additionally, when the beamsplitter drive is applied, the cavities are Stark shifted by differing amounts, again imparting differing phases to the mode operators. We find that the magnitude of these Stark shifts is also dependent on the ancilla state.

The degrees of freedom under our control, namely the pre- and post-delays, the frame updates on each cavity and the pump phase, are sufficient to account for the phases acquired during a single cSWAP. However, these same parameters will not correct the phases on a subsequent cSWAP---a new set of parameters is required.

To correct the phase for a series of $N$ cSWAPs, we could calibrate the delay required before and after every cSWAP pulse, as well as the frame rotations associated with each one. Alternatively however, a single delay before and after the entire sequence, as well as a single frame rotation on each cavity, is sufficient. Since this allows us to perform a long sequence while calibrating fewer parameters, this is what we opt to do in the experiment, as shown in Fig.~\ref{fig:cSWAP}d. In order to calibrate these delays and frame rotations, we use the same procedure outlined above, but replace the single cSWAP pulse with $N$ of them. Since the duration of the single delay before and after depends on the number of cSWAP pulses performed, the total duration of the sequence in not linear in $N$, as evidenced by the uneven top $x$-axis scale in Fig.~\ref{fig:cSWAP}f.

\section{Joint Wigner tomography on a Bell state in the coherent-state basis}
\label{app:jw}
\begin{figure}
\begin{center}
\includegraphics[width=1\linewidth]{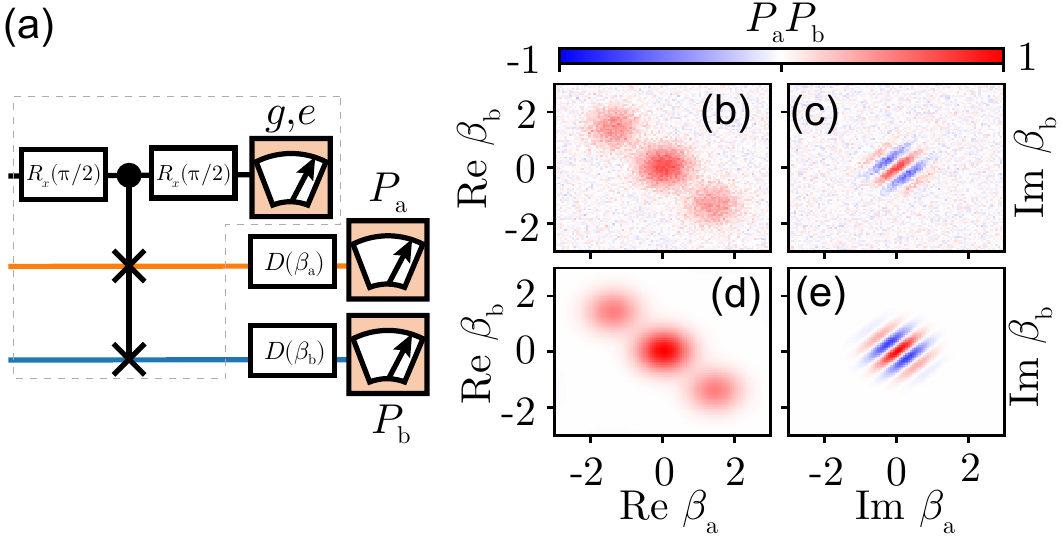}
\caption {\textbf{Joint Wigner tomography on $\Psi^+$}. (a) SWAP test sequence (in grey dashed lines) followed by joint Wigner sequence. The SWAP test is composed of a $\pi/2$ pulse on the control ancilla, followed by the cSWAP gate, the reverse $\pi/2$ pulse on the control ancilla, and its measurement. (b-e) Plane-cuts from the measured (b-c) and simulated (d-e) joint Wigner distribution of $\ket{\Psi^+}$. Correlations between the real quadratures of the two cavities are shown in (b,d) in the plane $\text{Im} \beta_\text{A} = 0 = \text{Im} \beta_\text{B}$, and correlations between the imaginary quadratures are shown in (c,e) in the plane $\text{Re} \beta_\text{A} = 0 = \text{Re} \beta_\text{B}$. 
}
\label{fig:jw}
\end{center}
\end{figure}

To demonstrate the quantum nature of the controlled-SWAP gate, we perform a SWAP test followed by joint Wigner tomography on the cavities. The sequence is shown in Fig.~\ref{fig:jw}a. Linecuts of the measured displaced parities along the $\text{Im} \beta_a = \text{Im} \beta_b = 0$ plane are shown in Fig.~\ref{fig:jw}b, and linecuts along the $\text{Re} \beta_a =\text{Re} \beta_b = 0$ plane are shown in Fig.~\ref{fig:jw}c. The measurements compare well with simulations of the ideal distributions (i.e. simulations without decoherence) shown in Fig.~\ref{fig:jw}d-e.

\section{Pauli Tomography}\label{Pauli_tomography}
\label{app:Paulibars}

For a single oscillator state, the results of Wigner tomography can be used to perform a maximum likelihood reconstruction of the density matrix. For many-oscillator states, though, the cost of this technique scales exponentially and to fully characterize even a two-oscillator density matrix would take several days.

If, however, we choose a qubit basis within the Hilbert space of each oscillator, we can obtain expectation values of all the Pauli operators by probing the Wigner function in just a handful of points in phase space \cite{wang:2016, vlastakis:2015}. In the coherent state basis $\{\ket{\alpha}, \ket{-\alpha}\}$, single-oscillator tomography can be simplified by conducting $4$ displaced parity measurements whose linear combinations approximate Pauli measurements.

To find the displacement $\beta$ needed to perform a Pauli measurement, we first apply the projection operator
\begin{equation}
  \hat{M} = \left(|\alpha\rangle\langle\alpha| + |-\alpha\rangle\langle-\alpha|\right),
\end{equation}
to the displaced parity operator $\hat{P}_\beta \equiv \hat{D}(\beta)\hat{P}\hat{D}(\beta)^\dagger$ to project it onto the coherent state basis 
\begin{equation}
  \hat{M}\hat{P_{\beta}}\hat{M}^\dagger = \begin{pmatrix}
                          \langle \alpha|P_{\beta}|\alpha\rangle & \langle \alpha|P_{\beta}|-\alpha\rangle \\
                          \langle -\alpha|P_{\beta}|\alpha\rangle & \langle -\alpha|P_{\beta}|-\alpha\rangle
                      \end{pmatrix}.
\end{equation}
We then find a set of displacements $\{\beta_i\}$ such that $\sum_i b_i \hat{M}\hat{P}_{\beta_i}\hat{M}^\dagger \in \{I, X, Y, Z\}$ for some coefficients $b_i$. (Here $I$, $X$, $Y$, and $Z$ are Pauli matrices; we omit the hats on these operators for brevity.)

When finding these displacements, the following relations prove useful:
\begin{align}
  \langle \beta |P_{0}|\beta\rangle = \langle \beta |-\beta \rangle &= e^{-2|\beta|^2}, \\
  \langle \beta |P_{0}|-\beta\rangle = \langle \beta|\beta \rangle &= 1, \\
  \langle \alpha|P_{\beta}|\alpha\rangle = \langle \alpha - \beta|-\alpha+\beta \rangle &= e^{-2|\alpha-\beta|^2}, \label{eqn:aba}\\
  \langle \alpha|P_{\beta}|-\alpha\rangle = \langle \alpha-\beta|\alpha+\beta\rangle &= e^{2(\alpha^*\beta - \beta^*\alpha)}\langle \beta|-\beta\rangle. \label{eqn:ab-a}
\end{align}
We note that these properties are identical to those quoted in Ref.~\cite{vlastakis:2015} except that in Eq.~(\ref{eqn:aba}) we do not conclude that $\langle \alpha|P_{\beta}|\alpha\rangle \ll 1$ (e.g. when $\beta = \alpha, \langle \alpha|P_{\beta}|\alpha\rangle=1$).

We now choose to write the identity matrix $I$ as the sum of $\big(\begin{smallmatrix}
 1 & 0 \\
 0 & 0
\end{smallmatrix}\big)$ and $\big(\begin{smallmatrix}
 0 & 0 \\
 0 & 1
\end{smallmatrix}\big)$, and Pauli $Z$ matrix as the difference.
Setting $\hat{M}\hat{P}_{\beta_1}\hat{M}^\dagger = \big(\begin{smallmatrix}
 1 & 0 \\
 0 & 0
\end{smallmatrix}\big),$
we must satisfy $\langle \alpha|\hat{P}_{\beta_1}|\alpha\rangle = 1$, which sets $\beta_1 = \alpha$.
We verify that $\langle \alpha|\hat{P}_{\alpha}|-\alpha\rangle = \langle -\alpha|\hat{P}_{\alpha}|\alpha\rangle \approx 0$ and $\langle -\alpha|\hat{P}_{\alpha}|-\alpha\rangle \approx 0$.
Similarly, setting $\hat{M}\hat{P}_{\beta_2}\hat{M}^\dagger = \big(\begin{smallmatrix}
 0 & 0 \\
 0 & 1
\end{smallmatrix}\big),$ we obtain $\beta_2 = -\alpha$.

$I$ can be constructed by summing the above two displaced parity measurements, 
\begin{equation}
  I = \hat{M}\hat{P}_{\alpha}\hat{M}^\dagger + \hat{M}\hat{P}_{-\alpha}\hat{M}^\dagger,
\end{equation}
while $Z$ can be constructed by taking the difference,
\begin{equation}
  Z = \hat{M}\hat{P}_{\alpha}\hat{M}^\dagger - \hat{M}\hat{P}_{-\alpha}\hat{M}^\dagger.
\end{equation}

We then find the displacement required to measure $X$ by setting $\hat{M}\hat{P}_{\beta_3}\hat{M}^\dagger = \big(\begin{smallmatrix}
 0 & 1 \\
 1 & 0
\end{smallmatrix}\big),$ which is satisfied by $\beta_3 = 0$.
Thus measuring photon number parity without any displacement is equivalent to a Pauli X measurement in this basis:
\begin{equation}
  X = \hat{M}\hat{P_{0}}\hat{M}^\dagger.
\end{equation}

Finally, we solve $\hat{M}\hat{P}_{\beta_4}\hat{M}^\dagger = \big(\begin{smallmatrix}
 0 & -i\\
 i & 0
\end{smallmatrix}\big)$ to obtain a Pauli $Y$ measurement. 
To ensure the correct relative phase between $\langle \alpha |P_{\beta_4}|-\alpha\rangle$ and $\langle -\alpha |P_{\beta_4}|\alpha\rangle$, we need $\beta_4 = \frac{ik\pi}{8\alpha}$ where $k$ is any odd integer.
However, choosing a larger $|\beta_4|$ results in smaller magnitude for $\langle \alpha |P_{\beta_4}|-\alpha\rangle$ and $\langle -\alpha |P_{\beta_4}|\alpha\rangle$. To maximize off-diagonal matrix elements, we thus choose $\beta_4 = \frac{i\pi}{8\alpha}$.

For $\alpha = \sqrt{2}$, this choice of $\beta_4$ yields off-diagonal elements with a magnitude of $\approx 0.86$, so the single-oscillator Pauli $Y$ measurement approximated by this displaced parity measurement has a maximum amplitude of $\approx 0.86$.
For general $\alpha$,
\begin{equation}
  \hat{M}\hat{P}_{\frac{i\pi}{8\alpha}}\hat{M}^\dagger= \left \langle \frac{i\pi}{8\alpha}\middle |-\frac{i\pi}{8\alpha}\right \rangle Y =e^{-\frac{\pi^2}{32\alpha^2}}Y.
\end{equation}

\begin{figure*}[!thb]
\begin{center}
\includegraphics[width=1\linewidth]{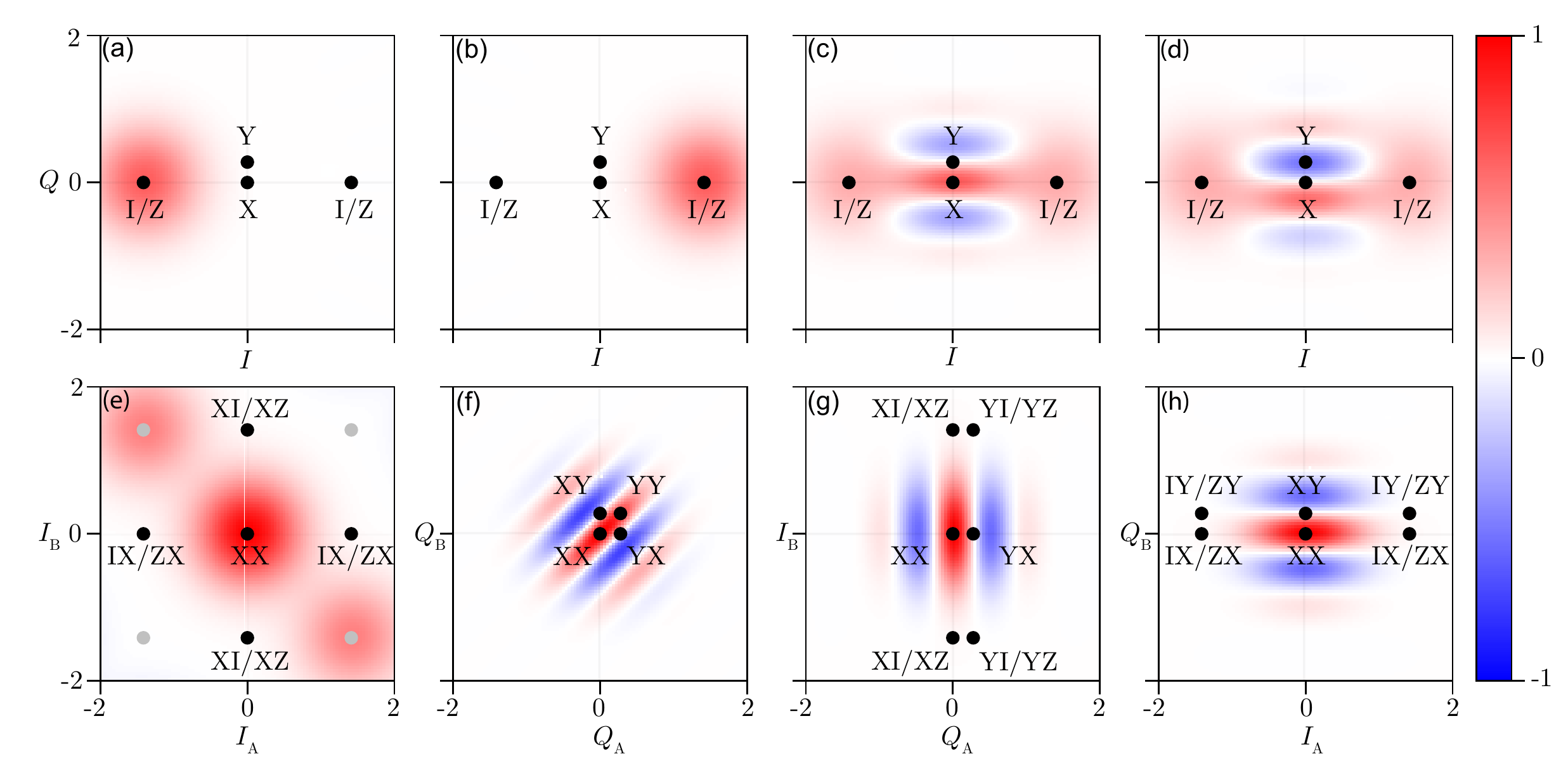}
\caption {\textbf{Local and Joint Wigner Plots} Top row shows displaced photon number parity of single oscillator states in the coherent state basis (a) $\ket{-Z} = \ket{-\alpha}$, (b) $\ket{+Z} = \ket{\alpha}$, (c) $\ket{+X} = \frac{\ket{\alpha} +\ket{-\alpha}}{N}$ and (d) $\ket{+Y} = \frac{\ket{\alpha} +i\ket{-\alpha}}{N}$ for $\alpha = \sqrt{2}$. The black dots indicate the positions at which the displaced parity is sampled to obtain measurements of the single oscillator Pauli operators. Bottom row shows slices of the displaced joint photon number parity of the two-oscillator state $\ket{\Psi^{+}}$ along (e) real axes of both oscillators, (f) imaginary axes of both oscillators, (g) imaginary axis of Alice and real axis of Bob and (h) vice versa. Here, the black dots indicate the positions at which displaced joint parity is sampled to obtain measurements of the two-oscillator Pauli operators. $II/ZZ/IZ/ZI$ measurements are obtained from linear combinations of measurements at the four grey dots.}
\label{fig:wigner_slices}
\end{center}
\end{figure*}

For a visual interpretation of how displaced parity measurements can be used to approximate Pauli measurements, we mark the displacements of interest on Wigner functions of the states $\ket{+Z}, \ket{-Z}, \ket{+X}, \ket{+Y}$ in Fig.~\ref{fig:wigner_slices}.

To move from a single-oscillator measurement to a two-oscillator measurement, we need all $16$ combinations of single-oscillator Pauli measurements $\{I, X, Y, Z\}$. 
The $16$ displacements whose linear combination make up the two-oscillator Pauli measurements are labeled in Fig.~\ref{fig:wigner_slices}e-h. Note that the two-oscillator measurement of YY has a maximum of amplitude of $e^{-\frac{\pi}{16\alpha^2}} \approx 0.73$ for $\alpha = \sqrt{2}$.

\section{cSWAP error budget}
\label{app:error_budget}

For both the Bell state preparation and the control experiment where we perform a SWAP-test on $\ket{\alpha, \alpha}$ (see Fig.~\ref{fig:cSWAP}c-f), we construct an error budget based on independent measurements of coherence times (summarized in Table. \ref{tab:time scale}), the duration of each part of the experimental sequence, and the sensitivity of each measurement to various error channels.

\begin{center}
\begin{table*}[h!]
\setlength{\tabcolsep}{0pt} 
\begin{tabular}{"m{2em}|m{10em}|m{12em}"m{7em}|m{7em}|m{7em}|m{7em}"}
\thickhline
\makecell{}&\makecell{error channel}& \makecell{probability of occurrence} & \makecell{estimated \\ (simulated) \\ effect on XX} &\makecell{estimated \\ (simulated) \\ effect on YY} & \makecell{estimated \\ (simulated) \\effect on II} &\makecell{estimated \\ (simulated) \\effect on ZZ}\\
\thickhline
\multirow{4}{*}[-8ex]{\rotatebox[origin=c]{90}{\makecell{cSWAP}}} & \makecell{oscillator decay \\ $\hat{a}$, $\hat{b}$} & \makecell{$\bar{n}\left(\frac{\tau_{cSWAP}}{T_{1cA}}+\frac{\tau_{cSWAP}}{T_{1cB}}\right)$ \\ $\approx 3.5\% \pm 0.1\%$} & \makecell{$3.5(1)\%$ \\ ($3.7\%$)} & \makecell{$2.57(7)\%$ \\ ($2.7\%$)} & \makecell{$0\%$ \\ ($0.2\%$)} & \makecell{ $0\%$ \\($0.3\%$)} \\ \cline{2-7}
& \makecell{oscillator dephasing \\ $\hat{a}^\dag\hat{a}$, $\hat{b}^\dag\hat{b}$} & \makecell{$2 \bar{n}\left(\frac{\tau_{cSWAP}}{T_{\phi cA}}+\frac{\tau_{cSWAP}}{T_{\phi cB}}\right)$ \\ $\approx 0.90\% \pm 0.15 \%$} & \makecell{$0\%$ \\ ($0.3\%$)} & \makecell{$0\%$ \\ ($0.1\%$)} & \makecell{$1.8(3)\%$ \\ ($1.7\%$)} & \makecell{$1.8(3)\%$ \\($1.8\%$)} \\ \cline{2-7}
& \makecell{ancilla decay \\ $|e\rangle \xrightarrow{} |g\rangle $} & \makecell{$\frac{1}{2}\frac{\tau_{cSWAP}}{T_{1qB}}$ \\ $\approx 1.17\% \pm 0.01 \%$} & \makecell{$1.17(1)\%$ \\ ($1.3\%$)} & \makecell{$0.859(7)\%$ \\ ($0.9\%$)} & \makecell{$1.17(1)\%$ \\ ($0.8\%$)} & \makecell{$1.17(1)\%$ \\ ($1.2\%$)}\\ \cline{2-7}
& \makecell{ancilla dephasing \\ $|e\rangle\langle e|$}& \makecell{$\frac{\tau_{cSWAP}}{2T_{\phi qB}}$ \\ $\approx 0.59\% \pm 0.03\%$} & \makecell{$1.18(6)\%$ \\ ($1.3\%$)} & \makecell{$0.87(4)\%$ \\($0.9\%$)}& \makecell{$0\%$ \\ ($0.1\%$)} & \makecell{$0\%$ \\($0.2\%$)}\\
\hline

\multirow{3}{*}[-4ex]{\rotatebox[origin=c]{90}{\makecell{cSWAP RO}}} & \makecell{oscillator decay \\ $\hat{a}$, $\hat{b}$} & \makecell{$\bar{n}\left(\frac{\tau_{RO}}{T_{1cA}}+\frac{\tau_{RO}}{T_{1cB}}\right)$ \\ $\approx 4.7\% \pm 0.2\%$} & \makecell{$9.4(3)\%$ \\ ($9.0\%$)} & \makecell{$6.9(2)\%$ \\ ($6.5\%$)} & \makecell{$0\%$ \\ ($0\%$)} & \makecell{$0\%$ \\ ($0\%$)} \\ \cline{2-7}
& \makecell{oscillator dephasing \\ $\hat{a}^\dag\hat{a}$, $\hat{b}^\dag\hat{b}$} & \makecell{$2 \bar{n}(\frac{\tau_{RO}}{T_{\phi cA}}+\frac{\tau_{RO}}{T_{\phi cB}})$ \\ $\approx 1.2\% \pm 0.2\%$} & \makecell{$0\%$ \\ ($0\%$)} & \makecell{$0\%$ \\ ($0\%$)} & \makecell{$2.4(4)\%$ \\ ($1.8\%$)} & \makecell{ $2.4(4)\%$ \\ ($1.9\%$)}\\ \cline{2-7}
& \makecell{ancilla decay \\ $|e\rangle \xrightarrow{} |g\rangle $} & \makecell{$\frac{\tau_{RO}}{T_{1qB}}$ \\ $\approx 3.15\% \pm 0.03\%$} & \makecell{$6.30(7)\%$ \\ ($6.0\%$)} & \makecell{$3.15(3)\%$ \\($2.9\%$)} & \makecell{$3.15(3)\%$ \\ ($2.1\%$)} & \makecell{$3.15(3)\%$ \\ ($3.1\%$)} \\ \cline{2-7}

\hline

\multirow{3}{*}[-4ex]{\rotatebox[origin=c]{90}{\makecell{parity map}}} & \makecell{oscillator decay \\ $\hat{a}$, $\hat{b}$} & \makecell{\text{XX/YY}: $\bar{n}\left(\frac{\tau_pA}{T_{1cA}}+\frac{\tau_pB}{T_{1cB}}\right)$\\ $\approx 1.20\% \pm 0.04\%$ \\ \text{II/ZZ}: $2\bar{n}\left(\frac{\tau_{pA}}{T_{1cA}}+\frac{\tau_{pB}}{T_{1cB}}\right)$ \\ $\approx 2.4\% \pm 0.08\%$} & \makecell{$1.20(4)\%$ \\ ($1.2\%$)} & \makecell{$0.88(3)\%$ \\ ($0.8\%$)}& \makecell{$0\%$ \\ ($0\%$)} & \makecell{$0\%$ \\ ($0\%$)}\\ \cline{2-7}
& \makecell{ancilla decay \\ $|e\rangle \xrightarrow{} |g\rangle $} & \makecell{$\frac{1}{2}\left(\frac{\tau_{pA}}{T_{1qA}}+\frac{\tau_{pB}}{T_{1qB}}\right)$ \\ $\approx 0.620\% \pm 0.005\%$} & \makecell{$0.620(5)\%$ \\ ($0.6\%$)} & \makecell{$0.455(4)\%$ \\ ($0.4\%$)} & \makecell{$0.620(5)\%$ \\ ($0.5\%$)} & \makecell{$0.620(5)\%$ \\ ($0.6\%$)}\\ \cline{2-7}
& \makecell{ancilla dephasing \\ $|e\rangle\langle e|$}& \makecell{$\frac{\tau_{pA}}{2T_{\phi qA}}+ \frac{\tau_{pB}}{2T_{\phi qB}}$ \\ $\approx 0.34\% \pm 0.01\%$} & \makecell{$0.68(3)\%$ \\ ($0.6\%$)} & \makecell{$0.50(2)\%$ \\ ($0.4\%$)} & \makecell{$0.68(3)\%$ \\ ($0.5\%$)} & \makecell{$0.68(3)\%$ \\ ($0.7\%$)}\\
\hline

\rotatebox[origin=c]{90}{\makecell{parity RO}} & \makecell{ancilla decay \\ $|e\rangle \xrightarrow{} |g\rangle $} &\makecell{$\frac{1}{2}\left(\frac{\tau_{RO}}{T_{1qA}}+\frac{\tau_{RO}}{T_{1qB}}\right)$ \\ $\approx 2.28\% \pm 0.02\%$} & \makecell{$4.57(4)\%$ \\ ($4.5\%$)} & \makecell{$3.36(3)\%$ \\ ($3.3\%$)} & \makecell{$4.57(4)\%$ \\ ($4.4\%$)} & \makecell{$4.57(4)\%$ \\ ($4.5\%$)}\\
\thickhline

\multicolumn{3}{"c"}{\multirow{2}{*}{estimated gate error per round}} & \makecell{$5.7(1)\%$} & \makecell{$4.2(1)\%$} & \makecell{$2.9(3)\%$}& \makecell{$2.9(3)\%$}\\ \cline{4-7}
\multicolumn{3}{"c"}{\multirow{2}{*}{}} & 
\multicolumn{4}{c"}{\boldmath{$4.0\% \pm 0.1\%$}}\\
\thickhline

\multicolumn{3}{"c"}{estimated infidelity per round, including $K$, $\chi^{'}$} & \multicolumn{4}{c"}{\boldmath{$4.2\% \pm 0.1\%$}}\\
\thickhline

\multicolumn{3}{"c"}{\multirow{2}{*}{estimated (simulated) Bell fidelity}} & \makecell{$74.2(4)\%$ \\($75.3\%)$} & \makecell{$56.5(3)\%$ \\ ($57.1\%$)} & \makecell{$86.5(6)\%$ \\ ($88.5\%$)} & \makecell{$86.5(6)\%$ \\ ($87.4\%$)}\\ \cline{4-7}
\multicolumn{3}{"c"}{\multirow{2}{*}{}} & \multicolumn{4}{c"}{\boldmath{$75.9\% \pm 0.2\%$}}\\
\thickhline

\multicolumn{3}{"c"}{\multirow{2}{*}{measured Bell fidelity}} & \makecell{$70.4(4)\%$} & \makecell{$51.5(5)\%$} & \makecell{$87(1)\%$} & \makecell{$87(1)\%$} \\ \cline{4-7}
\multicolumn{3}{"c"}{\multirow{2}{*}{}} & \multicolumn{4}{c"}{\makecell{\boldmath{$74.1\% \pm 0.4\%$}}}\\
\thickhline

\end{tabular}
\caption{\textbf{Error budget for Bell state preparation using cSWAP.}
Columns show the predicted decrease of the measured value of non-zero joint Pauli measurements for each error channel, whose probability of occurrence is estimated by the expression in column 2. The values in parentheses are the predictions based instead on a Lindblad master equation simulation.
$\bar{n} = 2$ is the average photon number in each oscillator. The time scales used are laid out in Table. \ref{tab:time scale}. The $YY$ column has a scale factor of $73.46\%$ applied, due to the quasi-orthogonality of the basis states (see App.~\ref{app:Paulibars}.)}
\label{tab:error_budget}
\end{table*}
\end{center}

The sequence in Fig.~\ref{fig:cSWAP}d can be divided into four parts: the controlled SWAP (including delays), the readout of the control transmon, the parity mapping sequence for joint Wigner tomography, and the final transmon measurements. 

The error channels considered are decay and dephasing on both ancillae and cavities. In this analysis, we neglect cavity no-jump evolution, errors during ancilla rotations and coupler errors since we expect them to have a negligible effect on fidelities.

The Bell state fidelity is comprised of the expectation values of four two-qubit Pauli operators, each of which is affected differently by errors. For example, $\langle XX \rangle$ is the measured joint photon number parity and so is only affected by errors that change the joint parity. Likewise, $\langle YY \rangle$ is measured near the origin of the joint Wigner function and is sensitive to parity jumps but less so to phase space rotations. Any error that takes the even Bell state to the odd Bell state results in a flip of the measured $\langle XX \rangle$ or $\langle YY \rangle$ value.

In contrast $\langle II \rangle$ and $\langle ZZ \rangle$ are identical for even and odd Bell states. They are however much more sensitive to rotations in phase space (coherent and incoherent).

Table \ref{tab:error_budget} summarizes the impact of errors on each Pauli measurement and how this contributes to the overall fidelity. Detailed justification for each error considered is provided below.

\subsection{Errors during cSWAP}

Following the cSWAP, the control-oscillator system will ideally be in the state
\begin{equation}
  \ket{\psi}_{\text{post-cSWAP}} = \frac{\ket{g} \ket{\alpha, -\alpha} + \ket{e}\ket{-\alpha, \alpha}}{\sqrt{2}}.
\end{equation}
The subsequent ancilla $-\pi/2$ rotation then generates a superposition of Bell states entangled with the ancilla,
\begin{equation}
  \ket{\psi}_{\text{post-rotation}} = \frac{\ket{g} \ket{\Psi^+} + \ket{e}\ket{\Psi^-}}{\sqrt{2}},
\end{equation}
at which point post-selecting on the outcome $\ket{g}$ projects us onto the desired Bell state.

Ancilla dephasing errors during the cSWAP however, can be mapped to ancilla bit flip errors immediately after the $-\pi/2$ rotation. The result is that our post-selection scheme will keep a shot in which we prepared $\ket{\Psi^-}$, for which $\langle XX \rangle$ and $\langle YY \rangle$ have the opposite sign. As discussed previously, $\langle II \rangle$ and $\langle ZZ \rangle$ are identical for both states $\ket{\Psi^+}$ and $\ket{\Psi^-}$, and so will be unaffected.

Oscillator decay transforms simply under the beamsplitter Hamiltonian, such that an oscillator decay event at any point during a SWAP is equivalent to a superposition of decay on Alice and Bob occurring immediately after. In our case, we have the state $(\cos{(gt_{\text{err}})}\hat{a}+i\sin{(gt_{\text{err}})}\hat{b})\ket{\psi}_{\text{post-cSWAP}}$, where $t_{\text{err}}$ describes when during the SWAP the error occurred.

Ancilla decay during the cSWAP does not transform as simply. In the Heisenberg picture where the evolution of cavity operators is visualised on a sphere, ancilla decay changes the rotation axis about which the Heisenberg operators rotate on the Bloch sphere partway through their trajectory. On average, this will scramble the cavity states, thus we expect this error to reduce all joint Pauli measurements.

Finally, oscillator dephasing errors should primarily affect the $II$ and $ZZ$ measurements. This is because these are measured by probing points far from the origin of the joint Wigner function in phase space, where even small phase space rotations have a large impact.

\subsection{Errors during cSWAP readout}

During the cSWAP readout, ancilla measurement (and post-selection on $\ket{g}$) ideally projects the oscillators onto an even Bell state. However, an oscillator decay event in either cavity during this time will flip the joint photon number parity of the state, and thus the measurement outcomes for both $XX$ and $YY$. Note that this happens regardless of when during the readout the photon loss occurs.

Ancilla decay during the readout leads not just to a projection on the incorrect Bell state, but also to a random rotation of Bob's phase due to the dispersive shift to the ancilla, depending on when during the readout the error occurs. The $XX$ measurement outcome, which is insensitive to phase, is therefore flipped, but the effect on $\langle YY \rangle$, $\langle II \rangle$ and $\langle ZZ \rangle$ is to reduce their amplitude uniformly.

Oscillator decay during the cSWAP readout will affect $\langle II \rangle$ and $\langle ZZ \rangle$ in the same manner as during the cSWAP itself.

\subsection{Errors during parity measurement}

Ancilla errors during the Wigner tomography, whether dephasing during the parity map or decay during the parity map and subsequent readout affect all joint Pauli measurements equally, flipping the sign of the result. The reduction in the measured value of each joint Pauli is thus reduced on average by twice the error probability.

Oscillator decay during the parity map affects the joint Pauli measurements depending on the oscillator state after the Wigner displacement. For the $XX$ measurement, this is (ideally) an eigenstate of joint parity and so decay will change the joint parity of the oscillator state. The probability of measuring $\pm 1$ depends on when during the parity map the error occurs, and so we can treat it as depolarizing the measurement results.

For the $II$ and $ZZ$ measurements, however, the states after displacement are superpositions of vacuum and a coherent state, e.g. $\ket{\psi} = \ket{0,0} + \ket{2\alpha, -2\alpha}$. Since $2|\alpha| \gg 1$, photon loss only negligibly affects the joint parity of this state.

\subsection{Effects of Kerr and $\chi^{'}$}

As the number of controlled SWAP operations $N$ increases in Fig.~\ref{fig:cSWAP}f, effects of higher-order non-linearities in the joint-cavity Hamiltonian become non-negligible. 
The most noticeable ones are cavity self-Kerrs and $\chi^{'}$.

Self-Kerrs of the cavities as shown in line $3$ of Eq.~\ref{eqn:H} result in detunings of the cavity frequencies that depend on cavity photon number quadratically.
This accumulation of photon number dependent phase manifests as rotation and smearing of the cavity states in phase space. 
What further complicates the story is the sixth-order non-linearity $\chi^{'}_\text{bc} \hat{b}^{\dagger^2}\hat{b}^2 \left(\ket{e}\bra{e}\right)_\text{c}$.
This parasitic term distorts Bob's cavity states the same ways as cavity self-Kerr, but only when the control transmon is excited. 
At the operating point for the cSWAP experiment, away from the flux point where self-Kerr is minimized, we measure cavity self-Kerrs $K_\text{a} \approx -0.8 ~\text{kHz}$, $K_\text{b} \approx -3.0 ~\text{kHz}$ and $\chi^{'}_\text{bc} \approx 0.9~\text{kHz}$.

The combined effect of the cavity self-Kerrs and $\chi^{'}$ is a photon number and control transmon dependent rotation and smearing of cavity states in phase space. The rotation is accounted for when the pre- and post-delays are calibrated for different $N$, but this cannot correct the smearing of the states, which is due to photon-number-dependent rotations. This smearing thus adds to the SPAM-corrected infidelities extracted from the slope in Fig.~\ref{fig:cSWAP}f.

Comparing the simulated fidelity up to $N=13$ rounds of cSWAP with and without self-Kerr terms and the $\chi^{'}$ term in the Hamiltonian shows an increase in the slope of $0.19\%\pm0.06\%$ per round for the Bell state preparation and $0.10\%\pm0.07\%$ per round for the control experiment. The error bars arise from uncertainty in the phase of the beamsplitter drive, which affects whether more photons are present in Alice (with less Kerr) or Bob (with more Kerr).

\subsection{Combining errors}

We estimate the combined fidelity as
\begin{equation}
  \mathcal{F} \approx \prod_{i}(1-p_i)+\sum_{i}q_{i}p_{i}\prod_{j\neq i}(1-p_j),
\end{equation}
where $p_i$ is the probability of an error in the $i^{th}$ channel and $q_i=0$ or $-1$ is the expected measured value in the event of the error occurring. This expression neglects contributions from multiple errors happening during the same run, which is a reasonable approximation given that $p_i\ll1~ \forall i$.

\subsection{Control experiment on identical oscillator states}
To validate our ability to predict errors, we conduct a control experiment where the cavities are initialized in $\ket{\alpha, \alpha}$ and perform a SWAP-test on these nominally identical states. For this experiment, we measure fidelity to the target state $\ket{\alpha,\alpha}$ by probing joint Wigner at $\ket{\alpha, \alpha}$. 

This control experiment is insensitive to oscillator decay. Since the oscillator states remain eigenstates of the respective collapse operators during the cSWAP, they are only affected by no-jump shrinking, which is negligible for the time scales we are measuring. Meanwhile, during the Wigner tomography, the displacement brings the oscillator state back to the origin where oscillator decay has no impact. For the same reason, oscillator dephasing has no impact during the Wigner tomography.

The joint Pauli measurements are also impervious to transmon dephasing during the SWAP test because all $|e\rangle$ measurements resulting from transmon dephasing are discarded.
So the only error channels we consider for this control experiment are ancilla decay and oscillator dephasing during the SWAP test, oscillator dephasing during the readout of the SWAP test, and transmon errors during the parity map and its readout.

\begin{center}
\begin{table*}[h!]
\setlength{\tabcolsep}{0pt} 
\begin{tabular}{"p{6em}|p{10em}|p{16em}"p{11em}"}
\thickhline
&\makecell{error channel}& \makecell{probability of occurrence} & \makecell{estimated (simulated)\\ effect on fidelity}\\
\thickhline

\multicolumn{1}{"c|}{\multirow{2}{*}{\makecell{cSWAP}}} & \makecell{ancilla decay \\ $|e\rangle\langle e|$} & \makecell{$\frac{\tau_cSWAP}{2T_{1qB}}$ \\ $\approx 1.17\% \pm 0.02\%$} & \makecell{$1.17\% \pm 0.02\%$ \\ ($1.0\%$)} \\
\cline{2-4}
& \makecell{oscillator dephasing \\ $\hat{a^\dag}\hat{a}$, $\hat{b^\dag}\hat{b}$} & \makecell{$2 \bar{n}(\frac{\tau_{cSWAP}}{T_{\phi cA}}+\frac{\tau_{cSWAP}}{T_{\phi cB}})$ \\ $\approx 0.9\% \pm 0.15\%$}& \makecell{$1.8\% \pm 0.3\%$ \\ ($1.3\%$)} \\
\hline

\makecell{cSWAP RO}& \makecell{oscillator dephasing \\ $\hat{a^\dag}\hat{a}$, $\hat{b^\dag}\hat{b}$} &\makecell{$2 \bar{n}(\frac{\tau_{RO}}{T_{\phi cA}}+\frac{\tau_{RO}}{T_{\phi cB}})$ \\ $\approx 1.2 \% \pm 0.2\%$} & \makecell{$2.4\% \pm 0.4\%$ \\ ($1.9\%$)} \\ 
\hline

\multicolumn{1}{"c|}{\multirow{2}{*}{\makecell{parity map}}} & \makecell{ancilla decay \\ $|e\rangle \xrightarrow{} |g\rangle $} & \makecell{$\frac{1}{2}\left(\frac{\tau_{pA}}{T_{1qA}}+\frac{\tau_{pB}}{T_{1qB}}\right)$ \\ $\approx 0.620\% \pm 0.005\%$} & \makecell{$0.620\% \pm 0.005\%$ \\ ($0.6\%$)} \\ \cline{2-4}
& \makecell{ancilla dephasing \\ $|e\rangle\langle e|$}& \makecell{$\frac{\tau_{pA}}{2T_{\phi qA}}+ \frac{\tau_{pB}}{2T_{\phi qB}}$ \\ $\approx 0.34\% \pm 0.01\%$} & \makecell{$0.68\% \pm 0.03\%$ \\ ($0.6\%$)} \\
\hline

\multicolumn{1}{"c|}{\multirow{1}{*}{\makecell{parity RO}}} & \makecell{ancilla decay \\ $|e\rangle \xrightarrow{} |g\rangle $} & \makecell{$\frac{1}{2}\left(\frac{\tau_{RO}}{T_{1qA}}+\frac{\tau_{RO}}{T_{1qB}}\right)$ \\ $\approx 2.28\% \pm 0.02\%$} & \makecell{$4.57\% \pm 0.04\%$ \\ ($4.5\%$)} \\
\thickhline
\multicolumn{3}{"c"}{estimated gate error per round} & \makecell{\boldmath{$3.0\% \pm 0.3\%$}} \\
\thickhline
\multicolumn{3}{"c"}{estimated infidelity per round, including $K$ and $\chi^{'}$} & \makecell{\boldmath{$3.1\% \pm 0.3\%$}} \\
\thickhline
\multicolumn{3}{"c"}{estimated (simulated) fidelity} & \makecell{\boldmath{$89.1\% \pm 0.7\%$} \\ ($90.6\%$)} \\
\thickhline
\multicolumn{3}{"c"}{measured fidelity} & \makecell{\boldmath{$89.5\%$}}\\
\thickhline
\end{tabular}
\caption{\textbf{Error budget for control experiment.} Estimated reduction in measured fidelity to target state due to each error channel. Values in parentheses are predicted infidelities from master equation simulations. The effects of Kerr and $\chi^{'}$ are described in App.~\ref{app:error_budget}4.}
\label{}
\end{table*}
\end{center}

\section{cSWAP with three bosonically-encoded qubits}
\label{app:3cav_cSWAP}
\begin{figure}[!thb]
\begin{center}
\includegraphics[width=1\linewidth]{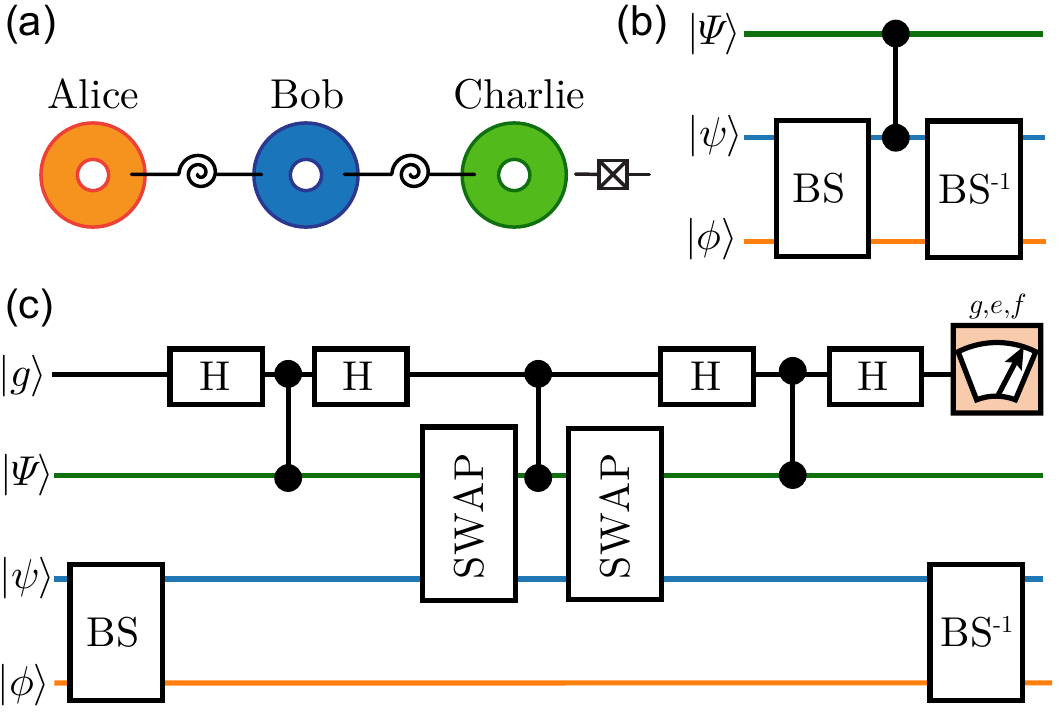}
\caption {\textbf{cSWAP with three bosonic qubits}. (a) Schematic of the hardware required for the proposed cSWAP on three bosonically encoded qubits: Alice, Bob, and Charlie are bosonic modes hosted in high-$Q$ stub cavities~\cite{reagor:2016} and linked by SNAIL couplers of the kind presented in this work. A transmon ancilla is inserted into Charlie's cavity; it provides the requisite nonlinearity. (b) Canonical construction of cSWAP from beamsplitter, controlled-phase shift, inverse beamsplitter. (c) Full sequence for the proposed cSWAP. SWAPs are uncoditioned on the state of the transmon, and only effect the bosonic modes---that is, they exchange the states of Alice and Charlie and do not affect the transmon rail.}  
\label{fig:3cav_cSWAP}
\end{center}
\end{figure}

A SWAP of two bosonically encoded qubits controlled on the state of a third bosonically encoded qubit can be realized with the hardware shown in Fig.~\ref{fig:3cav_cSWAP}a. The proposed sequence utilizes the canonical construction of cSWAP from beamsplitters and controlled phase-shifts (Fig.~\ref{fig:3cav_cSWAP}b). To realize the controlled phase-shift between Charlie and Alice, the transmon ancilla serves as an intermediary. After the beamsplitter, Charlie's state is mapped to the transmon. A SWAP then exchanges Alice and Charlie's states, such that the following controlled phase-shift between the transmon (now storing Charlie's original state) and Charlie (now storing Alice's original state) accomplishes the desired operation. The sequence is then reversed to disentangle the transmon. Measurement of the transmon rail at the conclusion of the gate can provide a flag to signal errors. 

We note that the efficacy of this sequence requires that the SWAP gates are unconditioned on the state of the transmon. One way to accomplish this is by pumping the beamsplitter interaction with $\delta = \chi/2$ while driving the transmon resonantly with Rabi-rate $\Omega \gg g_{\text{bs}}$~\cite{tsunoda:2022}. Fine tuning $\Omega$ such that $2 \Omega t_\text{bs} = \Omega t_\text{SWAP} = 2 n \pi$ for some integer $n$ ensures that the transmon state is also unchanged by this dynamical decoupling.

\section{Comparing continuous and Trotterized cSWAP sequences}
\label{app:cSWAP_comparison}
As illustrated in Fig.~\ref{fig:3cav_cSWAP}b, a cSWAP can be constructed from a 50:50 beamsplitter (unconditional on the ancilla state), a parity map (enacted by a delay of $\pi/\chi$), and another unconditional 50:50 beamsplitter, with opposite phase to the first. This differs from the approach taken in Sec.~\ref{sec:cond}, in which we let the ancilla and both cavities interact simultaneously. 

\begin{figure}[!thb]
\begin{center}
\includegraphics[width=1\linewidth]{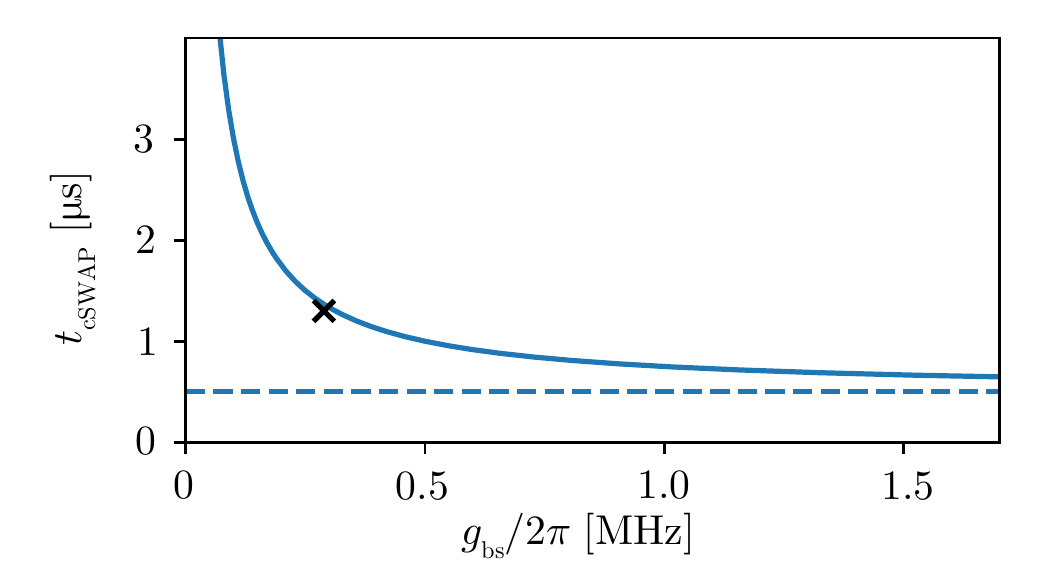}
\caption{\textbf{Comparing cSWAP approaches.} Blue solid line shows gate time as a function of achievable beamsplitter rate for a cSWAP composed of a a controlled phase shift sandwiched between two unconditional beamsplitter operations, with the dashed line showing the limit as the beamsplitter rate increases. The cross indicates the method implemented in this work. Both assume a dispersive shift to the transmon of $\chi/2\pi = -1.104$~MHz.}
\label{fig:cSWAP_comparison}
\end{center}
\end{figure}

Analogously to the unconditional SWAP discussed in App.~\ref{app:3cav_cSWAP}, the unconditional 50:50 beamsplitter can be performed by pumping the beamsplitter interaction with $\delta = \chi/2$ while resonantly driving the control transmon with Rabi-rate $\Omega \gg g_{\text{bs}}$, in a time $\pi/4g_{\text{bs}}$. Assuming no restriction on how fast the beamsplitter pulse can be ramped on or off, this gives a duration of $\pi/2g_{\text{bs}} + \pi/\chi$ for the full cSWAP. (In practice, ensuring the spectral content of the beamsplitter pulse is sufficiently narrow to avoid driving unwanted processes introduces some non-negligible ramp time.) In Fig.~\ref{fig:cSWAP_comparison}, this gate duration (solid blue line) is plotted as a function of $g_{\text{bs}}$ for $\chi/2\pi = -1.104$ MHz, the value observed in our system. Comparing this `Trotterized' method to the performance of the `continuous' scheme used in this paper (black cross), we see that they give similar gate duration, given the same beamsplitter rate. The flexibility of the `Trotterized' method to further reduce gate time as the achievable beamsplitter rate increases makes it a promising scheme for future cSWAP implementations.

\section{Experimental setup}
\label{app:exp_setup}
To help minimize phase drift between our cavity drives, we source the local oscillators for Alice and Bob from channels 1 and 2 of the same SignalCore SC5510A generator. For quantum limited amplification of our readout signal at the base of the refrigerator, we use a Josephson Array Mode Parametric Amplifier (JAMPA)~\cite{sivak:2020} and a lumped SNAIL Parametric Amplifier~\cite{frattini:2017} (LSPA). 

\begin{figure*}[!thb]
\begin{center}
\includegraphics[width=0.93\linewidth]{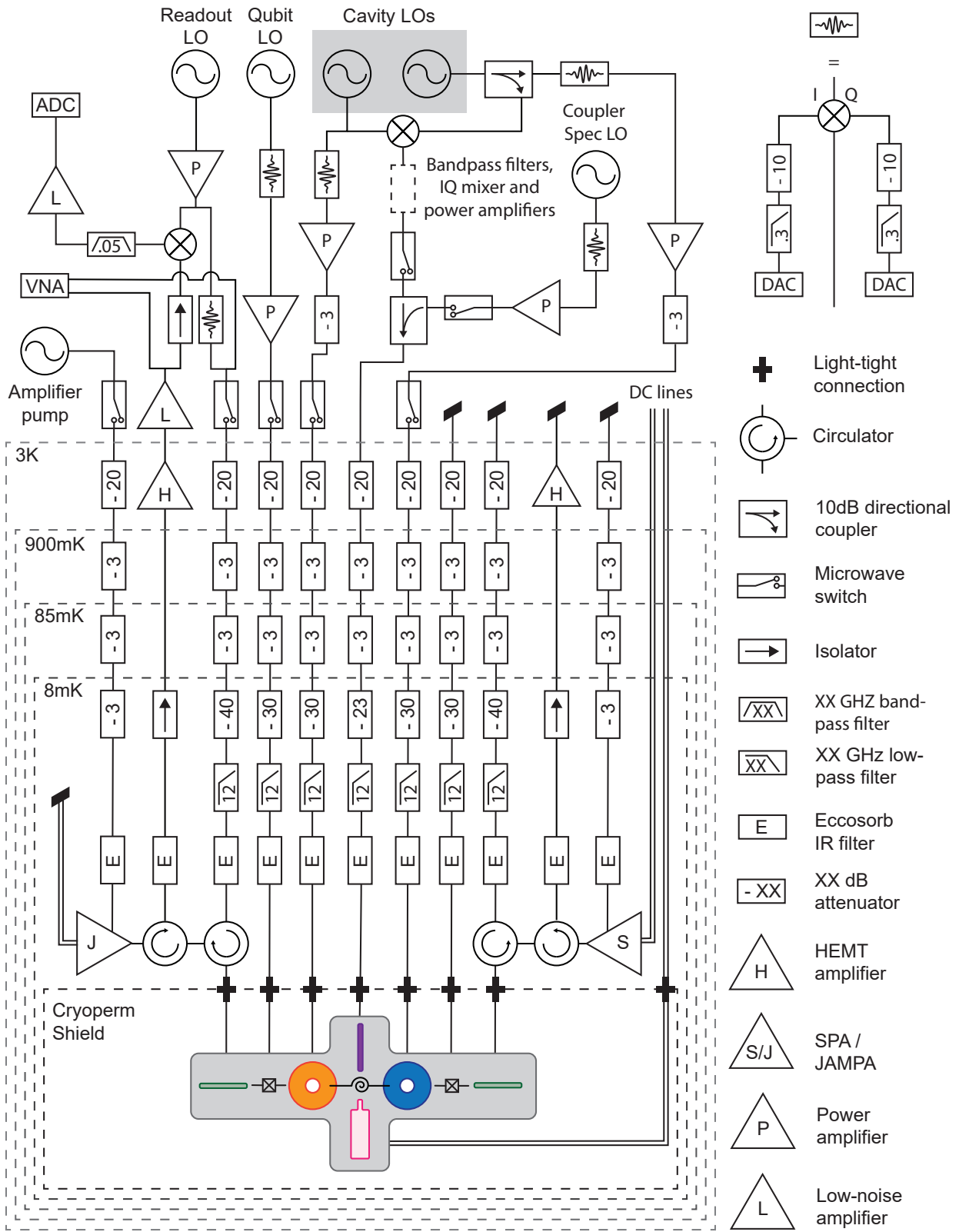}
\caption {\textbf{Wiring diagram}. The gray region within the Cryoperm shield shows the experimental package, containing two microwave cavities (orange and blue), a SNAIL coupler, two transmon ancillas, two readout resonators (green), a buffer mode (purple) and a flux transformer (pink). On chip Purcell filters (present in the experiment) are omitted from the diagram.}
\label{fig:wiring}
\end{center}
\end{figure*}

\begin{thebibliography}{69}%
\makeatletter
\providecommand \@ifxundefined [1]{%
 \@ifx{#1\undefined}
}%
\providecommand \@ifnum [1]{%
 \ifnum #1\expandafter \@firstoftwo
 \else \expandafter \@secondoftwo
 \fi
}%
\providecommand \@ifx [1]{%
 \ifx #1\expandafter \@firstoftwo
 \else \expandafter \@secondoftwo
 \fi
}%
\providecommand \natexlab [1]{#1}%
\providecommand \enquote  [1]{``#1''}%
\providecommand \bibnamefont  [1]{#1}%
\providecommand \bibfnamefont [1]{#1}%
\providecommand \citenamefont [1]{#1}%
\providecommand \href@noop [0]{\@secondoftwo}%
\providecommand \href [0]{\begingroup \@sanitize@url \@href}%
\providecommand \@href[1]{\@@startlink{#1}\@@href}%
\providecommand \@@href[1]{\endgroup#1\@@endlink}%
\providecommand \@sanitize@url [0]{\catcode `\\12\catcode `\$12\catcode
  `\&12\catcode `\#12\catcode `\^12\catcode `\_12\catcode `\%12\relax}%
\providecommand \@@startlink[1]{}%
\providecommand \@@endlink[0]{}%
\providecommand \url  [0]{\begingroup\@sanitize@url \@url }%
\providecommand \@url [1]{\endgroup\@href {#1}{\urlprefix }}%
\providecommand \urlprefix  [0]{URL }%
\providecommand \Eprint [0]{\href }%
\providecommand \doibase [0]{https://doi.org/}%
\providecommand \selectlanguage [0]{\@gobble}%
\providecommand \bibinfo  [0]{\@secondoftwo}%
\providecommand \bibfield  [0]{\@secondoftwo}%
\providecommand \translation [1]{[#1]}%
\providecommand \BibitemOpen [0]{}%
\providecommand \bibitemStop [0]{}%
\providecommand \bibitemNoStop [0]{.\EOS\space}%
\providecommand \EOS [0]{\spacefactor3000\relax}%
\providecommand \BibitemShut  [1]{\csname bibitem#1\endcsname}%
\let\auto@bib@innerbib\@empty
\bibitem [{\citenamefont {Cochrane}\ \emph {et~al.}(1999)\citenamefont
  {Cochrane}, \citenamefont {Milburn},\ and\ \citenamefont
  {Munro}}]{cochrane:1999}%
  \BibitemOpen
  \bibfield  {author} {\bibinfo {author} {\bibfnamefont {P.~T.}\ \bibnamefont
  {Cochrane}}, \bibinfo {author} {\bibfnamefont {G.~J.}\ \bibnamefont
  {Milburn}},\ and\ \bibinfo {author} {\bibfnamefont {W.~J.}\ \bibnamefont
  {Munro}},\ }\bibfield  {title} {\bibinfo {title} {{Macroscopically distinct
  quantum-superposition states as a bosonic code for amplitude damping}},\
  }\href@noop {} {\bibfield  {journal} {\bibinfo  {journal} {Physical Review
  A}\ }\textbf {\bibinfo {volume} {59}},\ \bibinfo {pages} {2631} (\bibinfo
  {year} {1999})}\BibitemShut {NoStop}%
\bibitem [{\citenamefont {Gottesman}\ \emph {et~al.}(2001)\citenamefont
  {Gottesman}, \citenamefont {Kitaev},\ and\ \citenamefont
  {Preskill}}]{gottesman:2001}%
  \BibitemOpen
  \bibfield  {author} {\bibinfo {author} {\bibfnamefont {D.}~\bibnamefont
  {Gottesman}}, \bibinfo {author} {\bibfnamefont {A.}~\bibnamefont {Kitaev}},\
  and\ \bibinfo {author} {\bibfnamefont {J.}~\bibnamefont {Preskill}},\
  }\bibfield  {title} {\bibinfo {title} {{Encoding a qubit in an oscillator}},\
  }\href@noop {} {\bibfield  {journal} {\bibinfo  {journal} {Physical Review
  A}\ }\textbf {\bibinfo {volume} {64}},\ \bibinfo {pages} {012310} (\bibinfo
  {year} {2001})}\BibitemShut {NoStop}%
\bibitem [{\citenamefont {Mirrahimi}\ \emph {et~al.}(2014)\citenamefont
  {Mirrahimi}, \citenamefont {Leghtas}, \citenamefont {Albert}, \citenamefont
  {Touzard}, \citenamefont {Schoelkopf}, \citenamefont {Jiang},\ and\
  \citenamefont {Devoret}}]{mirrahimi:2014}%
  \BibitemOpen
  \bibfield  {author} {\bibinfo {author} {\bibfnamefont {M.}~\bibnamefont
  {Mirrahimi}}, \bibinfo {author} {\bibfnamefont {Z.}~\bibnamefont {Leghtas}},
  \bibinfo {author} {\bibfnamefont {V.~V.}\ \bibnamefont {Albert}}, \bibinfo
  {author} {\bibfnamefont {S.}~\bibnamefont {Touzard}}, \bibinfo {author}
  {\bibfnamefont {R.~J.}\ \bibnamefont {Schoelkopf}}, \bibinfo {author}
  {\bibfnamefont {L.}~\bibnamefont {Jiang}},\ and\ \bibinfo {author}
  {\bibfnamefont {M.~H.}\ \bibnamefont {Devoret}},\ }\bibfield  {title}
  {\bibinfo {title} {{Dynamically protected cat-qubits: a new paradigm for
  universal quantum computation}},\ }\href@noop {} {\bibfield  {journal}
  {\bibinfo  {journal} {New Journal of Physics}\ }\textbf {\bibinfo {volume}
  {16}},\ \bibinfo {pages} {045014} (\bibinfo {year} {2014})}\BibitemShut
  {NoStop}%
\bibitem [{\citenamefont {Michael}\ \emph {et~al.}(2016)\citenamefont
  {Michael}, \citenamefont {Silveri}, \citenamefont {Brierley}, \citenamefont
  {Albert}, \citenamefont {Salmilehto}, \citenamefont {Jiang},\ and\
  \citenamefont {Girvin}}]{michael:2016}%
  \BibitemOpen
  \bibfield  {author} {\bibinfo {author} {\bibfnamefont {M.~H.}\ \bibnamefont
  {Michael}}, \bibinfo {author} {\bibfnamefont {M.}~\bibnamefont {Silveri}},
  \bibinfo {author} {\bibfnamefont {R.~T.}\ \bibnamefont {Brierley}}, \bibinfo
  {author} {\bibfnamefont {V.~V.}\ \bibnamefont {Albert}}, \bibinfo {author}
  {\bibfnamefont {J.}~\bibnamefont {Salmilehto}}, \bibinfo {author}
  {\bibfnamefont {L.}~\bibnamefont {Jiang}},\ and\ \bibinfo {author}
  {\bibfnamefont {S.~M.}\ \bibnamefont {Girvin}},\ }\bibfield  {title}
  {\bibinfo {title} {{New class of quantum error-correcting codes for a bosonic
  mode}},\ }\href@noop {} {\bibfield  {journal} {\bibinfo  {journal} {Physical
  Review X}\ }\textbf {\bibinfo {volume} {6}},\ \bibinfo {pages} {031006}
  (\bibinfo {year} {2016})}\BibitemShut {NoStop}%
\bibitem [{\citenamefont {Puri}\ \emph {et~al.}(2017)\citenamefont {Puri},
  \citenamefont {Boutin},\ and\ \citenamefont {Blais}}]{puri:2017}%
  \BibitemOpen
  \bibfield  {author} {\bibinfo {author} {\bibfnamefont {S.}~\bibnamefont
  {Puri}}, \bibinfo {author} {\bibfnamefont {S.}~\bibnamefont {Boutin}},\ and\
  \bibinfo {author} {\bibfnamefont {A.}~\bibnamefont {Blais}},\ }\bibfield
  {title} {\bibinfo {title} {{Engineering the quantum states of light in a
  Kerr-nonlinear resonator by two-photon driving}},\ }\href@noop {} {\bibfield
  {journal} {\bibinfo  {journal} {npj Quantum Information}\ }\textbf {\bibinfo
  {volume} {3}},\ \bibinfo {pages} {1} (\bibinfo {year} {2017})}\BibitemShut
  {NoStop}%
\bibitem [{\citenamefont {Chuang}\ \emph {et~al.}(1997)\citenamefont {Chuang},
  \citenamefont {Leung},\ and\ \citenamefont {Yamamoto}}]{Chuang:1997}%
  \BibitemOpen
  \bibfield  {author} {\bibinfo {author} {\bibfnamefont {I.~L.}\ \bibnamefont
  {Chuang}}, \bibinfo {author} {\bibfnamefont {D.~W.}\ \bibnamefont {Leung}},\
  and\ \bibinfo {author} {\bibfnamefont {Y.}~\bibnamefont {Yamamoto}},\
  }\bibfield  {title} {\bibinfo {title} {{Bosonic quantum codes for amplitude
  damping}},\ }\href {https://doi.org/10.1103/PhysRevA.56.1114} {\bibfield
  {journal} {\bibinfo  {journal} {Physical Review A}\ }\textbf {\bibinfo
  {volume} {56}},\ \bibinfo {pages} {1114} (\bibinfo {year}
  {1997})}\BibitemShut {NoStop}%
\bibitem [{\citenamefont {Reagor}\ \emph {et~al.}(2013)\citenamefont {Reagor},
  \citenamefont {Paik}, \citenamefont {Catelani}, \citenamefont {Sun},
  \citenamefont {Axline}, \citenamefont {Holland}, \citenamefont {Pop},
  \citenamefont {Masluk}, \citenamefont {Brecht}, \citenamefont {Frunzio} \emph
  {et~al.}}]{reagor:2013}%
  \BibitemOpen
  \bibfield  {author} {\bibinfo {author} {\bibfnamefont {M.}~\bibnamefont
  {Reagor}}, \bibinfo {author} {\bibfnamefont {H.}~\bibnamefont {Paik}},
  \bibinfo {author} {\bibfnamefont {G.}~\bibnamefont {Catelani}}, \bibinfo
  {author} {\bibfnamefont {L.}~\bibnamefont {Sun}}, \bibinfo {author}
  {\bibfnamefont {C.}~\bibnamefont {Axline}}, \bibinfo {author} {\bibfnamefont
  {E.}~\bibnamefont {Holland}}, \bibinfo {author} {\bibfnamefont {I.~M.}\
  \bibnamefont {Pop}}, \bibinfo {author} {\bibfnamefont {N.~A.}\ \bibnamefont
  {Masluk}}, \bibinfo {author} {\bibfnamefont {T.}~\bibnamefont {Brecht}},
  \bibinfo {author} {\bibfnamefont {L.}~\bibnamefont {Frunzio}}, \emph
  {et~al.},\ }\bibfield  {title} {\bibinfo {title} {{Reaching 10 ms single
  photon lifetimes for superconducting aluminum cavities}},\ }\href@noop {}
  {\bibfield  {journal} {\bibinfo  {journal} {Applied Physics Letters}\
  }\textbf {\bibinfo {volume} {102}},\ \bibinfo {pages} {192604} (\bibinfo
  {year} {2013})}\BibitemShut {NoStop}%
\bibitem [{\citenamefont {Romanenko}\ \emph {et~al.}(2020)\citenamefont
  {Romanenko}, \citenamefont {Pilipenko}, \citenamefont {Zorzetti},
  \citenamefont {Frolov}, \citenamefont {Awida}, \citenamefont {Belomestnykh},
  \citenamefont {Posen},\ and\ \citenamefont {Grassellino}}]{romanenko:2020}%
  \BibitemOpen
  \bibfield  {author} {\bibinfo {author} {\bibfnamefont {A.}~\bibnamefont
  {Romanenko}}, \bibinfo {author} {\bibfnamefont {R.}~\bibnamefont
  {Pilipenko}}, \bibinfo {author} {\bibfnamefont {S.}~\bibnamefont {Zorzetti}},
  \bibinfo {author} {\bibfnamefont {D.}~\bibnamefont {Frolov}}, \bibinfo
  {author} {\bibfnamefont {M.}~\bibnamefont {Awida}}, \bibinfo {author}
  {\bibfnamefont {S.}~\bibnamefont {Belomestnykh}}, \bibinfo {author}
  {\bibfnamefont {S.}~\bibnamefont {Posen}},\ and\ \bibinfo {author}
  {\bibfnamefont {A.}~\bibnamefont {Grassellino}},\ }\bibfield  {title}
  {\bibinfo {title} {{Three-dimensional superconducting resonators at T< 20 mK
  with photon lifetimes up to $\tau$= 2 s}},\ }\href@noop {} {\bibfield
  {journal} {\bibinfo  {journal} {Physical Review Applied}\ }\textbf {\bibinfo
  {volume} {13}},\ \bibinfo {pages} {034032} (\bibinfo {year}
  {2020})}\BibitemShut {NoStop}%
\bibitem [{\citenamefont {Cai}\ \emph {et~al.}(2021)\citenamefont {Cai},
  \citenamefont {Ma}, \citenamefont {Wang}, \citenamefont {Zou},\ and\
  \citenamefont {Sun}}]{Cai:2021}%
  \BibitemOpen
  \bibfield  {author} {\bibinfo {author} {\bibfnamefont {W.}~\bibnamefont
  {Cai}}, \bibinfo {author} {\bibfnamefont {Y.}~\bibnamefont {Ma}}, \bibinfo
  {author} {\bibfnamefont {W.}~\bibnamefont {Wang}}, \bibinfo {author}
  {\bibfnamefont {C.-L.}\ \bibnamefont {Zou}},\ and\ \bibinfo {author}
  {\bibfnamefont {L.}~\bibnamefont {Sun}},\ }\bibfield  {title} {\bibinfo
  {title} {{Bosonic quantum error correction codes in superconducting quantum
  circuits}},\ }\href
  {https://doi.org/https://doi.org/10.1016/j.fmre.2020.12.006} {\bibfield
  {journal} {\bibinfo  {journal} {Fundamental Research}\ }\textbf {\bibinfo
  {volume} {1}},\ \bibinfo {pages} {50} (\bibinfo {year} {2021})}\BibitemShut
  {NoStop}%
\bibitem [{\citenamefont {Blais}\ \emph {et~al.}(2004)\citenamefont {Blais},
  \citenamefont {Huang}, \citenamefont {Wallraff}, \citenamefont {Girvin},\
  and\ \citenamefont {Schoelkopf}}]{blais:2004}%
  \BibitemOpen
  \bibfield  {author} {\bibinfo {author} {\bibfnamefont {A.}~\bibnamefont
  {Blais}}, \bibinfo {author} {\bibfnamefont {R.-S.}\ \bibnamefont {Huang}},
  \bibinfo {author} {\bibfnamefont {A.}~\bibnamefont {Wallraff}}, \bibinfo
  {author} {\bibfnamefont {S.~M.}\ \bibnamefont {Girvin}},\ and\ \bibinfo
  {author} {\bibfnamefont {R.~J.}\ \bibnamefont {Schoelkopf}},\ }\bibfield
  {title} {\bibinfo {title} {{Cavity quantum electrodynamics for
  superconducting electrical circuits: An architecture for quantum
  computation}},\ }\href@noop {} {\bibfield  {journal} {\bibinfo  {journal}
  {Physical Review A}\ }\textbf {\bibinfo {volume} {69}},\ \bibinfo {pages}
  {062320} (\bibinfo {year} {2004})}\BibitemShut {NoStop}%
\bibitem [{\citenamefont {Leghtas}\ \emph {et~al.}(2013)\citenamefont
  {Leghtas}, \citenamefont {Kirchmair}, \citenamefont {Vlastakis},
  \citenamefont {Devoret}, \citenamefont {Schoelkopf},\ and\ \citenamefont
  {Mirrahimi}}]{leghtas:2013}%
  \BibitemOpen
  \bibfield  {author} {\bibinfo {author} {\bibfnamefont {Z.}~\bibnamefont
  {Leghtas}}, \bibinfo {author} {\bibfnamefont {G.}~\bibnamefont {Kirchmair}},
  \bibinfo {author} {\bibfnamefont {B.}~\bibnamefont {Vlastakis}}, \bibinfo
  {author} {\bibfnamefont {M.~H.}\ \bibnamefont {Devoret}}, \bibinfo {author}
  {\bibfnamefont {R.~J.}\ \bibnamefont {Schoelkopf}},\ and\ \bibinfo {author}
  {\bibfnamefont {M.}~\bibnamefont {Mirrahimi}},\ }\bibfield  {title} {\bibinfo
  {title} {{Deterministic protocol for mapping a qubit to coherent state
  superpositions in a cavity}},\ }\href@noop {} {\bibfield  {journal} {\bibinfo
   {journal} {Physical Review A}\ }\textbf {\bibinfo {volume} {87}},\ \bibinfo
  {pages} {042315} (\bibinfo {year} {2013})}\BibitemShut {NoStop}%
\bibitem [{\citenamefont {Heeres}\ \emph {et~al.}(2017)\citenamefont {Heeres},
  \citenamefont {Reinhold}, \citenamefont {Ofek}, \citenamefont {Frunzio},
  \citenamefont {Jiang}, \citenamefont {Devoret},\ and\ \citenamefont
  {Schoelkopf}}]{heeres:2017}%
  \BibitemOpen
  \bibfield  {author} {\bibinfo {author} {\bibfnamefont {R.~W.}\ \bibnamefont
  {Heeres}}, \bibinfo {author} {\bibfnamefont {P.}~\bibnamefont {Reinhold}},
  \bibinfo {author} {\bibfnamefont {N.}~\bibnamefont {Ofek}}, \bibinfo {author}
  {\bibfnamefont {L.}~\bibnamefont {Frunzio}}, \bibinfo {author} {\bibfnamefont
  {L.}~\bibnamefont {Jiang}}, \bibinfo {author} {\bibfnamefont {M.~H.}\
  \bibnamefont {Devoret}},\ and\ \bibinfo {author} {\bibfnamefont {R.~J.}\
  \bibnamefont {Schoelkopf}},\ }\bibfield  {title} {\bibinfo {title}
  {{Implementing a universal gate set on a logical qubit encoded in an
  oscillator}},\ }\href@noop {} {\bibfield  {journal} {\bibinfo  {journal}
  {Nature communications}\ }\textbf {\bibinfo {volume} {8}},\ \bibinfo {pages}
  {1} (\bibinfo {year} {2017})}\BibitemShut {NoStop}%
\bibitem [{\citenamefont {F{\"o}sel}\ \emph {et~al.}(2020)\citenamefont
  {F{\"o}sel}, \citenamefont {Krastanov}, \citenamefont {Marquardt},\ and\
  \citenamefont {Jiang}}]{fosel:2020}%
  \BibitemOpen
  \bibfield  {author} {\bibinfo {author} {\bibfnamefont {T.}~\bibnamefont
  {F{\"o}sel}}, \bibinfo {author} {\bibfnamefont {S.}~\bibnamefont
  {Krastanov}}, \bibinfo {author} {\bibfnamefont {F.}~\bibnamefont
  {Marquardt}},\ and\ \bibinfo {author} {\bibfnamefont {L.}~\bibnamefont
  {Jiang}},\ }\bibfield  {title} {\bibinfo {title} {{Efficient cavity control
  with SNAP gates}},\ }\href@noop {} {\bibfield  {journal} {\bibinfo  {journal}
  {arXiv preprint arXiv:2004.14256}\ } (\bibinfo {year} {2020})}\BibitemShut
  {NoStop}%
\bibitem [{\citenamefont {Kudra}\ \emph {et~al.}(2022)\citenamefont {Kudra},
  \citenamefont {Kervinen}, \citenamefont {Strandberg}, \citenamefont {Ahmed},
  \citenamefont {Scigliuzzo}, \citenamefont {Osman}, \citenamefont {Lozano},
  \citenamefont {Thol\'en}, \citenamefont {Borgani}, \citenamefont {Haviland},
  \citenamefont {Ferrini}, \citenamefont {Bylander}, \citenamefont {Kockum},
  \citenamefont {Quijandr\'{\i}a}, \citenamefont {Delsing},\ and\ \citenamefont
  {Gasparinetti}}]{kudra:2021}%
  \BibitemOpen
  \bibfield  {author} {\bibinfo {author} {\bibfnamefont {M.}~\bibnamefont
  {Kudra}}, \bibinfo {author} {\bibfnamefont {M.}~\bibnamefont {Kervinen}},
  \bibinfo {author} {\bibfnamefont {I.}~\bibnamefont {Strandberg}}, \bibinfo
  {author} {\bibfnamefont {S.}~\bibnamefont {Ahmed}}, \bibinfo {author}
  {\bibfnamefont {M.}~\bibnamefont {Scigliuzzo}}, \bibinfo {author}
  {\bibfnamefont {A.}~\bibnamefont {Osman}}, \bibinfo {author} {\bibfnamefont
  {D.~P.}\ \bibnamefont {Lozano}}, \bibinfo {author} {\bibfnamefont {M.~O.}\
  \bibnamefont {Thol\'en}}, \bibinfo {author} {\bibfnamefont {R.}~\bibnamefont
  {Borgani}}, \bibinfo {author} {\bibfnamefont {D.~B.}\ \bibnamefont
  {Haviland}}, \bibinfo {author} {\bibfnamefont {G.}~\bibnamefont {Ferrini}},
  \bibinfo {author} {\bibfnamefont {J.}~\bibnamefont {Bylander}}, \bibinfo
  {author} {\bibfnamefont {A.~F.}\ \bibnamefont {Kockum}}, \bibinfo {author}
  {\bibfnamefont {F.}~\bibnamefont {Quijandr\'{\i}a}}, \bibinfo {author}
  {\bibfnamefont {P.}~\bibnamefont {Delsing}},\ and\ \bibinfo {author}
  {\bibfnamefont {S.}~\bibnamefont {Gasparinetti}},\ }\bibfield  {title}
  {\bibinfo {title} {{Robust preparation of Wigner-negative States with
  optimized SNAP-displacement sequences}},\ }\href
  {https://doi.org/10.1103/PRXQuantum.3.030301} {\bibfield  {journal} {\bibinfo
   {journal} {PRX Quantum}\ }\textbf {\bibinfo {volume} {3}},\ \bibinfo {pages}
  {030301} (\bibinfo {year} {2022})}\BibitemShut {NoStop}%
\bibitem [{\citenamefont {Eickbusch}\ \emph {et~al.}(2022)\citenamefont
  {Eickbusch}, \citenamefont {Sivak}, \citenamefont {Ding}, \citenamefont
  {Elder}, \citenamefont {Jha}, \citenamefont {Venkatraman}, \citenamefont
  {Royer}, \citenamefont {Girvin}, \citenamefont {Schoelkopf},\ and\
  \citenamefont {Devoret}}]{eickbusch:2021}%
  \BibitemOpen
  \bibfield  {author} {\bibinfo {author} {\bibfnamefont {A.}~\bibnamefont
  {Eickbusch}}, \bibinfo {author} {\bibfnamefont {V.}~\bibnamefont {Sivak}},
  \bibinfo {author} {\bibfnamefont {A.~Z.}\ \bibnamefont {Ding}}, \bibinfo
  {author} {\bibfnamefont {S.~S.}\ \bibnamefont {Elder}}, \bibinfo {author}
  {\bibfnamefont {S.~R.}\ \bibnamefont {Jha}}, \bibinfo {author} {\bibfnamefont
  {J.}~\bibnamefont {Venkatraman}}, \bibinfo {author} {\bibfnamefont
  {B.}~\bibnamefont {Royer}}, \bibinfo {author} {\bibfnamefont {S.~M.}\
  \bibnamefont {Girvin}}, \bibinfo {author} {\bibfnamefont {R.~J.}\
  \bibnamefont {Schoelkopf}},\ and\ \bibinfo {author} {\bibfnamefont {M.~H.}\
  \bibnamefont {Devoret}},\ }\bibfield  {title} {\bibinfo {title} {{Fast
  universal control of an oscillator with weak dispersive coupling to a
  qubit}},\ }\href {https://doi.org/10.1038/s41567-022-01776-9} {\bibfield
  {journal} {\bibinfo  {journal} {Nature Physics}\ }\textbf {\bibinfo {volume}
  {18}},\ \bibinfo {pages} {1464} (\bibinfo {year} {2022})}\BibitemShut
  {NoStop}%
\bibitem [{\citenamefont {Rosenblum}\ \emph {et~al.}(2018)\citenamefont
  {Rosenblum}, \citenamefont {Gao}, \citenamefont {Reinhold}, \citenamefont
  {Wang}, \citenamefont {Axline}, \citenamefont {Frunzio}, \citenamefont
  {Girvin}, \citenamefont {Jiang}, \citenamefont {Mirrahimi}, \citenamefont
  {Devoret} \emph {et~al.}}]{rosenblum:2018}%
  \BibitemOpen
  \bibfield  {author} {\bibinfo {author} {\bibfnamefont {S.}~\bibnamefont
  {Rosenblum}}, \bibinfo {author} {\bibfnamefont {Y.~Y.}\ \bibnamefont {Gao}},
  \bibinfo {author} {\bibfnamefont {P.}~\bibnamefont {Reinhold}}, \bibinfo
  {author} {\bibfnamefont {C.}~\bibnamefont {Wang}}, \bibinfo {author}
  {\bibfnamefont {C.~J.}\ \bibnamefont {Axline}}, \bibinfo {author}
  {\bibfnamefont {L.}~\bibnamefont {Frunzio}}, \bibinfo {author} {\bibfnamefont
  {S.~M.}\ \bibnamefont {Girvin}}, \bibinfo {author} {\bibfnamefont
  {L.}~\bibnamefont {Jiang}}, \bibinfo {author} {\bibfnamefont
  {M.}~\bibnamefont {Mirrahimi}}, \bibinfo {author} {\bibfnamefont {M.~H.}\
  \bibnamefont {Devoret}}, \emph {et~al.},\ }\bibfield  {title} {\bibinfo
  {title} {{A CNOT gate between multiphoton qubits encoded in two cavities}},\
  }\href@noop {} {\bibfield  {journal} {\bibinfo  {journal} {Nature
  communications}\ }\textbf {\bibinfo {volume} {9}},\ \bibinfo {pages} {1}
  (\bibinfo {year} {2018})}\BibitemShut {NoStop}%
\bibitem [{\citenamefont {Reinhold}\ \emph {et~al.}(2020)\citenamefont
  {Reinhold}, \citenamefont {Rosenblum}, \citenamefont {Ma}, \citenamefont
  {Frunzio}, \citenamefont {Jiang},\ and\ \citenamefont
  {Schoelkopf}}]{reinhold:2020}%
  \BibitemOpen
  \bibfield  {author} {\bibinfo {author} {\bibfnamefont {P.}~\bibnamefont
  {Reinhold}}, \bibinfo {author} {\bibfnamefont {S.}~\bibnamefont {Rosenblum}},
  \bibinfo {author} {\bibfnamefont {W.-L.}\ \bibnamefont {Ma}}, \bibinfo
  {author} {\bibfnamefont {L.}~\bibnamefont {Frunzio}}, \bibinfo {author}
  {\bibfnamefont {L.}~\bibnamefont {Jiang}},\ and\ \bibinfo {author}
  {\bibfnamefont {R.~J.}\ \bibnamefont {Schoelkopf}},\ }\bibfield  {title}
  {\bibinfo {title} {{Error-corrected gates on an encoded qubit}},\ }\href@noop
  {} {\bibfield  {journal} {\bibinfo  {journal} {Nature Physics}\ }\textbf
  {\bibinfo {volume} {16}},\ \bibinfo {pages} {822} (\bibinfo {year}
  {2020})}\BibitemShut {NoStop}%
\bibitem [{\citenamefont {Puri}\ \emph {et~al.}(2019)\citenamefont {Puri},
  \citenamefont {Grimm}, \citenamefont {Campagne-Ibarcq}, \citenamefont
  {Eickbusch}, \citenamefont {Noh}, \citenamefont {Roberts}, \citenamefont
  {Jiang}, \citenamefont {Mirrahimi}, \citenamefont {Devoret},\ and\
  \citenamefont {Girvin}}]{puri:2019}%
  \BibitemOpen
  \bibfield  {author} {\bibinfo {author} {\bibfnamefont {S.}~\bibnamefont
  {Puri}}, \bibinfo {author} {\bibfnamefont {A.}~\bibnamefont {Grimm}},
  \bibinfo {author} {\bibfnamefont {P.}~\bibnamefont {Campagne-Ibarcq}},
  \bibinfo {author} {\bibfnamefont {A.}~\bibnamefont {Eickbusch}}, \bibinfo
  {author} {\bibfnamefont {K.}~\bibnamefont {Noh}}, \bibinfo {author}
  {\bibfnamefont {G.}~\bibnamefont {Roberts}}, \bibinfo {author} {\bibfnamefont
  {L.}~\bibnamefont {Jiang}}, \bibinfo {author} {\bibfnamefont
  {M.}~\bibnamefont {Mirrahimi}}, \bibinfo {author} {\bibfnamefont {M.~H.}\
  \bibnamefont {Devoret}},\ and\ \bibinfo {author} {\bibfnamefont {S.~M.}\
  \bibnamefont {Girvin}},\ }\bibfield  {title} {\bibinfo {title} {{Stabilized
  cat in a driven nonlinear cavity: A fault-tolerant error syndrome
  detector}},\ }\href {https://doi.org/10.1103/PhysRevX.9.041009} {\bibfield
  {journal} {\bibinfo  {journal} {PRX}\ }\textbf {\bibinfo {volume} {9}},\
  \bibinfo {pages} {041009} (\bibinfo {year} {2019})}\BibitemShut {NoStop}%
\bibitem [{\citenamefont {Grimm}\ \emph {et~al.}(2020)\citenamefont {Grimm},
  \citenamefont {Frattini}, \citenamefont {Puri}, \citenamefont {Mundhada},
  \citenamefont {Touzard}, \citenamefont {Mirrahimi}, \citenamefont {Girvin},
  \citenamefont {Shankar},\ and\ \citenamefont {Devoret}}]{grimm:2020}%
  \BibitemOpen
  \bibfield  {author} {\bibinfo {author} {\bibfnamefont {A.}~\bibnamefont
  {Grimm}}, \bibinfo {author} {\bibfnamefont {N.~E.}\ \bibnamefont {Frattini}},
  \bibinfo {author} {\bibfnamefont {S.}~\bibnamefont {Puri}}, \bibinfo {author}
  {\bibfnamefont {S.~O.}\ \bibnamefont {Mundhada}}, \bibinfo {author}
  {\bibfnamefont {S.}~\bibnamefont {Touzard}}, \bibinfo {author} {\bibfnamefont
  {M.}~\bibnamefont {Mirrahimi}}, \bibinfo {author} {\bibfnamefont {S.~M.}\
  \bibnamefont {Girvin}}, \bibinfo {author} {\bibfnamefont {S.}~\bibnamefont
  {Shankar}},\ and\ \bibinfo {author} {\bibfnamefont {M.~H.}\ \bibnamefont
  {Devoret}},\ }\bibfield  {title} {\bibinfo {title} {{Stabilization and
  operation of a Kerr-cat qubit}},\ }\href@noop {} {\bibfield  {journal}
  {\bibinfo  {journal} {Nature}\ }\textbf {\bibinfo {volume} {584}},\ \bibinfo
  {pages} {205} (\bibinfo {year} {2020})}\BibitemShut {NoStop}%
\bibitem [{\citenamefont {Hu}\ \emph {et~al.}(2019)\citenamefont {Hu},
  \citenamefont {Ma}, \citenamefont {Cai}, \citenamefont {Mu}, \citenamefont
  {Xu}, \citenamefont {Wang}, \citenamefont {Wu}, \citenamefont {Wang},
  \citenamefont {Song}, \citenamefont {Zou} \emph {et~al.}}]{hu:2019}%
  \BibitemOpen
  \bibfield  {author} {\bibinfo {author} {\bibfnamefont {L.}~\bibnamefont
  {Hu}}, \bibinfo {author} {\bibfnamefont {Y.}~\bibnamefont {Ma}}, \bibinfo
  {author} {\bibfnamefont {W.}~\bibnamefont {Cai}}, \bibinfo {author}
  {\bibfnamefont {X.}~\bibnamefont {Mu}}, \bibinfo {author} {\bibfnamefont
  {Y.}~\bibnamefont {Xu}}, \bibinfo {author} {\bibfnamefont {W.}~\bibnamefont
  {Wang}}, \bibinfo {author} {\bibfnamefont {Y.}~\bibnamefont {Wu}}, \bibinfo
  {author} {\bibfnamefont {H.}~\bibnamefont {Wang}}, \bibinfo {author}
  {\bibfnamefont {Y.~P.}\ \bibnamefont {Song}}, \bibinfo {author}
  {\bibfnamefont {C.-L.}\ \bibnamefont {Zou}}, \emph {et~al.},\ }\bibfield
  {title} {\bibinfo {title} {{Quantum error correction and universal gate set
  operation on a binomial bosonic logical qubit}},\ }\href@noop {} {\bibfield
  {journal} {\bibinfo  {journal} {Nature Physics}\ }\textbf {\bibinfo {volume}
  {15}},\ \bibinfo {pages} {503} (\bibinfo {year} {2019})}\BibitemShut
  {NoStop}%
\bibitem [{\citenamefont {Campagne-Ibarcq}\ \emph {et~al.}(2020)\citenamefont
  {Campagne-Ibarcq}, \citenamefont {Eickbusch}, \citenamefont {Touzard},
  \citenamefont {Zalys-Geller}, \citenamefont {Frattini}, \citenamefont
  {Sivak}, \citenamefont {Reinhold}, \citenamefont {Puri}, \citenamefont
  {Shankar}, \citenamefont {Schoelkopf} \emph {et~al.}}]{campagne:2020}%
  \BibitemOpen
  \bibfield  {author} {\bibinfo {author} {\bibfnamefont {P.}~\bibnamefont
  {Campagne-Ibarcq}}, \bibinfo {author} {\bibfnamefont {A.}~\bibnamefont
  {Eickbusch}}, \bibinfo {author} {\bibfnamefont {S.}~\bibnamefont {Touzard}},
  \bibinfo {author} {\bibfnamefont {E.}~\bibnamefont {Zalys-Geller}}, \bibinfo
  {author} {\bibfnamefont {N.~E.}\ \bibnamefont {Frattini}}, \bibinfo {author}
  {\bibfnamefont {V.~V.}\ \bibnamefont {Sivak}}, \bibinfo {author}
  {\bibfnamefont {P.}~\bibnamefont {Reinhold}}, \bibinfo {author}
  {\bibfnamefont {S.}~\bibnamefont {Puri}}, \bibinfo {author} {\bibfnamefont
  {S.}~\bibnamefont {Shankar}}, \bibinfo {author} {\bibfnamefont {R.~J.}\
  \bibnamefont {Schoelkopf}}, \emph {et~al.},\ }\bibfield  {title} {\bibinfo
  {title} {{Quantum error correction of a qubit encoded in grid states of an
  oscillator}},\ }\href@noop {} {\bibfield  {journal} {\bibinfo  {journal}
  {Nature}\ }\textbf {\bibinfo {volume} {584}},\ \bibinfo {pages} {368}
  (\bibinfo {year} {2020})}\BibitemShut {NoStop}%
\bibitem [{\citenamefont {Sivak}\ \emph {et~al.}(2022)\citenamefont {Sivak},
  \citenamefont {Eickbusch}, \citenamefont {Royer}, \citenamefont {Singh},
  \citenamefont {Tsioutsios}, \citenamefont {Ganjam}, \citenamefont {Miano},
  \citenamefont {Brock}, \citenamefont {Ding}, \citenamefont {Frunzio} \emph
  {et~al.}}]{sivak:2022}%
  \BibitemOpen
  \bibfield  {author} {\bibinfo {author} {\bibfnamefont {V.~V.}\ \bibnamefont
  {Sivak}}, \bibinfo {author} {\bibfnamefont {A.}~\bibnamefont {Eickbusch}},
  \bibinfo {author} {\bibfnamefont {B.}~\bibnamefont {Royer}}, \bibinfo
  {author} {\bibfnamefont {S.}~\bibnamefont {Singh}}, \bibinfo {author}
  {\bibfnamefont {I.}~\bibnamefont {Tsioutsios}}, \bibinfo {author}
  {\bibfnamefont {S.}~\bibnamefont {Ganjam}}, \bibinfo {author} {\bibfnamefont
  {A.}~\bibnamefont {Miano}}, \bibinfo {author} {\bibfnamefont {B.~L.}\
  \bibnamefont {Brock}}, \bibinfo {author} {\bibfnamefont {A.~Z.}\ \bibnamefont
  {Ding}}, \bibinfo {author} {\bibfnamefont {L.}~\bibnamefont {Frunzio}}, \emph
  {et~al.},\ }\bibfield  {title} {\bibinfo {title} {{Real-time quantum error
  correction beyond break-even}},\ }\href@noop {} {\bibfield  {journal}
  {\bibinfo  {journal} {arXiv preprint arXiv:2211.09116}\ } (\bibinfo {year}
  {2022})}\BibitemShut {NoStop}%
\bibitem [{\citenamefont {Ni}\ \emph {et~al.}(2022)\citenamefont {Ni},
  \citenamefont {Li}, \citenamefont {Deng}, \citenamefont {Cai}, \citenamefont
  {Zhang}, \citenamefont {Wang}, \citenamefont {Yang}, \citenamefont {Yu},
  \citenamefont {Yan}, \citenamefont {Liu} \emph {et~al.}}]{ni:2022}%
  \BibitemOpen
  \bibfield  {author} {\bibinfo {author} {\bibfnamefont {Z.}~\bibnamefont
  {Ni}}, \bibinfo {author} {\bibfnamefont {S.}~\bibnamefont {Li}}, \bibinfo
  {author} {\bibfnamefont {X.}~\bibnamefont {Deng}}, \bibinfo {author}
  {\bibfnamefont {Y.}~\bibnamefont {Cai}}, \bibinfo {author} {\bibfnamefont
  {L.}~\bibnamefont {Zhang}}, \bibinfo {author} {\bibfnamefont
  {W.}~\bibnamefont {Wang}}, \bibinfo {author} {\bibfnamefont {Z.-B.}\
  \bibnamefont {Yang}}, \bibinfo {author} {\bibfnamefont {H.}~\bibnamefont
  {Yu}}, \bibinfo {author} {\bibfnamefont {F.}~\bibnamefont {Yan}}, \bibinfo
  {author} {\bibfnamefont {S.}~\bibnamefont {Liu}}, \emph {et~al.},\ }\bibfield
   {title} {\bibinfo {title} {{Beating the break-even point with a
  discrete-variable-encoded logical qubit}},\ }\href@noop {} {\bibfield
  {journal} {\bibinfo  {journal} {arXiv preprint arXiv:2211.09319}\ } (\bibinfo
  {year} {2022})}\BibitemShut {NoStop}%
\bibitem [{\citenamefont {Wang}\ \emph {et~al.}(2016)\citenamefont {Wang},
  \citenamefont {Gao}, \citenamefont {Reinhold}, \citenamefont {Heeres},
  \citenamefont {Ofek}, \citenamefont {Chou}, \citenamefont {Axline},
  \citenamefont {Reagor}, \citenamefont {Blumoff}, \citenamefont {Sliwa} \emph
  {et~al.}}]{wang:2016}%
  \BibitemOpen
  \bibfield  {author} {\bibinfo {author} {\bibfnamefont {C.}~\bibnamefont
  {Wang}}, \bibinfo {author} {\bibfnamefont {Y.~Y.}\ \bibnamefont {Gao}},
  \bibinfo {author} {\bibfnamefont {P.}~\bibnamefont {Reinhold}}, \bibinfo
  {author} {\bibfnamefont {R.~W.}\ \bibnamefont {Heeres}}, \bibinfo {author}
  {\bibfnamefont {N.}~\bibnamefont {Ofek}}, \bibinfo {author} {\bibfnamefont
  {K.}~\bibnamefont {Chou}}, \bibinfo {author} {\bibfnamefont {C.}~\bibnamefont
  {Axline}}, \bibinfo {author} {\bibfnamefont {M.}~\bibnamefont {Reagor}},
  \bibinfo {author} {\bibfnamefont {J.}~\bibnamefont {Blumoff}}, \bibinfo
  {author} {\bibfnamefont {K.~M.}\ \bibnamefont {Sliwa}}, \emph {et~al.},\
  }\bibfield  {title} {\bibinfo {title} {{A Schr{\"o}dinger cat living in two
  boxes}},\ }\href@noop {} {\bibfield  {journal} {\bibinfo  {journal}
  {Science}\ }\textbf {\bibinfo {volume} {352}},\ \bibinfo {pages} {1087}
  (\bibinfo {year} {2016})}\BibitemShut {NoStop}%
\bibitem [{Note1()}]{Note1}%
  \BibitemOpen
  \bibinfo {note} {A third approach is to engineer a noise-biased coupler~\cite
  {pietikainen:2021}.}\BibitemShut {Stop}%
\bibitem [{\citenamefont {Lloyd}\ and\ \citenamefont
  {Braunstein}(1999)}]{lloyd:1999}%
  \BibitemOpen
  \bibfield  {author} {\bibinfo {author} {\bibfnamefont {S.}~\bibnamefont
  {Lloyd}}\ and\ \bibinfo {author} {\bibfnamefont {S.~L.}\ \bibnamefont
  {Braunstein}},\ }\bibfield  {title} {\bibinfo {title} {{Quantum computation
  over continuous variables}},\ }\href
  {https://doi.org/10.1103/PhysRevLett.82.1784} {\bibfield  {journal} {\bibinfo
   {journal} {Physical Review Letters}\ }\textbf {\bibinfo {volume} {82}},\
  \bibinfo {pages} {1784} (\bibinfo {year} {1999})}\BibitemShut {NoStop}%
\bibitem [{\citenamefont {Knill}\ \emph {et~al.}(2001)\citenamefont {Knill},
  \citenamefont {Laflamme},\ and\ \citenamefont {Milburn}}]{knill:2001}%
  \BibitemOpen
  \bibfield  {author} {\bibinfo {author} {\bibfnamefont {E.}~\bibnamefont
  {Knill}}, \bibinfo {author} {\bibfnamefont {R.}~\bibnamefont {Laflamme}},\
  and\ \bibinfo {author} {\bibfnamefont {G.~J.}\ \bibnamefont {Milburn}},\
  }\bibfield  {title} {\bibinfo {title} {{A scheme for efficient quantum
  computation with linear optics}},\ }\href@noop {} {\bibfield  {journal}
  {\bibinfo  {journal} {Nature}\ }\textbf {\bibinfo {volume} {409}},\ \bibinfo
  {pages} {46} (\bibinfo {year} {2001})}\BibitemShut {NoStop}%
\bibitem [{\citenamefont {Gao}\ \emph {et~al.}(2019)\citenamefont {Gao},
  \citenamefont {Lester}, \citenamefont {Chou}, \citenamefont {Frunzio},
  \citenamefont {Devoret}, \citenamefont {Jiang}, \citenamefont {Girvin},\ and\
  \citenamefont {Schoelkopf}}]{gao:2019}%
  \BibitemOpen
  \bibfield  {author} {\bibinfo {author} {\bibfnamefont {Y.~Y.}\ \bibnamefont
  {Gao}}, \bibinfo {author} {\bibfnamefont {B.~J.}\ \bibnamefont {Lester}},
  \bibinfo {author} {\bibfnamefont {K.~S.}\ \bibnamefont {Chou}}, \bibinfo
  {author} {\bibfnamefont {L.}~\bibnamefont {Frunzio}}, \bibinfo {author}
  {\bibfnamefont {M.~H.}\ \bibnamefont {Devoret}}, \bibinfo {author}
  {\bibfnamefont {L.}~\bibnamefont {Jiang}}, \bibinfo {author} {\bibfnamefont
  {S.~M.}\ \bibnamefont {Girvin}},\ and\ \bibinfo {author} {\bibfnamefont
  {R.~J.}\ \bibnamefont {Schoelkopf}},\ }\bibfield  {title} {\bibinfo {title}
  {{Entanglement of bosonic modes through an engineered exchange
  interaction}},\ }\href@noop {} {\bibfield  {journal} {\bibinfo  {journal}
  {Nature}\ }\textbf {\bibinfo {volume} {566}},\ \bibinfo {pages} {509}
  (\bibinfo {year} {2019})}\BibitemShut {NoStop}%
\bibitem [{\citenamefont {Gao}\ \emph {et~al.}(2018)\citenamefont {Gao},
  \citenamefont {Lester}, \citenamefont {Zhang}, \citenamefont {Wang},
  \citenamefont {Rosenblum}, \citenamefont {Frunzio}, \citenamefont {Jiang},
  \citenamefont {Girvin},\ and\ \citenamefont {Schoelkopf}}]{gao:2018}%
  \BibitemOpen
  \bibfield  {author} {\bibinfo {author} {\bibfnamefont {Y.~Y.}\ \bibnamefont
  {Gao}}, \bibinfo {author} {\bibfnamefont {B.~J.}\ \bibnamefont {Lester}},
  \bibinfo {author} {\bibfnamefont {Y.}~\bibnamefont {Zhang}}, \bibinfo
  {author} {\bibfnamefont {C.}~\bibnamefont {Wang}}, \bibinfo {author}
  {\bibfnamefont {S.}~\bibnamefont {Rosenblum}}, \bibinfo {author}
  {\bibfnamefont {L.}~\bibnamefont {Frunzio}}, \bibinfo {author} {\bibfnamefont
  {L.}~\bibnamefont {Jiang}}, \bibinfo {author} {\bibfnamefont {S.~M.}\
  \bibnamefont {Girvin}},\ and\ \bibinfo {author} {\bibfnamefont {R.~J.}\
  \bibnamefont {Schoelkopf}},\ }\bibfield  {title} {\bibinfo {title}
  {{Programmable interference between two microwave quantum memories}},\
  }\href@noop {} {\bibfield  {journal} {\bibinfo  {journal} {Physical Review
  X}\ }\textbf {\bibinfo {volume} {8}},\ \bibinfo {pages} {021073} (\bibinfo
  {year} {2018})}\BibitemShut {NoStop}%
\bibitem [{\citenamefont {Zhang}\ \emph {et~al.}(2019)\citenamefont {Zhang},
  \citenamefont {Lester}, \citenamefont {Gao}, \citenamefont {Jiang},
  \citenamefont {Schoelkopf},\ and\ \citenamefont {Girvin}}]{zhang:2019}%
  \BibitemOpen
  \bibfield  {author} {\bibinfo {author} {\bibfnamefont {Y.}~\bibnamefont
  {Zhang}}, \bibinfo {author} {\bibfnamefont {B.~J.}\ \bibnamefont {Lester}},
  \bibinfo {author} {\bibfnamefont {Y.~Y.}\ \bibnamefont {Gao}}, \bibinfo
  {author} {\bibfnamefont {L.}~\bibnamefont {Jiang}}, \bibinfo {author}
  {\bibfnamefont {R.~J.}\ \bibnamefont {Schoelkopf}},\ and\ \bibinfo {author}
  {\bibfnamefont {S.~M.}\ \bibnamefont {Girvin}},\ }\bibfield  {title}
  {\bibinfo {title} {{Engineering bilinear mode coupling in circuit QED: Theory
  and experiment}},\ }\href@noop {} {\bibfield  {journal} {\bibinfo  {journal}
  {Physical Review A}\ }\textbf {\bibinfo {volume} {99}},\ \bibinfo {pages}
  {012314} (\bibinfo {year} {2019})}\BibitemShut {NoStop}%
\bibitem [{\citenamefont {Chen}\ \emph {et~al.}(2014)\citenamefont {Chen},
  \citenamefont {Neill}, \citenamefont {Roushan}, \citenamefont {Leung},
  \citenamefont {Fang}, \citenamefont {Barends}, \citenamefont {Kelly},
  \citenamefont {Campbell}, \citenamefont {Chen}, \citenamefont {Chiaro} \emph
  {et~al.}}]{chen:2014}%
  \BibitemOpen
  \bibfield  {author} {\bibinfo {author} {\bibfnamefont {Y.}~\bibnamefont
  {Chen}}, \bibinfo {author} {\bibfnamefont {C.}~\bibnamefont {Neill}},
  \bibinfo {author} {\bibfnamefont {P.}~\bibnamefont {Roushan}}, \bibinfo
  {author} {\bibfnamefont {N.}~\bibnamefont {Leung}}, \bibinfo {author}
  {\bibfnamefont {M.}~\bibnamefont {Fang}}, \bibinfo {author} {\bibfnamefont
  {R.}~\bibnamefont {Barends}}, \bibinfo {author} {\bibfnamefont
  {J.}~\bibnamefont {Kelly}}, \bibinfo {author} {\bibfnamefont
  {B.}~\bibnamefont {Campbell}}, \bibinfo {author} {\bibfnamefont
  {Z.}~\bibnamefont {Chen}}, \bibinfo {author} {\bibfnamefont {B.}~\bibnamefont
  {Chiaro}}, \emph {et~al.},\ }\bibfield  {title} {\bibinfo {title} {{Qubit
  architecture with high coherence and fast tunable coupling}},\ }\href@noop {}
  {\bibfield  {journal} {\bibinfo  {journal} {Physical Review Letters}\
  }\textbf {\bibinfo {volume} {113}},\ \bibinfo {pages} {220502} (\bibinfo
  {year} {2014})}\BibitemShut {NoStop}%
\bibitem [{\citenamefont {Yan}\ \emph {et~al.}(2018)\citenamefont {Yan},
  \citenamefont {Krantz}, \citenamefont {Sung}, \citenamefont {Kjaergaard},
  \citenamefont {Campbell}, \citenamefont {Orlando}, \citenamefont
  {Gustavsson},\ and\ \citenamefont {Oliver}}]{yan:2018}%
  \BibitemOpen
  \bibfield  {author} {\bibinfo {author} {\bibfnamefont {F.}~\bibnamefont
  {Yan}}, \bibinfo {author} {\bibfnamefont {P.}~\bibnamefont {Krantz}},
  \bibinfo {author} {\bibfnamefont {Y.}~\bibnamefont {Sung}}, \bibinfo {author}
  {\bibfnamefont {M.}~\bibnamefont {Kjaergaard}}, \bibinfo {author}
  {\bibfnamefont {D.~L.}\ \bibnamefont {Campbell}}, \bibinfo {author}
  {\bibfnamefont {T.~P.}\ \bibnamefont {Orlando}}, \bibinfo {author}
  {\bibfnamefont {S.}~\bibnamefont {Gustavsson}},\ and\ \bibinfo {author}
  {\bibfnamefont {W.~D.}\ \bibnamefont {Oliver}},\ }\bibfield  {title}
  {\bibinfo {title} {{Tunable coupling scheme for implementing high-fidelity
  two-qubit gates}},\ }\href@noop {} {\bibfield  {journal} {\bibinfo  {journal}
  {Physical Review Applied}\ }\textbf {\bibinfo {volume} {10}},\ \bibinfo
  {pages} {054062} (\bibinfo {year} {2018})}\BibitemShut {NoStop}%
\bibitem [{\citenamefont {Pfaff}\ \emph {et~al.}(2017)\citenamefont {Pfaff},
  \citenamefont {Axline}, \citenamefont {Burkhart}, \citenamefont {Vool},
  \citenamefont {Reinhold}, \citenamefont {Frunzio}, \citenamefont {Jiang},
  \citenamefont {Devoret},\ and\ \citenamefont {Schoelkopf}}]{pfaff:2017}%
  \BibitemOpen
  \bibfield  {author} {\bibinfo {author} {\bibfnamefont {W.}~\bibnamefont
  {Pfaff}}, \bibinfo {author} {\bibfnamefont {C.~J.}\ \bibnamefont {Axline}},
  \bibinfo {author} {\bibfnamefont {L.~D.}\ \bibnamefont {Burkhart}}, \bibinfo
  {author} {\bibfnamefont {U.}~\bibnamefont {Vool}}, \bibinfo {author}
  {\bibfnamefont {P.}~\bibnamefont {Reinhold}}, \bibinfo {author}
  {\bibfnamefont {L.}~\bibnamefont {Frunzio}}, \bibinfo {author} {\bibfnamefont
  {L.}~\bibnamefont {Jiang}}, \bibinfo {author} {\bibfnamefont {M.~H.}\
  \bibnamefont {Devoret}},\ and\ \bibinfo {author} {\bibfnamefont {R.~J.}\
  \bibnamefont {Schoelkopf}},\ }\bibfield  {title} {\bibinfo {title}
  {Controlled release of multiphoton quantum states from a microwave cavity
  memory},\ }\href@noop {} {\bibfield  {journal} {\bibinfo  {journal} {Nature
  Physics}\ }\textbf {\bibinfo {volume} {13}},\ \bibinfo {pages} {882}
  (\bibinfo {year} {2017})}\BibitemShut {NoStop}%
\bibitem [{\citenamefont {Frattini}\ \emph {et~al.}(2017)\citenamefont
  {Frattini}, \citenamefont {Vool}, \citenamefont {Shankar}, \citenamefont
  {Narla}, \citenamefont {Sliwa},\ and\ \citenamefont
  {Devoret}}]{frattini:2017}%
  \BibitemOpen
  \bibfield  {author} {\bibinfo {author} {\bibfnamefont {N.~E.}\ \bibnamefont
  {Frattini}}, \bibinfo {author} {\bibfnamefont {U.}~\bibnamefont {Vool}},
  \bibinfo {author} {\bibfnamefont {S.}~\bibnamefont {Shankar}}, \bibinfo
  {author} {\bibfnamefont {A.}~\bibnamefont {Narla}}, \bibinfo {author}
  {\bibfnamefont {K.}~\bibnamefont {Sliwa}},\ and\ \bibinfo {author}
  {\bibfnamefont {M.~H.}\ \bibnamefont {Devoret}},\ }\bibfield  {title}
  {\bibinfo {title} {{3-wave mixing Josephson dipole element}},\ }\href@noop {}
  {\bibfield  {journal} {\bibinfo  {journal} {Applied Physics Letters}\
  }\textbf {\bibinfo {volume} {110}},\ \bibinfo {pages} {222603} (\bibinfo
  {year} {2017})}\BibitemShut {NoStop}%
\bibitem [{\citenamefont {Zhou}\ \emph {et~al.}(2021)\citenamefont {Zhou},
  \citenamefont {Lu}, \citenamefont {Praquin}, \citenamefont {Chien},
  \citenamefont {Kaufman}, \citenamefont {Cao}, \citenamefont {Xia},
  \citenamefont {Mong}, \citenamefont {Pfaff}, \citenamefont {Pekker} \emph
  {et~al.}}]{zhou:2021}%
  \BibitemOpen
  \bibfield  {author} {\bibinfo {author} {\bibfnamefont {C.}~\bibnamefont
  {Zhou}}, \bibinfo {author} {\bibfnamefont {P.}~\bibnamefont {Lu}}, \bibinfo
  {author} {\bibfnamefont {M.}~\bibnamefont {Praquin}}, \bibinfo {author}
  {\bibfnamefont {T.-C.}\ \bibnamefont {Chien}}, \bibinfo {author}
  {\bibfnamefont {R.}~\bibnamefont {Kaufman}}, \bibinfo {author} {\bibfnamefont
  {X.}~\bibnamefont {Cao}}, \bibinfo {author} {\bibfnamefont {M.}~\bibnamefont
  {Xia}}, \bibinfo {author} {\bibfnamefont {R.}~\bibnamefont {Mong}}, \bibinfo
  {author} {\bibfnamefont {W.}~\bibnamefont {Pfaff}}, \bibinfo {author}
  {\bibfnamefont {D.}~\bibnamefont {Pekker}}, \emph {et~al.},\ }\bibfield
  {title} {\bibinfo {title} {{A modular quantum computer based on a quantum
  state router}},\ }\href@noop {} {\bibfield  {journal} {\bibinfo  {journal}
  {arXiv preprint arXiv:2109.06848}\ } (\bibinfo {year} {2021})}\BibitemShut
  {NoStop}%
\bibitem [{\citenamefont {Zakka-Bajjani}\ \emph {et~al.}(2011)\citenamefont
  {Zakka-Bajjani}, \citenamefont {Nguyen}, \citenamefont {Lee}, \citenamefont
  {Vale}, \citenamefont {Simmonds},\ and\ \citenamefont
  {Aumentado}}]{zakka:2011}%
  \BibitemOpen
  \bibfield  {author} {\bibinfo {author} {\bibfnamefont {E.}~\bibnamefont
  {Zakka-Bajjani}}, \bibinfo {author} {\bibfnamefont {F.}~\bibnamefont
  {Nguyen}}, \bibinfo {author} {\bibfnamefont {M.}~\bibnamefont {Lee}},
  \bibinfo {author} {\bibfnamefont {L.~R.}\ \bibnamefont {Vale}}, \bibinfo
  {author} {\bibfnamefont {R.~W.}\ \bibnamefont {Simmonds}},\ and\ \bibinfo
  {author} {\bibfnamefont {J.}~\bibnamefont {Aumentado}},\ }\bibfield  {title}
  {\bibinfo {title} {Quantum superposition of a single microwave photon in two
  different `colour’ states},\ }\href@noop {} {\bibfield  {journal} {\bibinfo
   {journal} {Nature Physics}\ }\textbf {\bibinfo {volume} {7}},\ \bibinfo
  {pages} {599} (\bibinfo {year} {2011})}\BibitemShut {NoStop}%
\bibitem [{\citenamefont {Sirois}\ \emph {et~al.}(2015)\citenamefont {Sirois},
  \citenamefont {Castellanos-Beltran}, \citenamefont {DeFeo}, \citenamefont
  {Ranzani}, \citenamefont {Lecocq}, \citenamefont {Simmonds}, \citenamefont
  {Teufel},\ and\ \citenamefont {Aumentado}}]{sirois:2015}%
  \BibitemOpen
  \bibfield  {author} {\bibinfo {author} {\bibfnamefont {A.~J.}\ \bibnamefont
  {Sirois}}, \bibinfo {author} {\bibfnamefont {M.~A.}\ \bibnamefont
  {Castellanos-Beltran}}, \bibinfo {author} {\bibfnamefont {M.~P.}\
  \bibnamefont {DeFeo}}, \bibinfo {author} {\bibfnamefont {L.}~\bibnamefont
  {Ranzani}}, \bibinfo {author} {\bibfnamefont {F.}~\bibnamefont {Lecocq}},
  \bibinfo {author} {\bibfnamefont {R.~W.}\ \bibnamefont {Simmonds}}, \bibinfo
  {author} {\bibfnamefont {J.~D.}\ \bibnamefont {Teufel}},\ and\ \bibinfo
  {author} {\bibfnamefont {J.}~\bibnamefont {Aumentado}},\ }\bibfield  {title}
  {\bibinfo {title} {Coherent-state storage and retrieval between
  superconducting cavities using parametric frequency conversion},\ }\href
  {https://doi.org/10.1063/1.4919759} {\bibfield  {journal} {\bibinfo
  {journal} {Applied Physics Letters}\ }\textbf {\bibinfo {volume} {106}},\
  \bibinfo {pages} {172603} (\bibinfo {year} {2015})}\BibitemShut {NoStop}%
\bibitem [{\citenamefont {Lu}\ \emph {et~al.}(2023)\citenamefont {Lu},
  \citenamefont {Maiti}, \citenamefont {Garmon}, \citenamefont {Ganjam},
  \citenamefont {Zhang}, \citenamefont {Claes}, \citenamefont {Frunzio},
  \citenamefont {Girvin},\ and\ \citenamefont {Schoelkopf}}]{lu:2023}%
  \BibitemOpen
  \bibfield  {author} {\bibinfo {author} {\bibfnamefont {Y.}~\bibnamefont
  {Lu}}, \bibinfo {author} {\bibfnamefont {A.}~\bibnamefont {Maiti}}, \bibinfo
  {author} {\bibfnamefont {J.~W.}\ \bibnamefont {Garmon}}, \bibinfo {author}
  {\bibfnamefont {S.}~\bibnamefont {Ganjam}}, \bibinfo {author} {\bibfnamefont
  {Y.}~\bibnamefont {Zhang}}, \bibinfo {author} {\bibfnamefont
  {J.}~\bibnamefont {Claes}}, \bibinfo {author} {\bibfnamefont
  {L.}~\bibnamefont {Frunzio}}, \bibinfo {author} {\bibfnamefont
  {S.}~\bibnamefont {Girvin}},\ and\ \bibinfo {author} {\bibfnamefont {R.~J.}\
  \bibnamefont {Schoelkopf}},\ }\bibfield  {title} {\bibinfo {title} {A
  high-fidelity microwave beamsplitter with a parity-protected converter},\
  }\href@noop {} {\bibfield  {journal} {\bibinfo  {journal} {arXiv preprint
  arXiv:2303.00959}\ } (\bibinfo {year} {2023})}\BibitemShut {NoStop}%
\bibitem [{\citenamefont {Frattini}\ \emph {et~al.}(2018)\citenamefont
  {Frattini}, \citenamefont {Sivak}, \citenamefont {Lingenfelter},
  \citenamefont {Shankar},\ and\ \citenamefont {Devoret}}]{frattini:2018}%
  \BibitemOpen
  \bibfield  {author} {\bibinfo {author} {\bibfnamefont {N.~E.}\ \bibnamefont
  {Frattini}}, \bibinfo {author} {\bibfnamefont {V.~V.}\ \bibnamefont {Sivak}},
  \bibinfo {author} {\bibfnamefont {A.}~\bibnamefont {Lingenfelter}}, \bibinfo
  {author} {\bibfnamefont {S.}~\bibnamefont {Shankar}},\ and\ \bibinfo {author}
  {\bibfnamefont {M.~H.}\ \bibnamefont {Devoret}},\ }\bibfield  {title}
  {\bibinfo {title} {{Optimizing the nonlinearity and dissipation of a SNAIL
  parametric amplifier for dynamic range}},\ }\href@noop {} {\bibfield
  {journal} {\bibinfo  {journal} {Physical Review Applied}\ }\textbf {\bibinfo
  {volume} {10}},\ \bibinfo {pages} {054020} (\bibinfo {year}
  {2018})}\BibitemShut {NoStop}%
\bibitem [{\citenamefont {Liu}\ \emph {et~al.}(2017)\citenamefont {Liu},
  \citenamefont {Chien}, \citenamefont {Cao}, \citenamefont {Lanes},
  \citenamefont {Alpern}, \citenamefont {Pekker},\ and\ \citenamefont
  {Hatridge}}]{liu:2017}%
  \BibitemOpen
  \bibfield  {author} {\bibinfo {author} {\bibfnamefont {G.}~\bibnamefont
  {Liu}}, \bibinfo {author} {\bibfnamefont {T.~C.}\ \bibnamefont {Chien}},
  \bibinfo {author} {\bibfnamefont {X.}~\bibnamefont {Cao}}, \bibinfo {author}
  {\bibfnamefont {O.}~\bibnamefont {Lanes}}, \bibinfo {author} {\bibfnamefont
  {E.}~\bibnamefont {Alpern}}, \bibinfo {author} {\bibfnamefont
  {D.}~\bibnamefont {Pekker}},\ and\ \bibinfo {author} {\bibfnamefont
  {M.}~\bibnamefont {Hatridge}},\ }\bibfield  {title} {\bibinfo {title}
  {{Josephson parametric converter saturation and higher order effects}},\
  }\href@noop {} {\bibfield  {journal} {\bibinfo  {journal} {Applied Physics
  Letters}\ }\textbf {\bibinfo {volume} {111}},\ \bibinfo {pages} {202603}
  (\bibinfo {year} {2017})}\BibitemShut {NoStop}%
\bibitem [{\citenamefont {Sivak}\ \emph {et~al.}(2019)\citenamefont {Sivak},
  \citenamefont {Frattini}, \citenamefont {Joshi}, \citenamefont
  {Lingenfelter}, \citenamefont {Shankar},\ and\ \citenamefont
  {Devoret}}]{sivak:2019}%
  \BibitemOpen
  \bibfield  {author} {\bibinfo {author} {\bibfnamefont {V.~V.}\ \bibnamefont
  {Sivak}}, \bibinfo {author} {\bibfnamefont {N.~E.}\ \bibnamefont {Frattini}},
  \bibinfo {author} {\bibfnamefont {V.~R.}\ \bibnamefont {Joshi}}, \bibinfo
  {author} {\bibfnamefont {A.}~\bibnamefont {Lingenfelter}}, \bibinfo {author}
  {\bibfnamefont {S.}~\bibnamefont {Shankar}},\ and\ \bibinfo {author}
  {\bibfnamefont {M.~H.}\ \bibnamefont {Devoret}},\ }\bibfield  {title}
  {\bibinfo {title} {{Kerr-free three-wave mixing in superconducting quantum
  circuits}},\ }\href@noop {} {\bibfield  {journal} {\bibinfo  {journal}
  {Physical Review Applied}\ }\textbf {\bibinfo {volume} {11}},\ \bibinfo
  {pages} {054060} (\bibinfo {year} {2019})}\BibitemShut {NoStop}%
\bibitem [{Note2()}]{Note2}%
  \BibitemOpen
  \bibinfo {note} {We constrain that fit by probing the DC resistance of the
  coupler at room temperature, and using the Ambegaokar-Baratoff relation \cite
  {Ambegaokar:1963} to infer the kinetic inductance of the SNAIL from that
  measurement.}\BibitemShut {Stop}%
\bibitem [{\citenamefont {Frattini}(2021)}]{frattini:2021}%
  \BibitemOpen
  \bibfield  {author} {\bibinfo {author} {\bibfnamefont {N.~E.}\ \bibnamefont
  {Frattini}},\ }\emph {\bibinfo {title} {{Three-wave mixing in superconducting
  circuits: Stabilizing cats with SNAILs}}},\ \href@noop {} {Ph.D. thesis},\
  \bibinfo  {school} {Yale University} (\bibinfo {year} {2021})\BibitemShut
  {NoStop}%
\bibitem [{\citenamefont {Miano}\ \emph {et~al.}(2023)\citenamefont {Miano},
  \citenamefont {Joshi}, \citenamefont {Liu}, \citenamefont {Dai},
  \citenamefont {Parakh}, \citenamefont {Frunzio},\ and\ \citenamefont
  {Devoret}}]{miano:2023}%
  \BibitemOpen
  \bibfield  {author} {\bibinfo {author} {\bibfnamefont {A.}~\bibnamefont
  {Miano}}, \bibinfo {author} {\bibfnamefont {V.}~\bibnamefont {Joshi}},
  \bibinfo {author} {\bibfnamefont {G.}~\bibnamefont {Liu}}, \bibinfo {author}
  {\bibfnamefont {W.}~\bibnamefont {Dai}}, \bibinfo {author} {\bibfnamefont
  {P.}~\bibnamefont {Parakh}}, \bibinfo {author} {\bibfnamefont
  {L.}~\bibnamefont {Frunzio}},\ and\ \bibinfo {author} {\bibfnamefont
  {M.}~\bibnamefont {Devoret}},\ }\bibfield  {title} {\bibinfo {title}
  {Hamiltonian extrema of an arbitrary flux-biased josephson circuit},\
  }\href@noop {} {\bibfield  {journal} {\bibinfo  {journal} {arXiv preprint
  arXiv:2302.03155}\ } (\bibinfo {year} {2023})}\BibitemShut {NoStop}%
\bibitem [{\citenamefont {Venkatraman}\ \emph {et~al.}(2022)\citenamefont
  {Venkatraman}, \citenamefont {Xiao}, \citenamefont {Corti\~nas},
  \citenamefont {Eickbusch},\ and\ \citenamefont {Devoret}}]{Venkatraman:2022}%
  \BibitemOpen
  \bibfield  {author} {\bibinfo {author} {\bibfnamefont {J.}~\bibnamefont
  {Venkatraman}}, \bibinfo {author} {\bibfnamefont {X.}~\bibnamefont {Xiao}},
  \bibinfo {author} {\bibfnamefont {R.~G.}\ \bibnamefont {Corti\~nas}},
  \bibinfo {author} {\bibfnamefont {A.}~\bibnamefont {Eickbusch}},\ and\
  \bibinfo {author} {\bibfnamefont {M.~H.}\ \bibnamefont {Devoret}},\
  }\bibfield  {title} {\bibinfo {title} {{Static effective Hamiltonian of a
  rapidly driven nonlinear system}},\ }\href
  {https://doi.org/10.1103/PhysRevLett.129.100601} {\bibfield  {journal}
  {\bibinfo  {journal} {Physical Review Letters}\ }\textbf {\bibinfo {volume}
  {129}},\ \bibinfo {pages} {100601} (\bibinfo {year} {2022})}\BibitemShut
  {NoStop}%
\bibitem [{\citenamefont {Milburn}(1989)}]{milburn:1989}%
  \BibitemOpen
  \bibfield  {author} {\bibinfo {author} {\bibfnamefont {G.~J.}\ \bibnamefont
  {Milburn}},\ }\bibfield  {title} {\bibinfo {title} {{Quantum optical Fredkin
  gate}},\ }\href {https://doi.org/10.1103/PhysRevLett.62.2124} {\bibfield
  {journal} {\bibinfo  {journal} {Physical Review Letters}\ }\textbf {\bibinfo
  {volume} {62}},\ \bibinfo {pages} {2124} (\bibinfo {year}
  {1989})}\BibitemShut {NoStop}%
\bibitem [{\citenamefont {Giovannetti}\ \emph
  {et~al.}(2008{\natexlab{a}})\citenamefont {Giovannetti}, \citenamefont
  {Lloyd},\ and\ \citenamefont {Maccone}}]{giovannetti:2008a}%
  \BibitemOpen
  \bibfield  {author} {\bibinfo {author} {\bibfnamefont {V.}~\bibnamefont
  {Giovannetti}}, \bibinfo {author} {\bibfnamefont {S.}~\bibnamefont {Lloyd}},\
  and\ \bibinfo {author} {\bibfnamefont {L.}~\bibnamefont {Maccone}},\
  }\bibfield  {title} {\bibinfo {title} {{Quantum random access memory}},\
  }\href {https://doi.org/10.1103/PhysRevLett.100.160501} {\bibfield  {journal}
  {\bibinfo  {journal} {Physical Review Letters}\ }\textbf {\bibinfo {volume}
  {100}},\ \bibinfo {pages} {160501} (\bibinfo {year}
  {2008}{\natexlab{a}})}\BibitemShut {NoStop}%
\bibitem [{\citenamefont {Giovannetti}\ \emph
  {et~al.}(2008{\natexlab{b}})\citenamefont {Giovannetti}, \citenamefont
  {Lloyd},\ and\ \citenamefont {Maccone}}]{giovannetti:2008b}%
  \BibitemOpen
  \bibfield  {author} {\bibinfo {author} {\bibfnamefont {V.}~\bibnamefont
  {Giovannetti}}, \bibinfo {author} {\bibfnamefont {S.}~\bibnamefont {Lloyd}},\
  and\ \bibinfo {author} {\bibfnamefont {L.}~\bibnamefont {Maccone}},\
  }\bibfield  {title} {\bibinfo {title} {{Architectures for a quantum random
  access memory}},\ }\href {https://doi.org/10.1103/PhysRevA.78.052310}
  {\bibfield  {journal} {\bibinfo  {journal} {Physical Review A}\ }\textbf
  {\bibinfo {volume} {78}},\ \bibinfo {pages} {052310} (\bibinfo {year}
  {2008}{\natexlab{b}})}\BibitemShut {NoStop}%
\bibitem [{\citenamefont {Ekert}\ \emph {et~al.}(2002)\citenamefont {Ekert},
  \citenamefont {Alves}, \citenamefont {Oi}, \citenamefont {Horodecki},
  \citenamefont {Horodecki},\ and\ \citenamefont {Kwek}}]{ekert:2002}%
  \BibitemOpen
  \bibfield  {author} {\bibinfo {author} {\bibfnamefont {A.~K.}\ \bibnamefont
  {Ekert}}, \bibinfo {author} {\bibfnamefont {C.~M.}\ \bibnamefont {Alves}},
  \bibinfo {author} {\bibfnamefont {D.~K.}\ \bibnamefont {Oi}}, \bibinfo
  {author} {\bibfnamefont {M.}~\bibnamefont {Horodecki}}, \bibinfo {author}
  {\bibfnamefont {P.}~\bibnamefont {Horodecki}},\ and\ \bibinfo {author}
  {\bibfnamefont {L.~C.}\ \bibnamefont {Kwek}},\ }\bibfield  {title} {\bibinfo
  {title} {{Direct estimations of linear and nonlinear functionals of a quantum
  state}},\ }\href@noop {} {\bibfield  {journal} {\bibinfo  {journal} {Physical
  Review Letters}\ }\textbf {\bibinfo {volume} {88}},\ \bibinfo {pages}
  {217901} (\bibinfo {year} {2002})}\BibitemShut {NoStop}%
\bibitem [{\citenamefont {Abanin}\ and\ \citenamefont
  {Demler}(2012)}]{abanin:2012}%
  \BibitemOpen
  \bibfield  {author} {\bibinfo {author} {\bibfnamefont {D.~A.}\ \bibnamefont
  {Abanin}}\ and\ \bibinfo {author} {\bibfnamefont {E.}~\bibnamefont
  {Demler}},\ }\bibfield  {title} {\bibinfo {title} {{Measuring entanglement
  entropy of a generic many-body system with a quantum switch}},\ }\href@noop
  {} {\bibfield  {journal} {\bibinfo  {journal} {Physical Review Letters}\
  }\textbf {\bibinfo {volume} {109}},\ \bibinfo {pages} {020504} (\bibinfo
  {year} {2012})}\BibitemShut {NoStop}%
\bibitem [{\citenamefont {Nguyen}\ \emph {et~al.}(2021)\citenamefont {Nguyen},
  \citenamefont {Tseng}, \citenamefont {Maslennikov}, \citenamefont {Gan},\
  and\ \citenamefont {Matsukevich}}]{nguyen:2021}%
  \BibitemOpen
  \bibfield  {author} {\bibinfo {author} {\bibfnamefont {C.-H.}\ \bibnamefont
  {Nguyen}}, \bibinfo {author} {\bibfnamefont {K.-W.}\ \bibnamefont {Tseng}},
  \bibinfo {author} {\bibfnamefont {G.}~\bibnamefont {Maslennikov}}, \bibinfo
  {author} {\bibfnamefont {H.~C.~J.}\ \bibnamefont {Gan}},\ and\ \bibinfo
  {author} {\bibfnamefont {D.}~\bibnamefont {Matsukevich}},\ }\bibfield
  {title} {\bibinfo {title} {{Experimental SWAP test of infinite dimensional
  quantum states}},\ }\href@noop {} {\bibfield  {journal} {\bibinfo  {journal}
  {arXiv preprint arXiv:2103.10219}\ } (\bibinfo {year} {2021})}\BibitemShut
  {NoStop}%
\bibitem [{\citenamefont {Carrasco}\ \emph {et~al.}(2021)\citenamefont
  {Carrasco}, \citenamefont {Elben}, \citenamefont {Kokail}, \citenamefont
  {Kraus},\ and\ \citenamefont {Zoller}}]{carrasco:2021}%
  \BibitemOpen
  \bibfield  {author} {\bibinfo {author} {\bibfnamefont {J.}~\bibnamefont
  {Carrasco}}, \bibinfo {author} {\bibfnamefont {A.}~\bibnamefont {Elben}},
  \bibinfo {author} {\bibfnamefont {C.}~\bibnamefont {Kokail}}, \bibinfo
  {author} {\bibfnamefont {B.}~\bibnamefont {Kraus}},\ and\ \bibinfo {author}
  {\bibfnamefont {P.}~\bibnamefont {Zoller}},\ }\bibfield  {title} {\bibinfo
  {title} {{Theoretical and experimental perspectives of quantum
  verification}},\ }\href {https://doi.org/10.1103/PRXQuantum.2.010102}
  {\bibfield  {journal} {\bibinfo  {journal} {PRX Quantum}\ }\textbf {\bibinfo
  {volume} {2}},\ \bibinfo {pages} {010102} (\bibinfo {year}
  {2021})}\BibitemShut {NoStop}%
\bibitem [{\citenamefont {Patel}\ \emph {et~al.}(2016)\citenamefont {Patel},
  \citenamefont {Ho}, \citenamefont {Ferreyrol}, \citenamefont {Ralph},\ and\
  \citenamefont {Pryde}}]{patel:2016}%
  \BibitemOpen
  \bibfield  {author} {\bibinfo {author} {\bibfnamefont {R.~B.}\ \bibnamefont
  {Patel}}, \bibinfo {author} {\bibfnamefont {J.}~\bibnamefont {Ho}}, \bibinfo
  {author} {\bibfnamefont {F.}~\bibnamefont {Ferreyrol}}, \bibinfo {author}
  {\bibfnamefont {T.~C.}\ \bibnamefont {Ralph}},\ and\ \bibinfo {author}
  {\bibfnamefont {G.~J.}\ \bibnamefont {Pryde}},\ }\bibfield  {title} {\bibinfo
  {title} {{A quantum Fredkin gate}},\ }\href@noop {} {\bibfield  {journal}
  {\bibinfo  {journal} {Science advances}\ }\textbf {\bibinfo {volume} {2}},\
  \bibinfo {pages} {e1501531} (\bibinfo {year} {2016})}\BibitemShut {NoStop}%
\bibitem [{\citenamefont {Schwinger}(1965)}]{schwinger:1965}%
  \BibitemOpen
  \bibfield  {author} {\bibinfo {author} {\bibfnamefont {J.}~\bibnamefont
  {Schwinger}},\ }\href@noop {} {\emph {\bibinfo {title} {{On angular momentum
  1952, reprinted in Quantum theory of angular momentum, ed. L. C. Biedenharm
  and H. Van Dam}}}}\ (\bibinfo  {publisher} {Academic Press, NY},\ \bibinfo
  {year} {1965})\BibitemShut {NoStop}%
\bibitem [{\citenamefont {Tsunoda}\ \emph {et~al.}(2022)\citenamefont
  {Tsunoda}, \citenamefont {Teoh}, \citenamefont {Kalfus}, \citenamefont
  {de~Graaf}, \citenamefont {Chapman}, \citenamefont {Curtis}, \citenamefont
  {Thakur}, \citenamefont {Girvin},\ and\ \citenamefont
  {Schoelkopf}}]{tsunoda:2022}%
  \BibitemOpen
  \bibfield  {author} {\bibinfo {author} {\bibfnamefont {T.}~\bibnamefont
  {Tsunoda}}, \bibinfo {author} {\bibfnamefont {J.~D.}\ \bibnamefont {Teoh}},
  \bibinfo {author} {\bibfnamefont {W.~D.}\ \bibnamefont {Kalfus}}, \bibinfo
  {author} {\bibfnamefont {S.~J.}\ \bibnamefont {de~Graaf}}, \bibinfo {author}
  {\bibfnamefont {B.~J.}\ \bibnamefont {Chapman}}, \bibinfo {author}
  {\bibfnamefont {J.~C.}\ \bibnamefont {Curtis}}, \bibinfo {author}
  {\bibfnamefont {N.}~\bibnamefont {Thakur}}, \bibinfo {author} {\bibfnamefont
  {S.~M.}\ \bibnamefont {Girvin}},\ and\ \bibinfo {author} {\bibfnamefont
  {R.~J.}\ \bibnamefont {Schoelkopf}},\ }\href
  {https://doi.org/10.48550/ARXIV.2212.11196} {\bibinfo {title}
  {Error-detectable bosonic entangling gates with a noisy ancilla}} (\bibinfo
  {year} {2022})\BibitemShut {NoStop}%
\bibitem [{\citenamefont {Vlastakis}\ \emph {et~al.}(2015)\citenamefont
  {Vlastakis}, \citenamefont {Petrenko}, \citenamefont {Ofek}, \citenamefont
  {Sun}, \citenamefont {Leghtas}, \citenamefont {Sliwa}, \citenamefont {Liu},
  \citenamefont {Hatridge}, \citenamefont {Blumoff}, \citenamefont {Frunzio}
  \emph {et~al.}}]{vlastakis:2015}%
  \BibitemOpen
  \bibfield  {author} {\bibinfo {author} {\bibfnamefont {B.}~\bibnamefont
  {Vlastakis}}, \bibinfo {author} {\bibfnamefont {A.}~\bibnamefont {Petrenko}},
  \bibinfo {author} {\bibfnamefont {N.}~\bibnamefont {Ofek}}, \bibinfo {author}
  {\bibfnamefont {L.}~\bibnamefont {Sun}}, \bibinfo {author} {\bibfnamefont
  {Z.}~\bibnamefont {Leghtas}}, \bibinfo {author} {\bibfnamefont
  {K.}~\bibnamefont {Sliwa}}, \bibinfo {author} {\bibfnamefont
  {Y.}~\bibnamefont {Liu}}, \bibinfo {author} {\bibfnamefont {M.}~\bibnamefont
  {Hatridge}}, \bibinfo {author} {\bibfnamefont {J.}~\bibnamefont {Blumoff}},
  \bibinfo {author} {\bibfnamefont {L.}~\bibnamefont {Frunzio}}, \emph
  {et~al.},\ }\bibfield  {title} {\bibinfo {title} {{Characterizing
  entanglement of an artificial atom and a cavity cat state with Bell’s
  inequality}},\ }\href@noop {} {\bibfield  {journal} {\bibinfo  {journal}
  {Nature Communications}\ }\textbf {\bibinfo {volume} {6}},\ \bibinfo {pages}
  {8970} (\bibinfo {year} {2015})}\BibitemShut {NoStop}%
\bibitem [{\citenamefont {Khaneja}\ \emph {et~al.}(2005)\citenamefont
  {Khaneja}, \citenamefont {Reiss}, \citenamefont {Kehlet}, \citenamefont
  {Schulte-Herbr{\"u}ggen},\ and\ \citenamefont {Glaser}}]{khaneja2005}%
  \BibitemOpen
  \bibfield  {author} {\bibinfo {author} {\bibfnamefont {N.}~\bibnamefont
  {Khaneja}}, \bibinfo {author} {\bibfnamefont {T.}~\bibnamefont {Reiss}},
  \bibinfo {author} {\bibfnamefont {C.}~\bibnamefont {Kehlet}}, \bibinfo
  {author} {\bibfnamefont {T.}~\bibnamefont {Schulte-Herbr{\"u}ggen}},\ and\
  \bibinfo {author} {\bibfnamefont {S.~J.}\ \bibnamefont {Glaser}},\ }\bibfield
   {title} {\bibinfo {title} {{Optimal control of coupled spin dynamics: design
  of NMR pulse sequences by gradient ascent algorithms}},\ }\href@noop {}
  {\bibfield  {journal} {\bibinfo  {journal} {Journal of magnetic resonance}\
  }\textbf {\bibinfo {volume} {172}},\ \bibinfo {pages} {296} (\bibinfo {year}
  {2005})}\BibitemShut {NoStop}%
\bibitem [{\citenamefont {Geen}\ and\ \citenamefont
  {Freeman}(1991)}]{geen1991}%
  \BibitemOpen
  \bibfield  {author} {\bibinfo {author} {\bibfnamefont {H.}~\bibnamefont
  {Geen}}\ and\ \bibinfo {author} {\bibfnamefont {R.}~\bibnamefont {Freeman}},\
  }\bibfield  {title} {\bibinfo {title} {{Band-selective radiofrequency
  pulses}},\ }\href@noop {} {\bibfield  {journal} {\bibinfo  {journal} {Journal
  of Magnetic Resonance}\ }\textbf {\bibinfo {volume} {93}},\ \bibinfo {pages}
  {93} (\bibinfo {year} {1991})}\BibitemShut {NoStop}%
\bibitem [{\citenamefont {Teoh}\ \emph {et~al.}(2022)\citenamefont {Teoh},
  \citenamefont {Winkel}, \citenamefont {Babla}, \citenamefont {Chapman},
  \citenamefont {Claes}, \citenamefont {de~Graaf}, \citenamefont {Garmon},
  \citenamefont {Kalfus}, \citenamefont {Lu}, \citenamefont {Maiti},
  \citenamefont {Sahay}, \citenamefont {Thakur}, \citenamefont {Tsunoda},
  \citenamefont {Xue}, \citenamefont {Frunzio}, \citenamefont {Girvin},
  \citenamefont {Puri},\ and\ \citenamefont {Schoelkopf}}]{teoh:2022}%
  \BibitemOpen
  \bibfield  {author} {\bibinfo {author} {\bibfnamefont {J.~D.}\ \bibnamefont
  {Teoh}}, \bibinfo {author} {\bibfnamefont {P.}~\bibnamefont {Winkel}},
  \bibinfo {author} {\bibfnamefont {H.~K.}\ \bibnamefont {Babla}}, \bibinfo
  {author} {\bibfnamefont {B.~J.}\ \bibnamefont {Chapman}}, \bibinfo {author}
  {\bibfnamefont {J.}~\bibnamefont {Claes}}, \bibinfo {author} {\bibfnamefont
  {S.~J.}\ \bibnamefont {de~Graaf}}, \bibinfo {author} {\bibfnamefont
  {J.~W.~O.}\ \bibnamefont {Garmon}}, \bibinfo {author} {\bibfnamefont {W.~D.}\
  \bibnamefont {Kalfus}}, \bibinfo {author} {\bibfnamefont {Y.}~\bibnamefont
  {Lu}}, \bibinfo {author} {\bibfnamefont {A.}~\bibnamefont {Maiti}}, \bibinfo
  {author} {\bibfnamefont {K.}~\bibnamefont {Sahay}}, \bibinfo {author}
  {\bibfnamefont {N.}~\bibnamefont {Thakur}}, \bibinfo {author} {\bibfnamefont
  {T.}~\bibnamefont {Tsunoda}}, \bibinfo {author} {\bibfnamefont {S.~H.}\
  \bibnamefont {Xue}}, \bibinfo {author} {\bibfnamefont {L.}~\bibnamefont
  {Frunzio}}, \bibinfo {author} {\bibfnamefont {S.~M.}\ \bibnamefont {Girvin}},
  \bibinfo {author} {\bibfnamefont {S.}~\bibnamefont {Puri}},\ and\ \bibinfo
  {author} {\bibfnamefont {R.~J.}\ \bibnamefont {Schoelkopf}},\ }\href@noop {}
  {\bibinfo {title} {Dual-rail encoding with superconducting cavities}}
  (\bibinfo {year} {2022}),\ \Eprint {https://arxiv.org/abs/2212.12077}
  {arXiv:2212.12077 [quant-ph]} \BibitemShut {NoStop}%
\bibitem [{\citenamefont {Pietik{\"a}inen}\ \emph {et~al.}(2021)\citenamefont
  {Pietik{\"a}inen}, \citenamefont {{\v{C}}ernot{\'\i}k}, \citenamefont {Puri},
  \citenamefont {Filip},\ and\ \citenamefont {Girvin}}]{pietikainen:2021}%
  \BibitemOpen
  \bibfield  {author} {\bibinfo {author} {\bibfnamefont {I.}~\bibnamefont
  {Pietik{\"a}inen}}, \bibinfo {author} {\bibfnamefont {O.}~\bibnamefont
  {{\v{C}}ernot{\'\i}k}}, \bibinfo {author} {\bibfnamefont {S.}~\bibnamefont
  {Puri}}, \bibinfo {author} {\bibfnamefont {R.}~\bibnamefont {Filip}},\ and\
  \bibinfo {author} {\bibfnamefont {S.~M.}\ \bibnamefont {Girvin}},\ }\bibfield
   {title} {\bibinfo {title} {{Ancilla-error-transparent controlled beam
  splitter gate}},\ }\href@noop {} {\bibfield  {journal} {\bibinfo  {journal}
  {arXiv preprint arXiv:2112.04375}\ } (\bibinfo {year} {2021})}\BibitemShut
  {NoStop}%
\bibitem [{\citenamefont {Ambegaokar}\ and\ \citenamefont
  {Baratoff}(1963)}]{Ambegaokar:1963}%
  \BibitemOpen
  \bibfield  {author} {\bibinfo {author} {\bibfnamefont {V.}~\bibnamefont
  {Ambegaokar}}\ and\ \bibinfo {author} {\bibfnamefont {A.}~\bibnamefont
  {Baratoff}},\ }\bibfield  {title} {\bibinfo {title} {{Tunneling between
  superconductors}},\ }\href {https://doi.org/10.1103/PhysRevLett.10.486}
  {\bibfield  {journal} {\bibinfo  {journal} {Physical Review Letters}\
  }\textbf {\bibinfo {volume} {10}},\ \bibinfo {pages} {486} (\bibinfo {year}
  {1963})}\BibitemShut {NoStop}%
\bibitem [{\citenamefont {Tinkham}(2004)}]{tinkham:2004}%
  \BibitemOpen
  \bibfield  {author} {\bibinfo {author} {\bibfnamefont {M.}~\bibnamefont
  {Tinkham}},\ }\href@noop {} {\emph {\bibinfo {title} {{Introduction to
  superconductivity}}}}\ (\bibinfo  {publisher} {Courier Corporation},\
  \bibinfo {year} {2004})\BibitemShut {NoStop}%
\bibitem [{\citenamefont {Zimmerman}(1971)}]{zimmerman:1971}%
  \BibitemOpen
  \bibfield  {author} {\bibinfo {author} {\bibfnamefont {J.~E.}\ \bibnamefont
  {Zimmerman}},\ }\bibfield  {title} {\bibinfo {title} {{Sensitivity
  enhancement of superconducting quantum interference devices through the use
  of fractional‐turn loops}},\ }\href@noop {} {\bibfield  {journal} {\bibinfo
   {journal} {Journal of Applied Physics}\ }\textbf {\bibinfo {volume} {42}},\
  \bibinfo {pages} {4483} (\bibinfo {year} {1971})}\BibitemShut {NoStop}%
\bibitem [{\citenamefont {Martinis}\ \emph {et~al.}(2003)\citenamefont
  {Martinis}, \citenamefont {Nam}, \citenamefont {Aumentado}, \citenamefont
  {Lang},\ and\ \citenamefont {Urbina}}]{martinis:2003}%
  \BibitemOpen
  \bibfield  {author} {\bibinfo {author} {\bibfnamefont {J.~M.}\ \bibnamefont
  {Martinis}}, \bibinfo {author} {\bibfnamefont {S.}~\bibnamefont {Nam}},
  \bibinfo {author} {\bibfnamefont {J.}~\bibnamefont {Aumentado}}, \bibinfo
  {author} {\bibfnamefont {K.~M.}\ \bibnamefont {Lang}},\ and\ \bibinfo
  {author} {\bibfnamefont {C.}~\bibnamefont {Urbina}},\ }\bibfield  {title}
  {\bibinfo {title} {{Decoherence of a superconducting qubit due to bias
  noise}},\ }\href@noop {} {\bibfield  {journal} {\bibinfo  {journal} {Physical
  Review B}\ }\textbf {\bibinfo {volume} {67}},\ \bibinfo {pages} {094510}
  (\bibinfo {year} {2003})}\BibitemShut {NoStop}%
\bibitem [{\citenamefont {You}\ \emph {et~al.}(2019)\citenamefont {You},
  \citenamefont {Sauls},\ and\ \citenamefont {Koch}}]{you:2019}%
  \BibitemOpen
  \bibfield  {author} {\bibinfo {author} {\bibfnamefont {X.}~\bibnamefont
  {You}}, \bibinfo {author} {\bibfnamefont {J.~A.}\ \bibnamefont {Sauls}},\
  and\ \bibinfo {author} {\bibfnamefont {J.}~\bibnamefont {Koch}},\ }\bibfield
  {title} {\bibinfo {title} {{Circuit quantization in the presence of
  time-dependent external flux}},\ }\href
  {https://doi.org/10.1103/PhysRevB.99.174512} {\bibfield  {journal} {\bibinfo
  {journal} {Physical Review B}\ }\textbf {\bibinfo {volume} {99}},\ \bibinfo
  {pages} {174512} (\bibinfo {year} {2019})}\BibitemShut {NoStop}%
\bibitem [{\citenamefont {Sivak}\ \emph {et~al.}(2020)\citenamefont {Sivak},
  \citenamefont {Shankar}, \citenamefont {Liu}, \citenamefont {Aumentado},\
  and\ \citenamefont {Devoret}}]{sivak:2020}%
  \BibitemOpen
  \bibfield  {author} {\bibinfo {author} {\bibfnamefont {V.~V.}\ \bibnamefont
  {Sivak}}, \bibinfo {author} {\bibfnamefont {S.}~\bibnamefont {Shankar}},
  \bibinfo {author} {\bibfnamefont {G.}~\bibnamefont {Liu}}, \bibinfo {author}
  {\bibfnamefont {J.}~\bibnamefont {Aumentado}},\ and\ \bibinfo {author}
  {\bibfnamefont {M.~H.}\ \bibnamefont {Devoret}},\ }\bibfield  {title}
  {\bibinfo {title} {{Josephson array-mode parametric amplifier}},\ }\href
  {https://doi.org/10.1103/PhysRevApplied.13.024014} {\bibfield  {journal}
  {\bibinfo  {journal} {Physical Review Applied}\ }\textbf {\bibinfo {volume}
  {13}},\ \bibinfo {pages} {024014} (\bibinfo {year} {2020})}\BibitemShut
  {NoStop}%
\bibitem [{\citenamefont {Leghtas}\ \emph {et~al.}(2015)\citenamefont
  {Leghtas}, \citenamefont {Touzard}, \citenamefont {Pop}, \citenamefont {Kou},
  \citenamefont {Vlastakis}, \citenamefont {Petrenko}, \citenamefont {Sliwa},
  \citenamefont {Narla}, \citenamefont {Shankar}, \citenamefont {Hatridge}
  \emph {et~al.}}]{leghtas:2015}%
  \BibitemOpen
  \bibfield  {author} {\bibinfo {author} {\bibfnamefont {Z.}~\bibnamefont
  {Leghtas}}, \bibinfo {author} {\bibfnamefont {S.}~\bibnamefont {Touzard}},
  \bibinfo {author} {\bibfnamefont {I.~M.}\ \bibnamefont {Pop}}, \bibinfo
  {author} {\bibfnamefont {A.}~\bibnamefont {Kou}}, \bibinfo {author}
  {\bibfnamefont {B.}~\bibnamefont {Vlastakis}}, \bibinfo {author}
  {\bibfnamefont {A.}~\bibnamefont {Petrenko}}, \bibinfo {author}
  {\bibfnamefont {K.~M.}\ \bibnamefont {Sliwa}}, \bibinfo {author}
  {\bibfnamefont {A.}~\bibnamefont {Narla}}, \bibinfo {author} {\bibfnamefont
  {S.}~\bibnamefont {Shankar}}, \bibinfo {author} {\bibfnamefont {M.~J.}\
  \bibnamefont {Hatridge}}, \emph {et~al.},\ }\bibfield  {title} {\bibinfo
  {title} {{Confining the state of light to a quantum manifold by engineered
  two-photon loss}},\ }\href@noop {} {\bibfield  {journal} {\bibinfo  {journal}
  {Science}\ }\textbf {\bibinfo {volume} {347}},\ \bibinfo {pages} {853}
  (\bibinfo {year} {2015})}\BibitemShut {NoStop}%
\bibitem [{\citenamefont {Minev}\ \emph {et~al.}(2021)\citenamefont {Minev},
  \citenamefont {Leghtas}, \citenamefont {Mundhada}, \citenamefont
  {Christakis}, \citenamefont {Pop},\ and\ \citenamefont
  {Devoret}}]{minev:2021}%
  \BibitemOpen
  \bibfield  {author} {\bibinfo {author} {\bibfnamefont {Z.~K.}\ \bibnamefont
  {Minev}}, \bibinfo {author} {\bibfnamefont {Z.}~\bibnamefont {Leghtas}},
  \bibinfo {author} {\bibfnamefont {S.~O.}\ \bibnamefont {Mundhada}}, \bibinfo
  {author} {\bibfnamefont {L.}~\bibnamefont {Christakis}}, \bibinfo {author}
  {\bibfnamefont {I.~M.}\ \bibnamefont {Pop}},\ and\ \bibinfo {author}
  {\bibfnamefont {M.~H.}\ \bibnamefont {Devoret}},\ }\bibfield  {title}
  {\bibinfo {title} {{Energy-participation quantization of Josephson
  circuits}},\ }\href@noop {} {\bibfield  {journal} {\bibinfo  {journal} {npj
  Quantum Information}\ }\textbf {\bibinfo {volume} {7}},\ \bibinfo {pages} {1}
  (\bibinfo {year} {2021})}\BibitemShut {NoStop}%
\bibitem [{\citenamefont {Reagor}\ \emph {et~al.}(2016)\citenamefont {Reagor},
  \citenamefont {Pfaff}, \citenamefont {Axline}, \citenamefont {Heeres},
  \citenamefont {Ofek}, \citenamefont {Sliwa}, \citenamefont {Holland},
  \citenamefont {Wang}, \citenamefont {Blumoff}, \citenamefont {Chou},
  \citenamefont {Hatridge}, \citenamefont {Frunzio}, \citenamefont {Devoret},
  \citenamefont {Jiang},\ and\ \citenamefont {Schoelkopf}}]{reagor:2016}%
  \BibitemOpen
  \bibfield  {author} {\bibinfo {author} {\bibfnamefont {M.}~\bibnamefont
  {Reagor}}, \bibinfo {author} {\bibfnamefont {W.}~\bibnamefont {Pfaff}},
  \bibinfo {author} {\bibfnamefont {C.}~\bibnamefont {Axline}}, \bibinfo
  {author} {\bibfnamefont {R.~W.}\ \bibnamefont {Heeres}}, \bibinfo {author}
  {\bibfnamefont {N.}~\bibnamefont {Ofek}}, \bibinfo {author} {\bibfnamefont
  {K.}~\bibnamefont {Sliwa}}, \bibinfo {author} {\bibfnamefont
  {E.}~\bibnamefont {Holland}}, \bibinfo {author} {\bibfnamefont
  {C.}~\bibnamefont {Wang}}, \bibinfo {author} {\bibfnamefont {J.}~\bibnamefont
  {Blumoff}}, \bibinfo {author} {\bibfnamefont {K.}~\bibnamefont {Chou}},
  \bibinfo {author} {\bibfnamefont {M.~J.}\ \bibnamefont {Hatridge}}, \bibinfo
  {author} {\bibfnamefont {L.}~\bibnamefont {Frunzio}}, \bibinfo {author}
  {\bibfnamefont {M.~H.}\ \bibnamefont {Devoret}}, \bibinfo {author}
  {\bibfnamefont {L.}~\bibnamefont {Jiang}},\ and\ \bibinfo {author}
  {\bibfnamefont {R.~J.}\ \bibnamefont {Schoelkopf}},\ }\bibfield  {title}
  {\bibinfo {title} {{Quantum memory with millisecond coherence in circuit
  QED}},\ }\href {https://doi.org/10.1103/PhysRevB.94.014506} {\bibfield
  {journal} {\bibinfo  {journal} {Physical Review B}\ }\textbf {\bibinfo
  {volume} {94}},\ \bibinfo {pages} {014506} (\bibinfo {year}
  {2016})}\BibitemShut {NoStop}%
\end{thebibliography}
\end{document}